\documentclass{aa}
\usepackage {graphicx}

\def\vt  {$v_ {t}$}

\begin {document}


\title {First stars V - Abundance patterns from C to Zn 
and supernova yields in the early Galaxy
\thanks{Based on observations obtained in the 
frame of the ESO programme ID  165.N-0276(A). This work has made use of 
the SIMBAD database.}
}

\author {
R. Cayrel\inst{1}\and
E. Depagne\inst{1}\and 
M. Spite\inst{1}\and 
V. Hill\inst{1}\and
F. Spite\inst{1}\and
P. Fran\c cois\inst{1}\and
B. Plez\inst{2}\and
T. Beers\inst{3}\and
F. Primas\inst{4}\and
J. Andersen\inst{5,9}\and
B. Barbuy\inst{6}\and
P. Bonifacio\inst{7}\and
P. Molaro\inst{7}\and
B. Nordstr\"{o}m\inst{5,8}
}

\offprints {monique.spite@obspm.fr}

\institute {
  GEPI, Observatoire de Paris-Meudon, F-92125 Meudon Cedex, 
  France,
  \and
  GRAAL, Universit\'e de Montpellier II, F-34095 Montpellier Cedex 05, 
  France,
  \and
  Department of Physics \& Astronomy, Michigan State University, East Lansing,
  MI 48824 USA,
  \and
  European Southern Observatory (ESO), Karl Schwarschild-Str. 2, D-85749
  Garching b. M\"unchen, Germany,
  \and
  Astronomical Observatory, NBIfAFG, Juliane Maries Vej 30, DK-2100 Copenhagen,
  Denmark,
  \and
  IAG, Universidade de Sao Paulo, Departamento de Astronomia, CP 3386,
  01060-970 Sao Paulo, Brazil, 
  \and
  Osservatorio Astronomico di Trieste, INAF,
  Via G.B. Tiepolo 11, I-34131 Trieste, Italy,
  \and
  Lund Observatory, Box 43, SE-221 00 Lund, Sweden,
  \and
  Nordic Optical Telescope Scientific Association, Apartado 474, ES-38 700 Santa 
Cruz de La Palma, Spain.
}

\date {Received XXX; accepted XXX}
\titlerunning {Abundances from C to Zn in extremely metal-poor giants}
\authorrunning {Cayrel et al.}
\abstract{

In the framework of the ESO Large Programme ``First Stars'',
very high-quality spectra of some 70 very metal-poor dwarfs and giants
were obtained with the ESO VLT
and UVES spectrograph. These stars are likely to have descended from the first 
generation(s) of stars formed after the Big Bang, and their detailed composition 
provides constraints on issues such as the nature of the first supernovae, the 
efficiency of mixing processes in the early Galaxy, the formation and evolution 
of the halo of the Galaxy, and the possible sources of reionization of the 
Universe.  This paper presents the abundance analysis of an homogeneous sample 
of 35 giants selected from the HK survey of Beers et al. (1992; 1999), 
emphasizing stars of extremely low metallicity: 30 of our 35 stars are in the 
range $-2.7 <{\rm [Fe/H]}< -4.1$, and 22 stars have ${\rm [Fe/H]} < -3.0$. Our 
new VLT/UVES spectra, at a resolving power of $R\sim45,000$ and with signal-to-noise 
ratios of 100-200 per pixel over the wavelength range 330 -- 1000 nm, are 
greatly superior to those of the classic studies of McWilliam et al. (1995) and 
Ryan, Norris, \& Beers (1996).

The immediate objective of the work is to determine precise, comprehensive, and 
homogeneous element abundances for this large sample of the most metal-poor 
giants presently known. In the analysis we combine the spectral line modeling 
code ``Turbospectrum'' with OSMARCS model atmospheres, which treat continuum 
scattering correctly and thus allow proper interpretation of the blue regions of 
the spectra, where scattering becomes important relative to continuous 
absorption ($\lambda < 400$~nm). We obtain detailed information 
on the trends of elemental abundance ratios and the star-to-star scatter around 
those trends, enabling us to separate the relative contributions of cosmic 
scatter and observational/analysis errors. 

Abundances of 17 elements from C to Zn have been measured in all stars, 
including K and Zn, which have not previously been detected in stars with [Fe/H] 
$< -$3.0. Among the key results, we discuss the oxygen abundance (from the 
forbidden [OI] line), the different and sometimes complex trends of the 
abundance ratios with metallicity, the very tight relationship between the 
abundances of certain elements (e.g., Fe and Cr), and the high [Zn/Fe] ratio in 
the most metal-poor stars.  Within the error bars, the trends of the abundance 
ratios with metallicity are consistent with those found in earlier literature, 
but in many cases the scatter around the average trends is {\it much} smaller 
than found in earlier studies, which were limited to lower-quality spectra. We 
find that the cosmic scatter in several element ratios may be as low as 
0.05~dex.
 
The evolution of the abundance trends and scatter with declining metallicity 
provides strong constraints on the yields of the first supernovae and their 
mixing into the early ISM. The abundance ratios found in our sample {\it do not} 
match the predicted yields from pair-instability hypernovae, but are consistent
with element production by supernovae with progenitor masses up to 100
M$_{\sun}$. Moreover, the composition of the ejecta that have enriched the 
matter now contained in our very metal-poor stars appears surprisingly uniform 
over the range $-4.0 \le {\rm [Fe/H]} < -3.0$. This would indicate either that 
we are observing the products of very similar primordial bursts of high-mass 
stars, or that the mixing of matter from different bursts of early star 
formation was extremely rapid. In any case, it is unlikely that we observed the 
ejecta from individual (single) supernovae (as has often been concluded in 
previous work), as we do not see  scatter due to different progenitor masses. 
The abundance ratios at the lowest metallicities ($-4.0 \le {\rm [Fe/H]} \le 
-3.0$) are compatible with those found by McWilliam et al. (1995) and later 
studies. However,  when elemental ratios are plotted with respect to Mg, we
find no clear slopes below [Mg/H] = --3, but rather, a plateau-like
behaviour defining a set of initial yields.

}

\maketitle
%
\section {Introduction} 

The early chemical evolution of the Galaxy is recorded in the elemental
abundances in the atmospheres of its low-mass, extremely metal-poor (XMP) stars. 
In the present Galaxy such stars are quite rare, especially the most 
metal-deficient examples. In particular, no star completely without heavy metals 
(a Pop III star) has been observed to date, although the recent discovery of a 
star with [Fe/H] = --5.3 (HE~0107-5240; Christlieb et al. 2002) proves that 
extreme examples of Pop II stars can still be found.

One simple explanation for the present lack of true zero-metallicity stars
would be the early production of substantial amounts of metals by very massive,
primitive zero-metal objects (Pop III stars). The lack of metals in such objects 
suggests that they should have formed with an Initial Mass Function (IMF) very
different from that observed at present, either biased towards higher masses 
(e.g. Omukai \& Nishi \cite{ON99}; Bromm, Coppi, \& Larson \cite{BCL99}), or 
with a bimodal shape (Nakamura \& Umemura \cite{NU00}). The existence of 
zero-metal, very massive stars is postulated because such objects are able to 
avoid the huge radiation pressure-driven mass loss predicted for very massive 
stars with significant metal content (e.g., Larson \cite{L99}; Abel, Bryan, \& 
Norman \cite{ABN00}; Baraffe, Heger \& Woosley \cite{HW02}). Such stars are
expected to play a role not only in the first episodes of heavy-element
nucleosynthesis, but also in the early reionization of the Universe (Kogut et
al. \cite{KSB03}). It remains possible, however, that the first stars included
substantial numbers of more classical O-type stars (up to $30-60 M_{\odot}$).

According to existing models, the heavy-element yields produced by these two 
varieties of progenitor stars, and their ability to expel these elements into 
the ISM, differ from one another in a number of ways. 
For example, a strong ``odd/even'' effect and low [Zn/Fe] ratios are expected
in the ejecta of very massive objects exploding as pair-instability
supernovae (PISN), contrary to what is expected to emerge from lower mass,
classical supernovae. Hence, precise elemental abundance ratios in extremely
metal-poor stars should provide a powerful means to discriminate between these
two kinds of ``first stars.''

A main aim of the present programme is to obtain precise determinations of the
elemental abundances in extremely metal-poor stars, since these abundances
reflect the yields of the first supernovae -- perhaps even of a single one 
according to Audouze \& Silk (\cite{AS95}) and Ryan et al. (\cite{RNB96});
see also Shigeyama \& Tsujimoto (\cite{ST98}), Nakamura et al. (\cite{NUN99}),
Chieffi \& Limongi (\cite{CL02}), and Umeda \& Nomoto (\cite{UN02}). However, 
whether or not they are associated with single supernovae, precise abundances 
provide very useful constraints on model yields of the first supernovae, which 
are not yet well understood.

The most reliable information is clearly to be obtained from a homogeneous and
systematic determination of elemental abundances in large samples of such stars,
so that reliable trends of the abundance ratios with metallicity may
be determined. Such trends may then be interpreted in terms of variable (or 
constant) yields as a function of time, of the progenitor masses, and/or of
the metallicity of the ISM in the early Galaxy. Moreover, high-quality data and
a careful, consistent analysis reduce the contribution of systematic and random
errors to the star-to-star scatter of the derived abundance ratios, enabling a 
much better estimate of their intrinsic (cosmic) scatter and thereby 
constraining the efficiency of the mixing processes in the primitive halo. 

Even after decades of dedicated searches, the number of XMP stars that are
sufficiently bright to be studied at sufficiently high spectral resolution, even
with large telescopes, remains small. The present paper reports observations of 
the first half of a sample of roughly 70 XMP candidates, including both turn-off 
and giant stars and selected from the HK survey of Beers and colleagues (Beers 
et al. \cite{BPS92}; Beers 1999).

Several papers have already been published on particularly interesting 
individual stars from this programme: Hill et al. (\cite{HPC02}), Depagne et al. 
(\cite{DHS02}), Fran\c cois et al. (\cite{FDH03}), and Sivarani et al. (\cite{SBM03}). 
In contrast, we discuss here the derived element abundances, from carbon to 
zinc, for our entire homogeneous sample of 35 very metal-poor giants. Among 
these, 22 have metallicities ${{\rm [Fe/H]} < -3.0}$ and thus qualify as XMP 
stars. 

We have carried out an analysis in a systematic and homogeneous
way, based on the highest-quality data obtained to date. Compared to previous
work (e.g. McWilliam et al. \cite{MPS95}; Ryan et al. 1996), our spectra cover a 
substantially larger wavelength range at much higher spectral resolution and S/N 
ratios, allowing for a significant leap forward in the accuracy of the derived 
elemental abundances (Sect. 2). These abundances were derived with particular 
care from the spectra, supplemented by new photometric data in several colours 
and using state-of-the-art model atmospheres (Sect. 3). Moreover, we study 
important elements, such as O, K, and Zn which were not analyzed in previous 
works. The derived elemental abundances and abundance ratios are presented in 
Sect. 4, the results are discussed in section 5, and conclusions are drawn in 
section 6.

\begin {table*}[t]
\caption {Log of the UVES observations. The S/N ratio per pixel is 
given for 
three representative wavelengths.  (Due to the large number of pixels
in each resolution element, the S/N ratios of the table need to be
multiplied by a factor of 1.3 in order to compare them to the values
available in the literature)}
\begin {center}
\begin {tabular}{lrrcrrrrrr}
\label {tab-Obs}
Star Name & & & Slit&\multicolumn{3}{c} {Total Exposure Time} \\
        &   & Date of& Width&     Blue  & Yellow& Red   & S/N& S/N& S/N\\
        & V & Observation & " &396nm & 573nm & 850nm & 400nm&510nm&630nm\\
\hline
~1 HD~2796       & 8.51& Oct 2000 & 1.0 & 1800  &  1300 & 400  & 250& 
390& 550\\
~2 HD~122563     & 6.20& Jul 2000 & 1.0 &       &       &      & 250& 
430& 670\\
~3 HD~186478     & 9.18& Oct 2000 & 1.0 &  800  &   400 & 400  &    
&    &    \\
~4 BD~+17:3248   & 9.37& Oct 2000 & 1.0 & 2700  &  2700 & 1200 & 160& 
290& 310\\
~~ ~~~~~~--      &     & Jun 2001 &  &       &       &      &    &    
&    \\
~~ ~~~~~~--      &     & Sep 2001 &  &       &       &      &    &    
&    \\   
~5 BD~--18:5550  & 9.35& Oct 2000 & 1.0 & 1800  &  1200 & 600  & 220& 
410& 630\\
~~ ~~~~~~--      &     & Sep 2001 &  &       &       &      &    &    
&    \\ 
~6 CD~--38:245   &12.01& Jul 2000 & 1.0 & 7200  &  3600 & 3600 & 150& 
150& 200\\
~~ ~~~~~~--      &     & Aug 2000 &  &       &       &      &    &    
&    \\ 
~7 BS~16467--062 &14.09& Jun 2001 & 1.0 & 3600  &  3600 &      &  90& 
140& 170\\
~~ ~~~~~~--      &     & Jul 2001 &  & 7200  &  3600 & 3600 &    &    
&    \\
~8 BS~16477--003 &14.22& Jun 2001 & 1.0 & 14400 &  7200 & 7200 &  90& 
130& 170\\
~9 BS~17569--049 &13.36& Jun 2001 & 1.0 & 9600  &  6600 & 3000 & 120& 
170& 260\\
10 CS~22169--035 &12.88& Oct 2000 & 1.0 & 7200  &  3600 & 3600 & 150& 
210& 280\\
11 CS~22172--002 &12.73& Oct 2000 & 1.0 & 7494  &  3600 & 3900 & 130& 
200& 330\\
12 CS~22186--025 &14.24& Oct 2001 & 1.0 &10800  &  7200 & 3600 &  95& 
140& 190\\
13 CS~22189--009 &14.04& Oct 2000 & 1.0 & 7200  &  3600 & 3600 &  90& 
150& 120\\
14 CS~22873--055 &12.65& May 2001 & 1.0 & 7200  &  3600 & 3600 & 140& 
150& 200\\
~~ ~~~~~~--      &     & Sep 2001 &  &       &       &      &    &    
&    \\
15 CS~22873--166 &11.82& Oct 2000 & 1.0 & 5400  &  2700 & 2700 & 160& 
240& 300\\
16 CS~22878--101 &13.73& Jul 2000 & 1.0 & 14400 &  7200 & 7200 &  85& 
100& 120\\
17 CS~22885--096 &13.33& Jul 2000 & 1.0 & 15835 &  9184 & 6600 & 160& 
250& 410\\
~~ ~~~~~~--      &     & Aug 2000 &  &       &       &      &    &    
&    \\ 
18 CS~22891--209 &12.17& Oct 2000 & 1.0 &  5400 &  2700 & 2700 & 160& 
200& 350\\
19 CS~22892--052 &13.18& Sep 2001 & 1.0 &  7200 &  3600 & 3600 & 140& 
130& 190\\
20 CS~22896--154 &13.64& Oct 2000 & 1.0 & 12600 &  7200 & 5400 & 110& 
230& 200\\
21 CS~22897--008 &13.33& Oct 2000 & 1.0 & 10800 &  5400 & 5400 & 100& 
170& 180\\
22 CS~22948--066 &13.47& Sep 2001 & 1.0 &  7200 &  3600 & 3600 & 100& 
130& 130\\
23 CS~22949--037 &14.36& Aug 2000 & 1.0 & 30000 & 19200 &10800 & 110& 
180& 170\\
~~ ~~~~~~--      &     & Sep 2001 &  &       &       &      &    &    
&    \\
24 CS~22952--015 &13.28& Oct 2000 & 1.0 & 10200 &  4800 & 5400 & 150& 
220& 250\\
25 CS~22953--003 &13.72& Sep 2001 & 1.0 & 13500 &  9900 & 3600 & 140& 
160& 210\\
26 CS~22956--050 &14.27& Sep 2001 & 1.0 &  9000 &  5400 & 3600 &  
75&  95& 130\\
27 CS~22966--057 &14.32& Sep 2001 & 1.0 &  9000 &  5400 & 3600 &  80& 
105& 120\\
28 CS~22968--014 &13.72& Oct 2000 & 1.0 & 14100 &  8700 & 5400 & 150& 
220& 240\\
29 CS~29491--053 &12.92& Oct 2001 & 1.0 &  5800 &  2900 & 2900 & 140& 
205& 230\\
30 CS~29495--041 &13.34& Jun 2001 & 1.0 &  7200 &  3600 & 3600 & 115& 
130& 170\\
~~ ~~~~~~--      &     & Sep 2001 &  &       &       &      &    &    
&    \\
31 CS~29502--042 &12.71& Oct 2000 & 1.0 & 13500 &  9900 & 3600 & 290& 
310& 330\\
~~ ~~~~~~--      &     & Sep 2001 &  &       &       &      &    &    
&    \\
32 CS-29516--024 &13.59& Jun 2001 & 1.0 &  3600 &  3600 &      & 140& 
205& 230\\
33 CS~29518--051 &13.02& Oct 2000 & 1.0 &  7200 &  3600 & 3600 & 100& 
150& 190\\
34 CS~30325--094 &12.33& Jul 2000 & 1.0 &  7200 &  6300 & 3600 & 110& 
220& 280\\
~~ ~~~~~~--      &     & Aug 2000 &  &       &       &      &    &    
&    \\
35 CS~31082--001 &11.70& Aug 2000 & 1.0 &  2400 &  1200 & 1200 & -* & 
-* & -* \\
~~ ~~~~~~--      &     & Aug 2000 & 0.45&  6000 &  3000 & 3000 &    
&    &    \\
~~ ~~~~~~--      &     & Oct 2000 & 0.45& 25200 & 10800 & 14400&    
&    &    \\ 
\hline
\multicolumn{8}{l}{* for CS~31082--001 the details of the 
observations 
are given in Hill et al.(2002)}\\
\end {tabular}
\end {center}
\end {table*}

\section {Observations and reductions}

The observations were performed during several runs from April 2000 to November
2001 with the VLT-UT2 and the high-resolution spectrograph UVES (Dekker et al.
\cite{DD00}). Details are presented in Table \ref{tab-Obs}. Accurate coordinates
for the brighter stars can be found in the SIMBAD database (http:
//simbad.u-strasbg.fr/); those for other stars are given in Table
\ref{tab-coor}. In this paper the names of the stars have been shortened to, for
example, CS~XXXXX--XXX instead of BPS~CS~XXXXX--XXX, where BPS is the SIMBAD
abbreviation for the catalogue of Beers, Preston, \& Shectman.
Several stars of our sample have duplicate names; the second name is
indicated in Table \ref{tab-coor} in italics .

\begin {table}[t]
\caption {Precise coordinates of the BPS programme stars.
Four of our stars have duplicate names. The second name
is given in italics. 
The coordinates of the HD and BD stars
of our sample can be found in SIMBAD.}
\begin {center}
\begin {tabular}{crrrr}
\label {tab-coor}
   & Star Name     & $\alpha (2000)$ & $\delta (2000)$\\
\hline
7  & BS~16467--062  & 13:42:00.63 & $+$17:48:40.8 \\
   & {\it BS~16934--060}  &  --~~~~~~   &     --~~~~~   \\
8  & BS~16477--003  & 14:32:56.91 & $+$06:46:06.9 \\
   & {\it CS 30317--084}  &  --~~~~~~   &     --~~~~~   \\
9  & BS~17569--049  & 22:04:58.36 & $+$04:01:32.1 \\
10 & CS~22169--035  & 04:12:13.88 & $-$12:05:05.0 \\
11 & CS~22172--002  & 03:14:20.84 & $-$10:35:11.2 \\

12 & CS~22186--025  & 04:24:32.80 & $-$37:09:02.5 \\
13 & CS~22189--009  & 02:41:42.37 & $-$13:28:10.5 \\
14 & CS~22873--055  & 19:53:49.78 & $-$59:40:00.1 \\
15 & CS~22873--166  & 20:19:22.02 & $-$61:30:14.9 \\
16 & CS~22878--101  & 16:45:31.44 & $+$08:14:45.4 \\

17 & CS~22885--096  & 20:20:51.17 & $-$39:53:30.1 \\
18 & CS~22891--209  & 19:42:02.16 & $-$61:03:44.6 \\
19 & CS~22892--052  & 22:17:01.65 & $-$16:39:27.1 \\
20 & CS~22896--154  & 19:42:26.88 & $-$56:58:34.0 \\
21 & CS~22897--008  & 21:03:11.85 & $-$65:05:08.8 \\

22 & CS~22948--066  & 21:44:51.17 & $-$37:27:54.9 \\
   & {\it CS 30343--064}  &  --~~~~~~   &     --~~~~~   \\
23 & CS~22949--037  & 23:26:29.80 & $-$02:39:57.9 \\
24 & CS~22952--015  & 23:37:28.69 & $-$05:47:56.6 \\
25 & CS~22953--003  & 01:02:15.85 & $-$61:43:45.8 \\
26 & CS~22956--050  & 21:58:05.83 & $-$65:13:27.1 \\

27 & CS~22966--057  & 23:48:57.76 & $-$29:39:22.8 \\
28 & CS~22968--014  & 03:06:29.50 & $-$54:30:32.5 \\
29 & CS~29491--053  & 22:36:56.30 & $-$28:31:06.4 \\
30 & CS~29495--041  & 21:36:33.27 & $-$28:18:48.5 \\
31 & CS~29502--042  & 22:21:48.82 & $+$02:28:44.8 \\
   & {\it CS 29516--041}  &  --~~~~~~   &     --~~~~~   \\
32 & CS~29516--024  & 22:26:15.35 & $+$02:51:46.2 \\
33 & CS~29518--051  & 01:24:10.01 & $-$28:15:21.0 \\
34 & CS~30325--094  & 14:54:39.27 & $+$04:21:38.0 \\
35 & CS~31082--001  & 01:29:31.13 & $-$16:00:45.4 \\
\hline
\end {tabular}
\end {center}
\end {table}

A dichroic beam-splitter was used for all of the observations, permitting the
use of both arms of the spectrograph simultaneously; the blue arm was centered
at 396nm and the red arm at either 573 or 850nm. The resulting spectral coverage
is almost complete from 330 nm to 1000 nm. The entrance slit, 1" on the sky,
yielded a resolving power of $R \approx 47,000$ at 400 nm and 43,000 at 630 nm. 
The S/N ratios per pixel at different wavelengths are summarized in Table
\ref{tab-Obs}. Since there are $\approx$ 5 pixels per resolution element, these
values should be multiplied by a factor 2.2 in order to obtain the S/N ratios
per resolution element (and by 1.3 when comparing them to S/N values in the 
literature, as most other spectrographs have only 3 pixels per resolution 
element). 

Norris et al. (\cite{NRB01}) defined a ``figure of merit,'' F, which is
useful for comparing the quality of high-resolution spectroscopic observations,
assuming the integrated signal from observed spectral features is made with the
same number of pixels. They suggest that, in order to achieve significant
progress in issues of importance for Galactic chemical evolution, spectra should
ideally be obtained with F larger than 500. The observations presented herein 
have figures of merit, F, in the blue (400 nm) between 850 and 3250, and in the 
red (630 nm) between 650 and 2350 (F is much higher for the two bright stars 
HD~122563 and BD-18:5550, which have been analysed several times in the 
literature and were observed with particular care to check for possible 
systematic errors).

The $r$-process enhanced, very metal-poor star CS~31082--001 was observed with 
slightly different settings and slit widths to obtain higher spectral resolution 
and complete coverage in the blue. The details of the observations for this star 
are given in Hill et al. (\cite{HPC02}).

\begin {figure*}
\begin {center}
\resizebox  {16.cm}{5.7cm} 
{\includegraphics {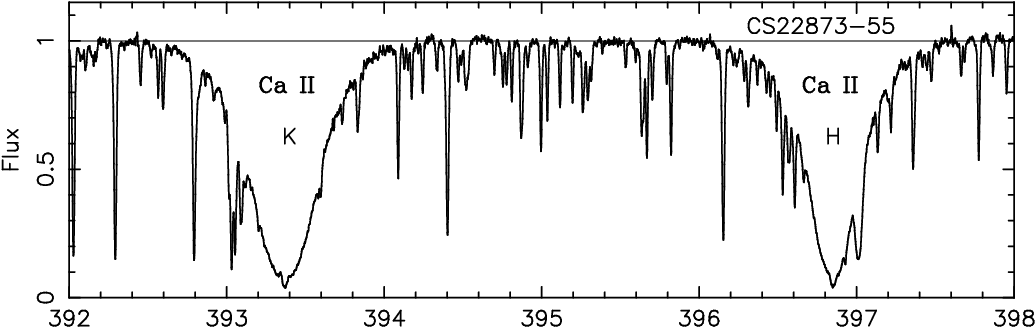} }
\resizebox  {16.cm}{5.7cm} 
{\includegraphics {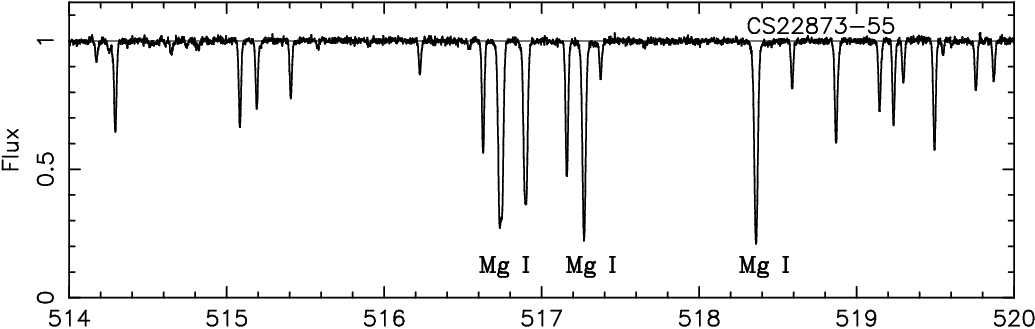} }
\resizebox  {16.cm}{5.7cm} 
{\includegraphics {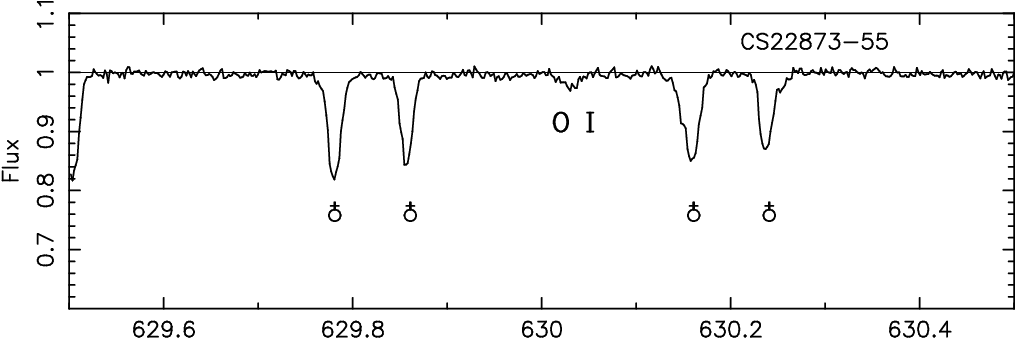} }
\caption {An example of the reduced spectra in the region of the Ca II H and K lines, the Mg I triplet, and in the region of the forbidden oxygen line. The abscissa is the wavelength in nm.  Telluric lines are indicated.}
\label {spectra}
\end {center}
\end {figure*}

The spectra were reduced using the UVES context (Ballester et al. \cite{BMB00}) within MIDAS, which performs
bias and inter-order background subtraction
(object and flat-field), optimal extraction of the object (rejecting cosmic ray
hits), division by a flat-field frame extracted with the same optimally weighted
profile as the object, wavelength calibration and rebinning to a constant value,
and merging of all overlapping orders. The spectra were then co-added and
finally normalized to unity in the continuum. For the reddest spectra (centered
at 850 nm), instead of correcting the image by the extracted flat-field, the
object frame was divided by the flat-field frame pixel-by-pixel (in 2D, before
extraction), which yields a better correction of the interference fringes that
appear in these frames.
An example of the spectra is given in Fig. \ref{spectra}.

\subsection {Equivalent widths}

In most cases the equivalent widths (EWs) of individual lines were measured
by Gaussian fitting and then employed to determine the abundances of the 
different elements. The equivalent widths of the lines for each star are 
available as an electronic file. In the cases of elements which suffer from 
hyperfine structure and/or molecular bands and blends, the abundances have been 
directly determined by spectral synthesis.

\begin {figure}
\begin {center}
\resizebox  {6.cm}{6.0cm} 
{\includegraphics {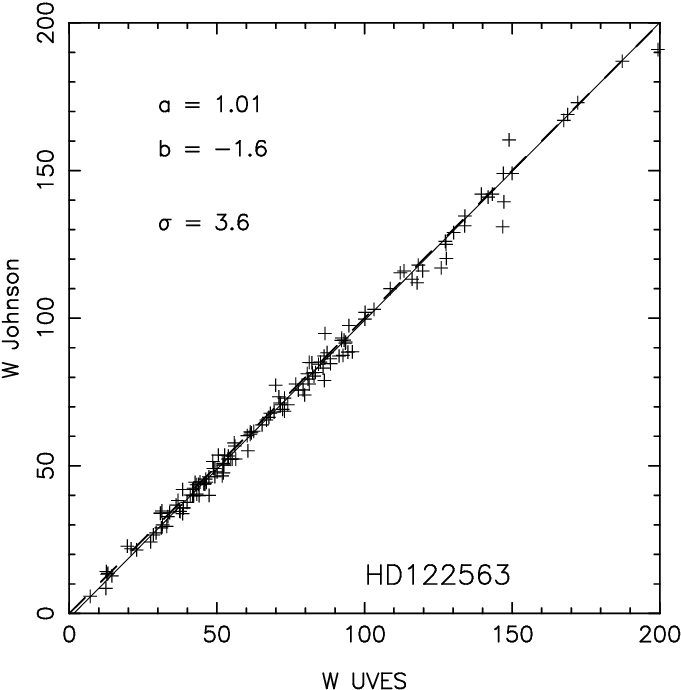}} 
\resizebox  {6.cm}{6.0cm} 
{\includegraphics {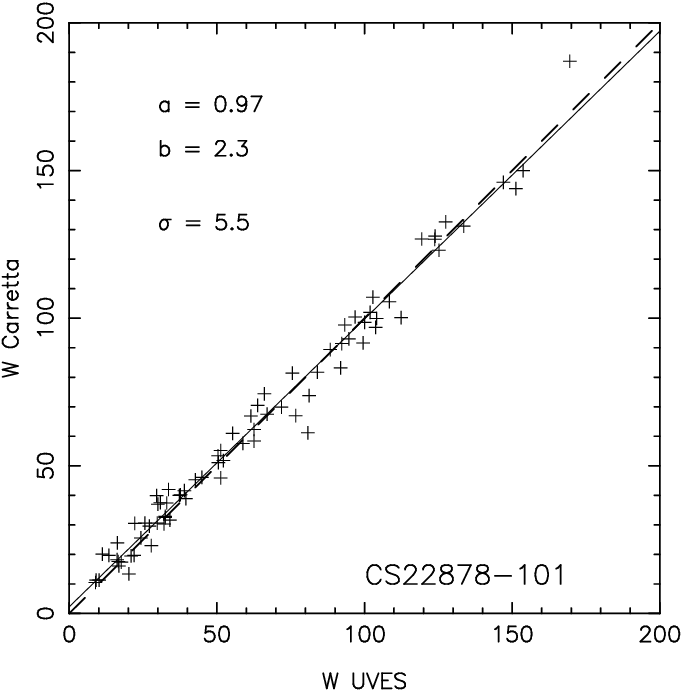}} 
\resizebox  {6.cm}{6.0cm} 
{\includegraphics {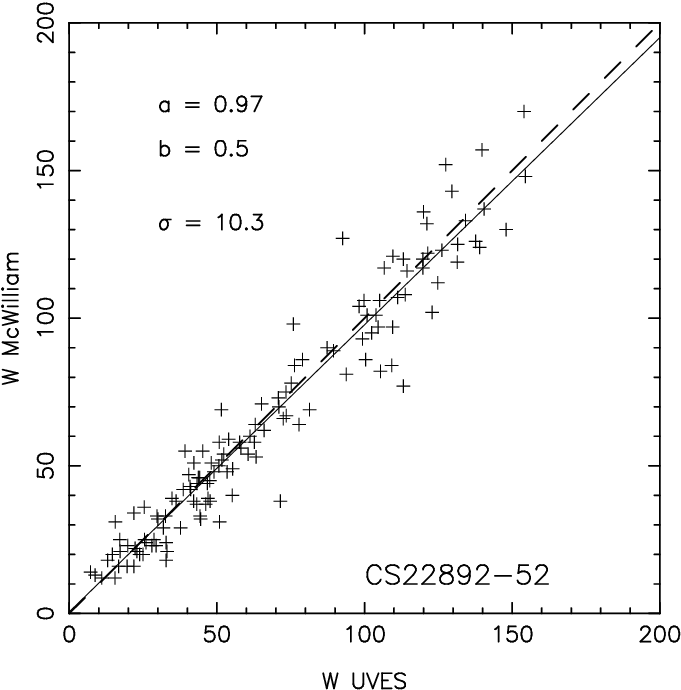}}
\caption {Comparison between our EWs and those of   Johnson (2002),
 Carretta et al. (2002), and McWilliam {\it et al.} (1995).
The one-to-one relation is shown by a dashed line, the mean 
curve (least squares) by a full thin line. The value of the slope (a) 
and zero point shift (b) of the regression are given, as well as the 
standard deviation of the fit ($\sigma$). }
\label {compw}
\end {center}
\end {figure}

In Fig. \ref{compw} we compare our measured EWs for stars in common with
several recent spectroscopic studies, e.g., McWilliam et al. (\cite{MPS95}),
Carretta et al. (\cite{CGC02}), and Johnson (\cite{JJ02}). The quality of
Johnson's spectra is similar to ours, and the agreement between the two sets of
measurements is excellent (standard deviation 3.6\,m\AA\ for HD~122563, and only
2.2\,m\AA\ when restricting the comparison to lines with EW $< 30$\,m\AA). The
agreement with the data of Carretta et al. is also quite good (standard 
deviation 5.5\,m\AA\ for CS~22878--101). When our data are compared to the 
equivalent widths of McWilliam et al., however, the standard deviation is 
larger, 10\,m\AA\ for CS~22892--052, presumably due to the much lower resolution 
and S/N ratio of the McWilliam et al. spectra ($R=22,000$ and $S/N=36$). 

The mean difference between our EWs and those reported in the literature is generally 
quite small; the regression line between our data and those of Johnson, Carretta 
et al., or McWilliam et al. has a slope close to one, with deviations always 
less than 3\% and a very small zero-point shift.

The expected uncertainty in the measured equivalent widths can be estimated from
Cayrel's formula (\cite{Cay88}) : $$ \sigma_{W} = {1.5 \over {S/N} } \sqrt {FWHM
* \delta x} $$ where S/N is the signal-to-noise ratio per pixel, FWHM is the
full width of the line at half maximum, and $\delta x$ the pixel size. The
predicted accuracy, $\sigma_{W}$, is 0.4m\AA\ for a typical S/N ratio of 150 and
only 0.3m\AA\ for a S/N ratio of 200. These are also the weakest lines which can
be detected in the spectra. However, it should be noted that this formula
neglects the uncertainty on the continuum placement, as well as the uncertainty
in the determination of the FWHM of the lines. 

We estimate that, using
homogeneous procedures for the determination of the continua and the line
widths, the statistical error for weak lines is of the order of 0.6--1.0m\AA,
depending on the S/N ratio of the spectrum and the level of crowding. Since
the lines used in our abundance analysis are generally very weak, the error on
the abundance determination depends almost linearly on the error of the measured
equivalent widths.

\section{Analysis and determination of the stellar parameters}

\begin {table*}[t]
\caption {Photometry and derived temperatures for the programme stars. 
The {\it V-I} and {\it V-R} colours are on the Johnson system,
while {\it J-K} and {\it V-K} have been transformed to the TCS system  
(see Alonso et al. 2001).}
\begin {center}
\begin {tabular}{lrrrrrrrrrrrrrrrrr}
\label {tab-Phot}
           &       &        & Teff&        & Teff&        & 
Teff&        &Teff&        & Teff& Adopted\\
~~~~Star Name &$E_{B-V}$& B-V$_{o}$&B-V&  V-R$_{o}$&V-R& 
J-K$_{o}$&J-K& V-K$_{o}$&V-K& V-I$_{o}$&V-I& Teff\\ 
\hline
~1 HD~2796        &  0.03&   0.71& 5072&   0.68& 4999&   0.54&  
4907&   2.23& 4902&      &     &  4950\\
~2 HD~122563      &  0.00&   0.90& 4653&   0.81& 4586&   0.61&  
4657&   2.51& 4574&      &     &  4600\\
~3 HD~186478      &  0.09&   0.84& 4726&   0.76& 4690&   0.58&  
4757&   2.26& 4811&      &     &  4700\\
 & \\
~4 BD+17:3248    &  0.06&   0.60& 5386&   0.59& 5238&   0.47&  
5142&   1.89& 5240&      &     &  5250\\
~5 BD-18:5550    &  0.08&   0.77& 4823&   0.76& 4709&   0.58&  
4745&   2.58& 4520&      &     &  4750\\
~6 CD-38:245     &  0.00&   0.76& 4841&   0.73& 4806&   0.58&  
4739&   2.36& 4712&  1.30& 4700&  4800\\
 & \\
~7 BS~16467--062   &  0.00&   0.60& 5352&   0.60& 5234&   0.44&  
5278&   1.89& 5284&  1.08& 5120&  5200\\
~8 BS~16477--003   &  0.01&   0.75& 4869&   0.68& 5004&   0.53&  
4937&   2.24& 4878&  1.23& 4848&  4900\\
~9 BS~17569--049 &  0.03&   0.86& 4718&       &     &   0.58&  
4732&   2.42& 4662&      &     &  4700\\
 & \\
10 CS~22169--035 &  0.02&   0.87& 4706&       &     &   0.59&  
4717&   2.47& 4617&  1.43& 4501&  4700\\
11 CS~22172--002 &  0.06&   0.75& 4854&       &     &   0.50&  
5034&   2.24& 4846&  1.26& 4770&  4800\\
12 CS~22186--025 &  0.01&   0.73& 4880&   0.70& 4887&   0.49&  
5087&   2.16& 4935&  1.21& 4855&  4900\\
 & \\
13 CS~22189--009 &  0.02&   0.70& 4917&   0.68& 4968&   0.53&  
4901&   2.18& 4913&      &     &  4900\\
14 CS~22873--055 &  0.03&   0.90& 4670&   0.81& 4577&   0.58&  
4738&   2.56& 4537&  1.45& 4473&  4550\\
15 CS~22873--166 &  0.03&   0.94& 4623&   0.83& 4542&   0.63&  
4595&   2.61& 4495&      &     &  4550\\
 & \\
16 CS~22878--101 &  0.06&   0.78& 4816&   0.69& 4933&   0.55&  
4848&   2.32& 4757&  1.25& 4792&  4800\\
17 CS~22885--096 &  0.03&   0.66& 5146&   0.67& 4987&   0.48&  
5114&   2.15& 4949&      &     &  5050\\
18 CS~22891--209 &  0.05&       &     &   0.76& 4732&   0.54&  
4894&   2.42& 4667&      &     &  4700\\
 & \\
19 CS~22892--052 &  0.00&   0.78& 4816&   0.69& 4921&   0.52&  
4967&   2.25& 4837&  1.27& 4761&  4850\\
20 CS~22896--154 &  0.04&   0.58& 5416&   0.60& 5238&   0.47&  
5167&   1.99& 5161&      &     &  5250\\
21 CS~22897--008 &  0.00&   0.69& 5052&   0.69& 4917&   0.55&  
4860&   2.33& 4750&      &     &  4900\\
 & \\
22 CS~22948--066 &  0.00&   0.63& 5243&   0.63& 5117&   0.50&  
5047&   2.02& 5111&  1.14& 4986&  5100\\
23 CS~22949--037 &  0.02&   0.72& 4887&   0.70& 4901&   0.52&  
4964&   2.22& 4874&      &     &  4900\\
24 CS~22952--015 &  0.01&   0.77& 4845&   0.73& 4820&   0.56&  
4845&   2.34& 4763&      &     &  4800\\
 & \\
25 CS~22953--003 &  0.00&   0.67& 5114&   0.64& 5088&   0.49&  
5084&   2.07& 5046&      &     &  5100\\
26 CS~22956--050 &  0.00&   0.68& 5083&       &     &   0.53&  
4911&   2.25& 4836&      &     &  4900\\
27 CS~22966--057 &  0.00&   0.61& 5314&   0.57& 5346&   0.42&  
5371&   1.78& 5427&  1.04& 5202&  5300\\
 & \\
28 CS~22968--014 &  0.00&   0.72& 4887&   0.69& 4917&   0.51&  
4981&   2.27& 4812&      &     &  4850\\
29 CS~29491--053 &  0.00&   0.84& 4742&       &     &   0.57&  
4787&   2.39& 4683&      &     &  4700\\
30 CS~29495--041 &  0.00&   0.81& 4778&       &     &   0.54&  
4866&   2.35& 4721&      &     &  4800\\
 & \\
31 CS~29502--042 &  0.00&   0.68& 5083&       &     &   0.45&  
5217&   2.05& 5073&      &     &  5100\\
32 CS~29516--024 &  0.06&   0.89& 4751&   0.88& 4430&   0.58&  
4744&   2.45& 4636&  1.44& 4486&  4650\\
33 CS~29518--051 &  0.00&   0.64& 5223&   0.61& 5208&   0.46&  
5173&   1.97& 5182&  1.09& 5099&  5200\\
 & \\
34 CS~30325--094 &  0.02&   0.70& 4919&        &     &   0.48& 
5100&   2.15& 4951&  1.17& 4931&  4950\\
35 CS~31082--001 &  0.00&   0.77& 4822&  0.69  & 4932&   0.54& 
4877&   2.21& 4882&  1.23& 4826&  4825\\
\hline
\end {tabular}
\end {center}
\end {table*}

The abundance analysis was performed using the LTE spectral line analysis code
``Turbospectrum'' together with OSMARCS model atmospheres. The OSMARCS models 
were originally developed by Gustafsson et al. (\cite{GBE75}) and have been
constantly improved and updated through the years by Plez et al. (\cite{PBN92}),
Edvardsson et al. (\cite{EAG93}), and Asplund et al. (\cite{AGK97}).
For a description of the most recent improvements, and the coming grid, see Gustafsson et al. \cite{GEE03}. 
Turbospectrum is described by Alvarez \& Plez (\cite{AP98}), and has recently 
been improved, partly for this work, especially through the addition of a
module for abundance determinations from measured equivalent widths.

The abundances of the different elements have been determined mainly from the
measured equivalent widths of isolated, weak lines. Synthetic spectra have only 
been used to determine the abundance of C and N from the CH and CN molecular
bands, or when lines were severely blended (e.g., for silicon) or required 
corrections for hyperfine splitting  (e.g., for manganese).

\subsection{Determination of stellar parameters }\label{sec-pardet}

The temperatures of the programme stars were estimated from the observed colour
indices using Alonso et al.'s calibration for giants (Alonso et al. 
\cite{AAM99}). This calibration is based on the Infrared Flux Method (IRFM), 
which provides the coefficients used to convert colours to effective 
temperatures (T$_{eff}$).  

The colours of the stars have been taken from Beers et al. (2003, in
preparation), Alonso et al. (\cite{AAM98}), the 2MASS catalogue (J,H,K) 
(Skrutskie et al. \cite{2MASS97}; Finlator et al. \cite{2MASS00}), or the DENIS 
catalogue (I,J,K) (Epchtein et al. \cite {DENIS99}). 

The indices {\it V-R} and {\it V-I}, originally on the Cousins system, have been
transformed (Bessell \cite{Bes83}) onto the Johnson system adopted by Alonso et 
al. for use with the relations T$_{eff}$ vs. {\it V-R} and {\it V-I}. The 2MASS 
{\it J-K} and {\it V-K} indices were transformed onto the TCS (Telescopio Carlos 
Sanchez) system
through the ESO system (Carpenter \cite{Car01}), since Alonso et al. use this
system for these colours. For the ``CS'' or ``BS'' stars the colour
indices have been corrected for reddening following Beers et al. (\cite{Be99}),
who used Burstein \& Heiles (\cite{BH82}) values, corrected for distance. This $E_{(B-V)}$
value is systematically smaller than the value computed from the Schlegel et al.
(\cite{SFD98}) map by about 0.02 mags. Arce \& Goodman (\cite{AG99}) found that
the Schlegel et al. map overestimates the extinction in some regions, in
particular when the extinction is large. For the bright stars of the sample,
reddening was taken from Pilachowski et al. (\cite{PSK96}).

The adopted values of $E_{(B-V)}$ and the dereddened colours are listed in
Table \ref{tab-Phot}, along with the corresponding derived temperatures. These
values would be about 50K higher if the Schlegel et al. (\cite{SFD98}) values
for reddening had been adopted. We note that for CS~31082--001 the
temperature deduced from {\it (V-R)} is higher than the temperature found in 
Hill et al. (\cite{HPC02}), because in the latter paper the transformation {\it 
V-R} vs. temperature had been taken from McWilliam et al. (\cite{MPS95}).

The final temperatures adopted for our analysis are listed in Table
\ref{tab-Phot} (column 13); 82\% of the temperatures deduced from the different
colours are located inside the interval $\mathrm{T(adopted)\pm 100K}$. This
corresponds to a random error of about 80K ($1~\sigma$).

The microturbulent velocity, \vt, was derived from Fe~I lines in the traditional
manner, requiring that the abundance derived for individual lines be independent
of the equivalent width of the line. Finally, the surface gravity, log $g$, was 
determined by requiring that the Fe and Ti abundances derived from Fe~I and 
Fe~II, resp. Ti~I, Ti~II lines be identical. 

It should be noted that our log $g$ values may be affected by NLTE effects
(overionization) and by uncertainties in the oscillator strengths of the Fe and
Ti lines. Carretta et al. (\cite{CGC02}) used another method: they deduced the
gravity from isochrones (Yi et al. \cite{YDK01}), and found that the abundances
of iron deduced from Fe~I or Fe~II lines show ionization equilibrium within
0.2~dex (cf. their Table 1). For the one star we have in common with these
authors (CS~22878--101), we adopted the same effective temperature
(T$_{eff}$=4800K), and the agreement for log $g$ is excellent; in both cases 
log $g$ = 1.3 (Table \ref{tab-parmod}). Hence, these two independent methods 
provide similar results. 


The final model atmosphere parameters adopted for the stars are given in Table 
\ref{tab-parmod}.

\begin {table}[t]
\caption {Adopted model parameters (T$_{eff}$, log $g$, \vt, [Fe/H]$_{m}$) and 
final iron abundances [Fe/H]$_{c}$ for the programme stars.}
\label {tab-parmod}
\begin {center}
\begin {tabular}{lccccc}
Star     & T$_{eff}$ & log g& \vt   &[Fe/H]$_{m}$&
 [Fe/H]$_{c}$\\
\hline
~1 HD~2796        & 4950 & 1.5 & 2.1  &-2.4& -2.47\\
~2 HD~122563      & 4600 & 1.1 & 2.0  &-2.8& -2.82\\
~3 HD~186478      & 4700 & 1.3 & 2.0  &-2.6& -2.59\\

~4 BD~+17:3248    & 5250 & 1.4 & 1.5  &-2.0& -2.07\\
~5 BD~--18:5550   & 4750 & 1.4 & 1.8  &-3.0& -3.06\\
~6 CD~--38:245    & 4800 & 1.5 & 2.2  &-4.0& -4.19\\

~7 BS~16467--062  & 5200 & 2.5 & 1.6  &-4.0& -3.77\\
~8 BS~16477--003  & 4900 & 1.7 & 1.8  &-3.4& -3.36\\
~9 BS~17569--049  & 4700 & 1.2 & 1.9  &-3.0& -2.88\\

10 CS~22169--035  & 4700 & 1.2 & 2.2  &-3.0& -3.04\\
11 CS~22172--002  & 4800 & 1.3 & 2.2  &-4.0& -3.86\\
12 CS~22186--025  & 4900 & 1.5 & 2.0  &-3.0& -3.00\\

13 CS~22189--009  & 4900 & 1.7 & 1.9  &-3.5& -3.49\\
14 CS~22873--055  & 4550 & 0.7 & 2.2  &-3.0& -2.99\\
15 CS~22873--166  & 4550 & 0.9 & 2.1  &-3.0& -2.97\\

16 CS~22878--101  & 4800 & 1.3 & 2.0  &-3.0& -3.25\\
17 CS~22885--096  & 5050 & 2.6 & 1.8  &-4.0& -3.78\\
18 CS~22891--209  & 4700 & 1.0 & 2.1  &-3.0& -3.29\\

19 CS~22892--052  & 4850 & 1.6 & 1.9  &-3.0& -3.03\\
20 CS~22896--154  & 5250 & 2.7 & 1.2  &-2.7& -2.69\\
21 CS~22897--008  & 4900 & 1.7 & 2.0  &-3.5& -3.41\\

22 CS~22948--066  & 5100 & 1.8 & 2.0  &-3.0& -3.14\\
23 CS~22949--037  & 4900 & 1.5 & 1.8  &-4.0& -3.97\\
24 CS~22952--015  & 4800 & 1.3 & 2.1  &-3.4& -3.43\\

25 CS~22953--003  & 5100 & 2.3 & 1.7  &-3.0& -2.84\\
26 CS~22956--050  & 4900 & 1.7 & 1.8  &-3.3& -3.33\\
27 CS~22966--057  & 5300 & 2.2 & 1.4  &-2.6& -2.62\\

28 CS~22968--014  & 4850 & 1.7 & 1.9  &-3.5& -3.56\\
29 CS~29491--053  & 4700 & 1.3 & 2.0  &-3.0& -3.04\\
30 CS~29495--041  & 4800 & 1.5 & 1.8  &-2.8& -2.82\\

31 CS~29502--042  & 5100 & 2.5 & 1.5  &-3.0& -3.19\\
32 CS~29516--024  & 4650 & 1.2 & 1.7  &-3.0& -3.06\\
33 CS~29518--051  & 5100 & 2.4 & 1.4  &-2.8& -2.78\\

34 CS~30325--094  & 4950 & 2.0 & 1.5  &-3.4& -3.30\\
35 CS~31082--001  & 4825 & 1.5 & 1.8  &-2.9& -2.91\\
\hline
\end {tabular}
\end {center}
\end {table}

\begin {figure}
\begin {center}
\resizebox  {7.cm}{4.0cm} 
{\includegraphics {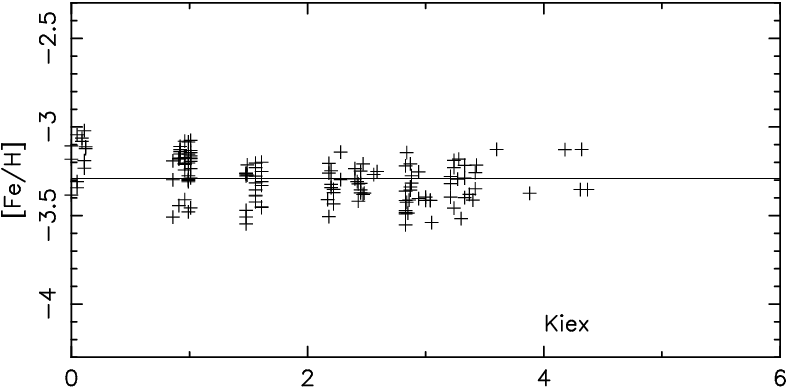}}
\resizebox  {7.cm}{4.0cm} 
{\includegraphics {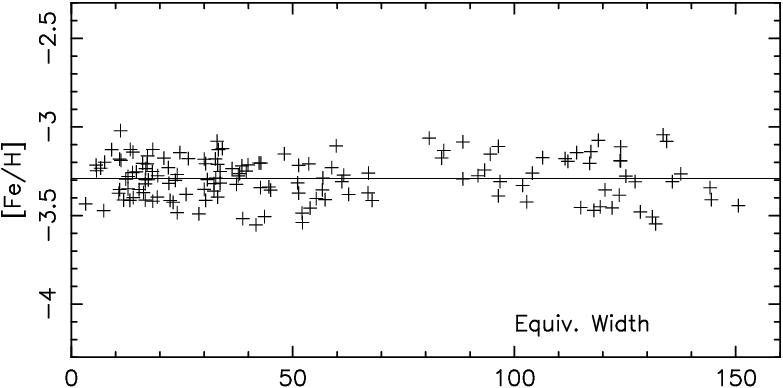}} 
\resizebox  {7.cm}{4.0cm} 
{\includegraphics {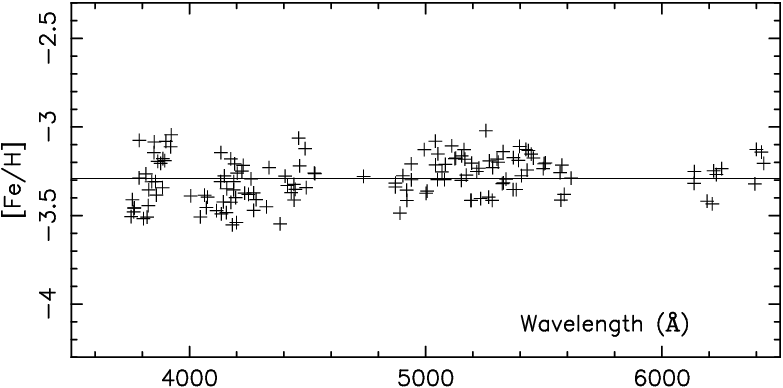}} 
\caption {Comparison of FeI abundance for CS~22878--101 vs. 
excitation potential, equivalent width, and wavelength.  The 
parameters adopted for the model are T$_{eff}$ = 4800K, log $g$ = 1.3, 
\vt = 2.0 km s$^{-1}$, and [Fe/H]$_m = -3.0$.  The line at 
[Fe/H] = --3.29 represents the mean value 
of the iron abundance deduced from the Fe~I lines.
}
\label {checkmod}
\end {center}
\end {figure}

\begin {table}[t]
\caption {Abundance uncertainties linked to stellar parameters.}
\label {errTemp1}
\begin {center}
\begin {tabular}{lrrrr}
\hline
\multicolumn {2}{l}{HD~122563}\\
\multicolumn {5}{c}{A: T$_{eff}$=4600K, log g=1.0 dex, vt=2.0 
km s$^{-1}$}\\
\multicolumn {5}{c}{B: T$_{eff}$=4600K, log g=0.9 dex, vt=2.0 
km s$^{-1}$}\\
\multicolumn {5}{c}{C: T$_{eff}$=4600K, log g=1.0 dex, vt=1.8 
km s$^{-1}$}\\
\multicolumn {5}{c}{D: T$_{eff}$=4500K, log g=1.0 dex, vt=2.0 
km s$^{-1}$}\\
\multicolumn {5}{c}{E: T$_{eff}$=4500K, log g=0.6 dex, vt=1.8 
km s$^{-1}$}\\
\hline
El.   & $\Delta_{B-A} $ & $\Delta_{C-A} $& $\Delta_{D-A} $& 
$\Delta_{E-A} $\\
\hline
[Fe/H]      &-0.00 & 0.06 &-0.09 &-0.06\\    
$[$O I/Fe]  &-0.03 &-0.06 & 0.04 &-0.12\\
$[$Na I/Fe] & 0.04 & 0.05 &-0.16 & 0.03\\
$[$Mg I/Fe] & 0.03 &-0.01 &-0.04 & 0.07\\
$[$Al I/Fe] & 0.04 & 0.04 &-0.13 & 0.08\\
$[$Si I/Fe] & 0.02 & 0.04 &-0.05 & 0.08\\
$[$K I/Fe]  & 0.02 &-0.04 &-0.02 & 0.01\\
$[$Ca I/Fe] & 0.02 &-0.03 & 0.00 & 0.05\\
$[$Sc II/Fe]&-0.02 & 0.02 & 0.04 &-0.01\\
$[$Ti I/Fe] & 0.02 &-0.01 &-0.09 &-0.00\\
$[$Ti II/Fe]&-0.02 & 0.02 & 0.04 & 0.01\\
$[$Cr I/Fe] & 0.02 & 0.05 &-0.13 & 0.03\\
$[$Mn I/Fe] & 0.03 & 0.07 &-0.18 &-0.03\\
$[$Fe I/Fe] & 0.03 & 0.03 &-0.11 & 0.03\\
$[$Fe II/Fe]&-0.03 &-0.04 & 0.11 &-0.03\\
$[$Co I/Fe] & 0.02 & 0.08 &-0.14 & 0.05\\
$[$Ni I/Fe] & 0.03 & 0.07 &-0.13 & 0.05\\
$[$Zn I/Fe] &-0.03 &-0.07 & 0.04 &-0.03\\
$[$O I/Mg I]&-0.06 &-0.05 &+0.07 &-0.19\\
$[$O/Mg]$^*$&-0.01 &+0.02 &-0.14 &-0.13\\
\hline
\multicolumn{5}{l}{Note: [O/Mg]$^*$ = [O~I/Fe~II]-[Mg~I/Fe~I]}\\
\end {tabular}
\end {center}
\end {table}

\begin {table}[t]
\caption {Abundance uncertainties linked to stellar parameters.}
\label {errTemp2}
\begin {center}
\begin {tabular}{lrrrr}
\hline
\multicolumn {2}{l}{CS~22948--066}\\
\multicolumn {5}{c}{A: T$_{eff}$=5100K, log g=1.8 dex, vt=2.0 
km s$^{-1}$}\\
\multicolumn {5}{c}{B: T$_{eff}$=5100K, log g=1.7 dex, vt=2.0 
km s$^{-1}$}\\
\multicolumn {5}{c}{C: T$_{eff}$=5100K, log g=1.8 dex, vt=1.8 
km s$^{-1}$}\\
\multicolumn {5}{c}{D: T$_{eff}$=5000K, log g=1.8 dex, vt=2.0 
km s$^{-1}$}\\
\multicolumn {5}{c}{E: T$_{eff}$=5000K, log g=1.5 dex, vt=2.0 
km s$^{-1}$}\\
\hline
El.   & $\Delta_{B-A} $ & $\Delta_{C-A} $& $\Delta_{D-A} $& 
$\Delta_{E-A} $\\
\hline
[Fe/H]      &-0.02 & 0.02 &-0.05 &-0.11\\
$[$O I/Fe]  &-0.01 &-0.02 &-0.02 &-0.05\\
$[$Na I/Fe] & 0.03 & 0.08 &-0.04 & 0.05\\
$[$Mg I/Fe] & 0.04 & 0.06 &-0.01 & 0.10\\
$[$Al I/Fe] & 0.03 & 0.05 &-0.04 & 0.04\\
$[$Si I/Fe] & 0.03 & 0.05 &-0.05 & 0.04\\
$[$K I/Fe]  & 0.02 &-0.02 &-0.03 & 0.05\\
$[$Ca I/Fe] & 0.02 &-0.01 &-0.02 & 0.05\\
$[$Sc II/Fe]&-0.01 & 0.02 & 0.00 &-0.03\\
$[$Ti I/Fe] & 0.02 &-0.02 &-0.07 &-0.01\\
$[$Ti II/Fe]&-0.01 & 0.03 & 0.00 &-0.03\\
$[$Cr I/Fe] & 0.03 & 0.03 &-0.06 & 0.02\\
$[$Mn I/Fe] & 0.03 & 0.05 &-0.06 & 0.02\\
$[$Fe I/Fe] & 0.02 & 0.03 &-0.06 & 0.02\\
$[$Fe II/Fe]&-0.02 &-0.02 & 0.06 &-0.01\\
$[$Co I/Fe] & 0.03 & 0.02 &-0.07 & 0.01\\
$[$Ni I/Fe] & 0.02 & 0.10 &-0.07 & 0.01\\
$[$Zn I/Fe] & 0.01 &-0.02 & 0.00 & 0.03\\
$[$O I/Mg I]&-0.05 &-0.08 &-0.01 &-0.15\\
$[$O/Mg]$^*$&-0.01 &-0.03 &-0.13 &-0.12\\
\hline
\multicolumn{5}{l}{Note: [O/Mg]$^*$ = [O~I/Fe~II]-[Mg~I/Fe~I]}\\
\end {tabular}
\end {center}
\end {table} 


\subsection{Validity checks}

To check the validity of the model parameters (T$_{eff}$, log g, \vt) we have 
plotted for all the Fe~I lines in each star (see Fig. \ref{checkmod}): {\it (i)}
the iron abundance as a function of the excitation potential of the line (to
check the adopted temperature and the importance of NLTE effects); {\it (ii)}
the abundance vs. the equivalent width of the line (to check on the
microturbulence velocity); {\it (iii)} the abundance vs. wavelength (as a
consistency check, which can shed light on problems linked to the synthetic
spectra computations).

Using the photometrically derived T$_{eff}$ we find no trend with the
excitation of Fe I lines (at least when only the lines with $\chi_{ex}>1$eV are
taken into account), contrary to what was reported by Johnson (2002) using
Kurucz models. Johnson also found, for the most metal-poor stars of her sample,
a trend of increasing abundance with decreasing wavelength. In a first approach
we saw the same effect, using a spectrum synthesis code that treats continuum
scattering as if it were absorption. The Turbospectrum code takes proper account
of continuum scattering, with the source function written as
$S_{\nu}=(\kappa_{\nu}\times B_{\nu} + \sigma_{\nu} \times J_{\nu})/
(\kappa_{\nu}+\sigma_{\nu})$, consistent with the OSMARCS code. With proper
treatment of continuum scattering we find no or much less correlation of
abundance with wavelength. 

Indeed, it turns out to be crucial to not approximate scattering by absorption. 
In our model with T$_{\rm eff}$ = 4600~K, log $g$ = 1.0, and [Fe/H] = --3, the 
ratio $\sigma/\kappa$ in the continuum opacity at the $\tau_{\lambda}=1$ level 
is 5.2 at $\lambda$ = 350 nm, whereas it is only 0.08 at $\lambda$ = 500 nm. At 
$\tau_{\lambda}=0.1$, these
numbers are 57 and 3.2 respectively. In the presence of significant scattering,
radiation in the continuum reflects the physical conditions of deeper, hotter
layers than those at $\tau=1$ (the $J_{\nu}$ part of the source function).
Neglecting this, which is equivalent to including scattering in the
absorption coefficient, results in too low a flux in the continuum, and thus too
weak spectral lines. 

In fact, for the model above, calculations show that the continuum flux with 
scattering included in absorption is only 55\% of its value with scattering at 
350 nm, 93\% at 500 nm). This forces the derived abundances towards higher 
values at the (short) wavelengths where scattering is important. {\it It is thus 
especially important to properly account for continuum scattering when studying 
metal-poor stellar spectra}! 

For completeness, we note that in some cases we even find a slight opposite 
trend (decreasing abundance at shorter wavelength), which we interpret as
due to NLTE effects in lines which are at least 
partly formed by scattering.

\subsection{Stellar parameter uncertainties and associated abundance 
uncertainties}\label{sec-errors}

For a given stellar temperature, the ionization equilibrium provides an estimate
of the stellar gravity with an internal accuracy of about 0.1 dex in log $g$,
and the microturbulence velocity \vt~ can be constrained within 0.2 km s$^{-1}$.
The largest uncertainties in the abundance determination arise in fact from the
uncertainty in the temperature of the stars.

First, the different indices give different temperature estimates; from Sect.
\ref{sec-pardet} we estimate that the corresponding error is about 80 K.
Another source of error is the estimation of the reddening. The error on
$E_{(B-V)}$ is about 0.02 mags, corresponding to a temperature error of about
60 K. Overall, we estimate that the total error on the adopted temperatures is 
on the order of 100 K. 
 
Tables \ref{errTemp1} and \ref{errTemp2} list the abundance uncertainties
arising from each of these three sources (log $g$, \vt, and T$_{eff}$) {\it
individually} (columns 2 to 4, from the comparison of models B, C and D to the
nominal model labeled A) for two stars which cover much of the parameter space 
of our sample: HD~122563 (T$_{eff}$ = 4600K, log $g$ = 1.0, \vt = 2.0 km s$^{-
1}$, and [Fe/H] = --2.8, Barbuy et al. \cite{BMS03}) and CS~22948--066
(T$_{eff}$ = 5100K, log $g$ = 1.8, \vt = 2.0 km s$^{-1}$, and [Fe/H] = --3.1).

Because gravity is determined from the ionization equilibrium, a variation of 
T$_{eff}$ will change log $g$ and also sometimes slightly influence \vt~ (as the 
strongest lines are statistically also those with the smaller excitation 
potentials, \vt~ is not totally independent of the adopted temperature). Hence, 
the total error budget is not the quadratic sum of the various sources of 
uncertainties, but contains significant covariance terms. 

As an illustration of the total expected uncertainty, we have computed the 
abundances in HD~122563 and CS~22948--066 with two models, one with the nominal 
temperature, gravity, and microturbulent velocity (model A) and another with a 
100K lower temperature, determining the corresponding ``best'' gravity and 
microturbulence values (model E). In HD~122563, log $g$ decreased by 0.4
dex and the \vt~ by 0.2 km s$^{-1}$, whereas for
CS~22948--066, log $g$ decreased by 0.3 dex while \vt~ required no change.

Tables \ref {errTemp1} and \ref {errTemp2} (column 5) show that the difference
in [Fe/H] between these two models amounts to $\sim$0.09 dex (0.06 dex for
HD~122563 and 0.11 dex for CS~22948--066), but the differences in the abundance
ratios are small ($< 0.07$ dex). In general, the model changes induce similar 
effects in the abundances of other elements and Fe, so that they largely cancel 
out in the ratio [X/Fe]. 

The lines of O and Mg behave rather differently from Fe, and large changes are 
found for the ratios [O/Fe] and [Mg/Fe]. Moreover, since the lines of Mg and O 
react in an opposite way to changes in the stellar parameters (gravity in 
particular), the ratio [O/Mg] as determined directly is particularly sensitive 
to these changes and hence is not a very robust result. However, using a 
slightly different definition of the [O/Mg] ratio (denoted as [O/Mg]$^*$ in 
Tables \ref{errTemp1} and \ref{errTemp2}), normalising O to Fe~II and Mg to Fe
I, i. e. [O/Mg]$^*$= [O/Fe~II] -- [Mg/Fe~I], makes it more robust against
uncertainties in the stellar gravity (column 2), but not in temperature
(column 4), so the overall uncertainty on the [O/Mg] ratio is still high (column 
5), up to 0.2 dex. Similar remarks apply to the ratio [O/Ca].

\section{Abundance results from C to Zn} \label{CZn-Fe}

The abundances of elements from C to Zn are presented for all the programme
stars in Tables \ref{tab-abund-a} to  \ref{tab-abund-e}. The 
abundances of elements heavier than Zn (such as Sr, Ba, etc.) will 
be discussed in a forthcoming paper.

\subsection{Carbon and Nitrogen}

In the course of normal stellar evolution, carbon is essentially all produced by
He burning. In zero-metal massive stars, primary nitrogen can be synthesized in
a H-burning layer where fresh carbon built in the helium burning core is
injected by mixing (e.g., induced by rotation).

The carbon abundance for our stars is determined by fitting the computed CH 
AX electronic transition band at 422.4 nm (the G-band) to the observed 
spectrum. 

In our sample the mean value of the ratio [C/Fe] is close to zero. In very
metal-poor stars it has been found that 10-15\% of stars with [Fe/H]$<-2.5$ are
carbon rich, increasing to 20-25\% for stars with [Fe/H]$<-3.0$ (Norris, Ryan, 
\& Beers 1997; Rossi et al. \cite {RBS99}; Christlieb 2003). However, for our 
sample we selected stars without anomalously strong G-bands, with only two
exceptions: CS~22892--052 (Sneden et al. \cite {SMP96}; \cite{SCI00}; 
\cite{SNE03} and CS~22949--037 (Depagne et al. \cite {DHS02}). As a consequence, 
our sample is biased against carbon-rich objects and cannot be used to constrain 
the full dispersion of carbon abundances at the lowest metallicities.

\begin {figure}
\begin {center}
\resizebox  {8.cm}{4cm} 
{\includegraphics {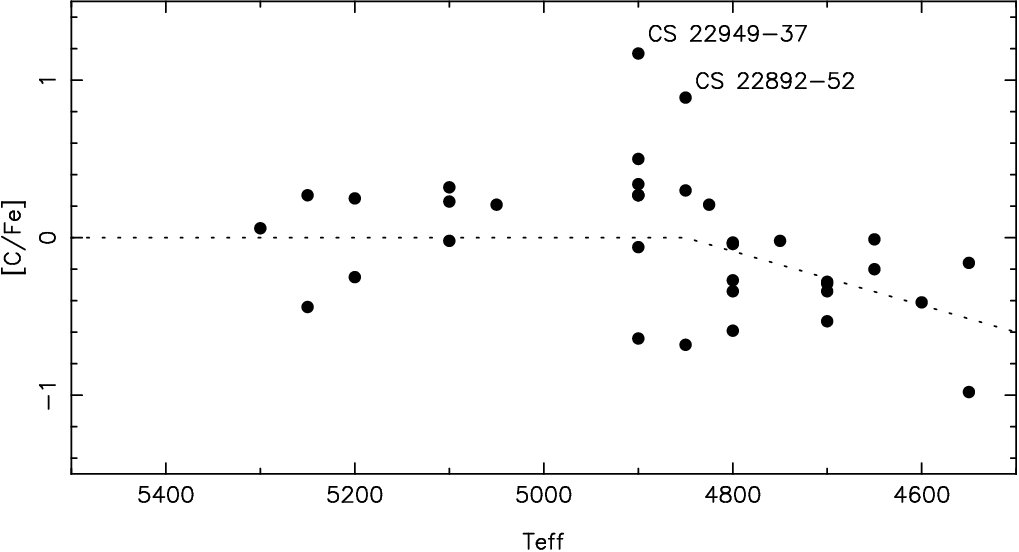} } 
\caption {[C/Fe] plotted vs. effective temperature. 
In the coolest stars (T$_{eff}< 4800$~K), the [C/Fe] ratio decreases due to 
mixing, which has likely brought processed material to the surface from deep 
layers where C is converted into N.
}
\label {C-Temp}
\end {center}
\end {figure}

Another word of caution concerns mixing episodes in these evolved stars. In
giants it is possible that material from deep layers, where carbon is converted
into nitrogen, has been brought to the surface  during previous mixing episodes.
This phenomenon is well known in globular cluster stars (e.g., Langer et al.
\cite{LKC86}; Kraft \cite{Kr94}). To check for this effect, Fig. \ref{C-Temp} 
shows the ratio 
[C/Fe] as a function of the estimated temperature of all our stars. There is an
indication of a decline in [C/Fe] at temperatures below 4800~K, consistent with
the expectation from deep mixing of processed material. 

It would also be interesting to probe mixing by plotting the [C/N] ratio vs. the
temperature of the star, but unfortunately nitrogen could be measured only in a
few of our programme stars. Nitrogen is best measured from the CN BX electronic
transition band at 388.8 nm. For most of our sample stars this CN band is
not visible. Indeed, it is detected in only six stars -- the two ``C-rich'' 
stars CS~22892--052 (Sneden et al. \cite {SMP96}) and CS~22949--037 (Depagne et 
al. \cite {DHS02}), which both present strong nitrogen enhancements -- probably 
linked to the carbon enhancement -- and four other stars with slight nitrogen 
enhancements. As expected if these enhancements are due to mixing episodes, 
these four stars also show carbon abundances below
the mean value. However, it should be noted that one of these stars is
hotter 4800 K (BD+17:3248, T$_{eff}$ = 5250 K).

\begin {figure}
\begin {center}
\resizebox  {8.cm}{4.cm} 
{\includegraphics {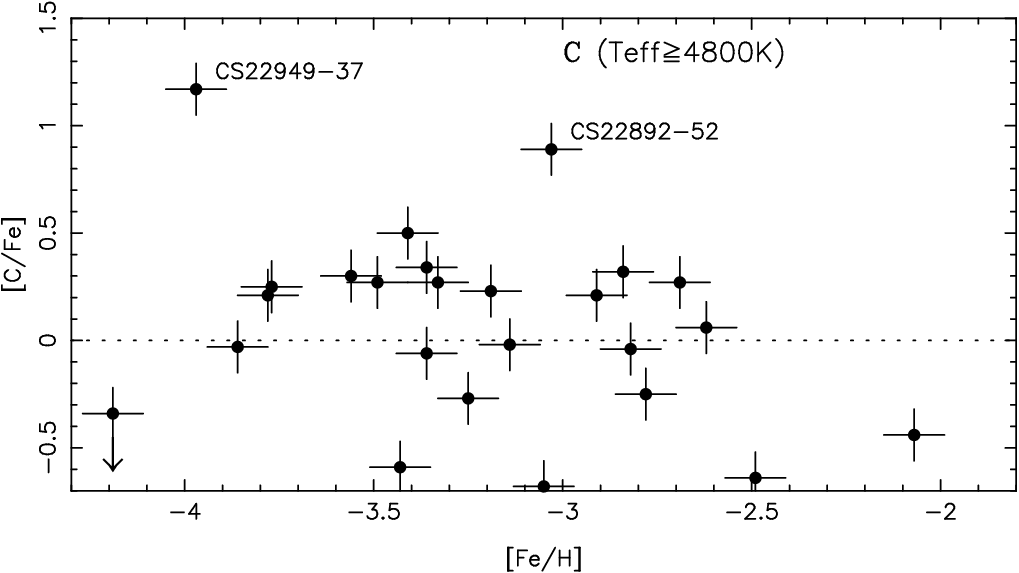} } 
\caption {[C/Fe] plotted vs. [Fe/H]. The ``C-rich'' objects CS~22892--052 and  
CS~22949--037 were excluded when computing the regression line. The slope of the 
line is not significant; the mean value of [C/Fe] is $\approx$0.2, and the 
dispersion is 0.37 dex, Note that the C abundance for CD-38:245 ([Fe/H) = 
--4.19) is only an upper limit.
}
\label {ab-C}
\end {center}
\end {figure}

To avoid potential difficulties with mixing, we selected only the stars with
temperatures higher than 4800 K to study the trend of the relation [C/Fe] vs.
metallicity (Fig. \ref{ab-C}).  In the interval $\mathrm {-4.1 < [Fe/H] < -2.5}$ 
the average ratio [C/Fe] is close to zero.  However, the dispersion is very
large (0.37 dex) and the slope of the regression line is not significant (the 
two CH-strong stars were excluded from these computations).  Obviously, a study 
of a more representative sample of low metallicity stars, including stars with 
large carbon enhancements, may well change these results.

\subsection {Oxygen}

During normal stellar evolution, oxygen is produced during the
central helium-burning phase, with some contribution from neon burning. In
massive stars large amounts of oxygen can be produced via explosive
nucleosynthesis (see Depagne et al. \cite {DHS02}). 

Oxygen is the most abundant heavy element throughout the cosmos.
However, it is well known that the oxygen abundance in stars is difficult to
determine, since the four O features in stellar spectra (the
forbidden lines at 630.0-636.4 nm, the permitted triplet at 777.2, 777.4, and
777.5 nm, the near IR vibration-rotation bands and the near-UV OH electronic
transition bands) often provide discrepant values. In the very metal-poor giants
over the range of wavelengths studied here (350 -- 1000 nm) the only line
available is the forbidden [O~I] line at 630.031 nm, generally admitted to be 
the most reliable one (Kraft \cite{Kr01}; Cayrel et al. \cite{Ca01}; Nissen et 
al. \cite{NPA02}).  This line is apparently not sensitive to non-LTE effects
(Kiselman \cite {Ki01}), but following Nissen et al. (\cite{NPA02}) it seems 
important to take into account hydrodynamical (3D) effects. 

Allende Prieto et al. (\cite{ALA01}) recently computed the solar abundance of
oxygen from the forbidden oxygen line, using synthetic spectra based on 3-D
hydrodynamical simulations of the solar atmosphere. Moreover, they subtracted
the contribution from a weak Ni~I line which blends with the solar oxygen
line, and computed a new very precise value of the transition probability of the
forbidden line from a new computation of the magnetic dipole (Storey  \& Zeippen 
\cite {SZ01}) and electric quadrupole contributions  (Galavis et al., \cite 
{GMZ97}). They found a $ \mathrm {log\; gf_{[OI]630.031}=-9.72}$, and an oxygen 
abundance $\mathrm {log \epsilon_{\odot}(O)=8.69}$ (with a 1D model the solar 
oxygen abundance would be $\mathrm {log \epsilon_{\odot}(O)=8.74}$, following 
Nissen et al. \cite  {NPA02}). We have also adopted $\mathrm {log \; 
gf_{[OI]630.031}=-9.72}$. As the solar reference value we assumed $\mathrm {log 
\epsilon_{\odot}(O)=8.74}$ for our initial (1D) computation of [O/H] and [O/Fe]. 
We also attempted to correct these 1-D computations for 3-D effects (see below) 
and in that case, the corresponding reference solar value $\mathrm {log 
\epsilon_{\odot}(O)=8.69}$ was used.


\begin {figure}
\begin {center}
\resizebox  {8.cm}{4.cm} 
{\includegraphics {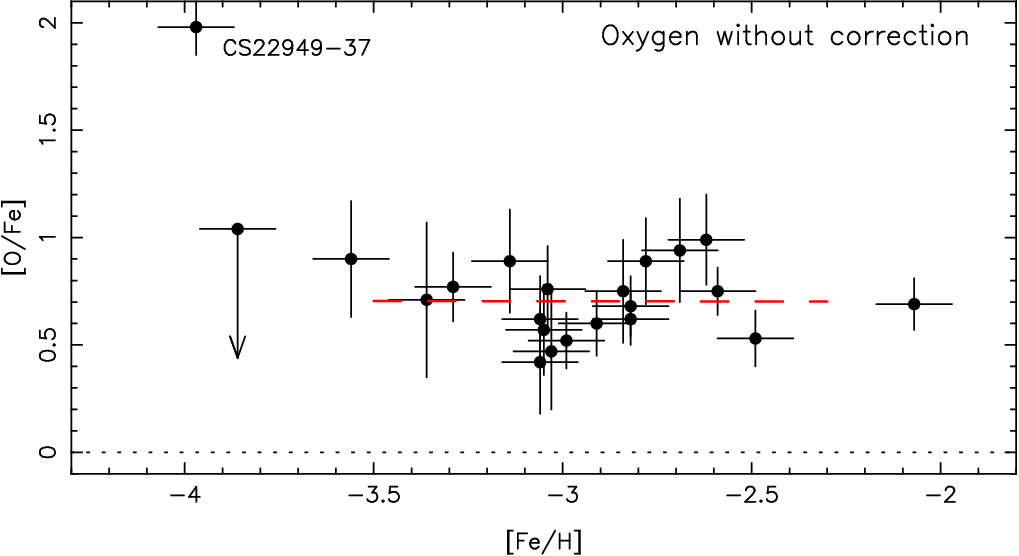} }
\resizebox  {8.cm}{4.cm} 
{\includegraphics {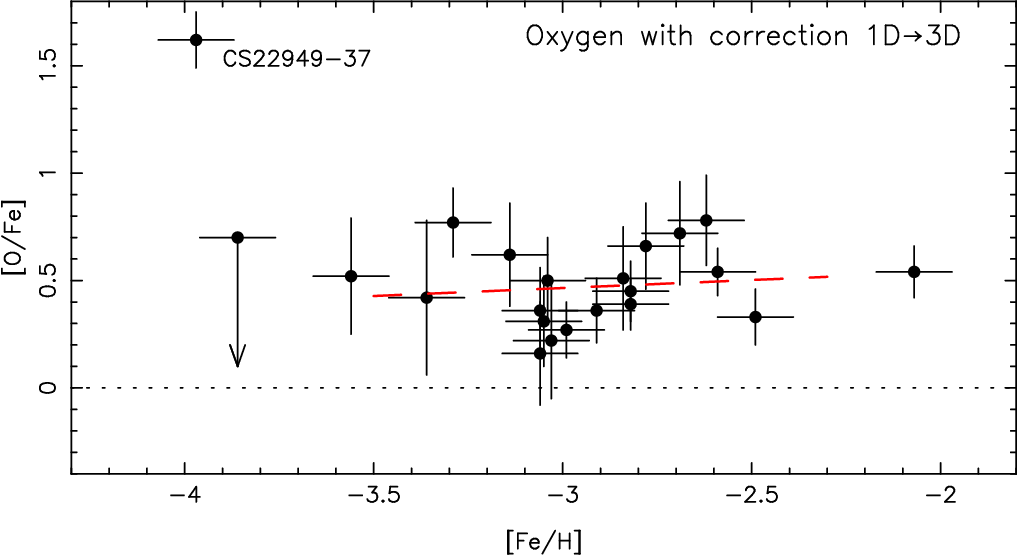} }
\caption {[O/Fe] plotted vs. [Fe/H], without and with a correction for stellar 
surface inhomogeneities (Nissen et al. \cite {NPA02}). The correction is 
uncertain since it has been computed for dwarfs. The slope of the regression 
line (dashed) is small and not significant. }
\label {ab-o}
\end {center}
\end {figure}

The forbidden oxygen line is very weak, especially in the most metal-poor
stars, where the line is generally below the limit of detection for [Fe/H] 
$<-3.5$. Hence, it has not been possible to determine oxygen abundances for 
{\em all} stars of the sample.  However, the high quality of the spectra allowed 
a precise measurement to be made for most of our stars. 

The oxygen line is
located in a region where the S/N ratio of the spectra is the highest.
Unfortunately, weak interference fringes from the CCD detector sometimes appear 
in this region and make the definition of the continuum more uncertain. 
We recall that the depression at the center of a 1m\AA\ line corresponds to 
only about 0.6\% of the continuum, so the measurement of lines with an 
equivalent width below 1 m\AA\ is
often difficult in this region of the spectrum. Moreover, in some unfortunate
cases, the stellar oxygen line is superimposed on strong telluric lines in
absorption or emission, and cannot be measured with sufficient precision,
even after correction for night-sky emission or absorption lines (sky 
subtraction or division by the spectrum of a fast-rotating hot star).  

We first computed the oxygen abundance using 1-D OSMARCS models. Plotting the
[O/Fe] ratio as a function of [Fe/H] reveals no significant slope; the mean
value is [O/Fe] $\approx 0.7 $ (Fig. \ref{ab-o}, upper panel). A large 
dispersion, on the order of 0.17 dex, is found.

Abundance corrections have been computed with 3-D radiative hydrodynamical
codes by Nissen et al. (\cite{NPA02}) for metal-deficient dwarfs, but not for
giant stars. However, they note that the sign of the correction is unlikely to
change, and that therefore the [O/Fe] ratio based on [O~I] lines should always
be smaller in 3-D than in the 1-D computations. Thus, we assumed as a first
approximation that the correction computed by Nissen et al. (\cite{NPA02}) is
valid also for metal-deficient giants. The result of this exercise is shown in
Fig. \ref{ab-o} (lower panel): The [O/Fe] still follows a ``plateau,'' which now
lies at about [O/Fe] = 0.47.

Quite recently, Johansson et al. (\cite{JLL03}) have made a new
determination of the $gf$ value for the Ni I 6300.34 line, itself actually a
blend of $^{58}$Ni and $^{60}$Ni lines. This blend does not affect our own
determinations of oxygen abundance, thanks to the smaller relative contribution
of Ni to the blend in oxygen-enhanced stars. But it {\it does} affect our
derived [O/Fe] values through a change of the solar oxygen abundance. Assuming
that we are in the linear domain for line depths smaller than 5\%, the new
oscillator strength would increase the contribution of the Ni blend in the Sun 
from 29\% to 43\%, inducing a correction of $\log(57/71) \approx 0.1$ dex to the
solar oxygen abundance and increasing our [O/Fe] values by the same amount.
However, the superb fit obtained by Allende Prieto et al. (\cite{ALA01}), would
likely also suffer from this significant enhancement of the Ni~I contribution. 

We finally note that the extremely metal-poor star CS~22949--037 has an
exceptionally high O abundance according to Depagne et al. (\cite{DHS02}). It
is, however, rather peculiar, displaying also very high abundances of Mg and 
several other elements, and should not be considered as representative of XMP
stars in general. Indeed, the forbidden oxygen line is not detectable in any 
other star with [Fe/H]$ \approx -4$ in our sample. If these stars have the same 
[O/Fe] ratio as the other XMP stars ([O/Fe] = 0.71 from 1-D models), then the 
computed equivalent widths of their [O~I] lines would be less than 0.5m {\rm 
\AA}, a value which is generally below our detection limit.

\begin {table*}
\caption {Coefficients  of the adopted regression lines.  In columns 2 and 3 
are given the coefficients $a$ and $b$ of the relation [X/Fe] = $a$ [Fe/H] 
+ $b$. In column 4 is given the scatter measured in the total interval, and in 
column 5 and 6 the scatter in two different intervals of 
metallicity. Column 7 lists an estimate of the scatter expected from 
measurement errors only.  For some elements the total scatter is hardly larger 
than the expected error.}
\begin {center}
\label {tab-dispel}
\begin {tabular}{cccccccc}
\hline

        &     &     &$-4.1<$ &$-4.1<$ &$-3.1<$&\\
        &     &     &[Fe/H]  &[Fe/H]  &[Fe/H] & \\
        &
\multicolumn{2}{c}{Regression Line} &$<-2.0$ &$<-3.1$ &$<-2.1$&\\
  &a & b&$\sigma_{\mathrm{reg}}$ &$\sigma_{\mathrm{reg}}1$
  &$\sigma_{\mathrm{reg}}2$&$\sigma_{\mathrm{mes}}$\\
\hline
$\mathrm{Na}$& 0.403$ \pm $0.010 & 1.420$ \pm $0.101 & 0.25 &0.32 
&0.18 &0.10\\
$\mathrm{Mg}$& 0.035$ \pm $0.003 & 0.380$ \pm $0.029 & 0.13 &0.11 
&0.15 &0.09\\
$\mathrm{Al}$& 0.047$ \pm $0.005 &-0.534$ \pm $0.052 & 0.18 &0.14 
&0.21 &0.10\\
$\mathrm{Si}$& 0.032$ \pm $0.004 & 0.541$ \pm $0.036 & 0.15 &0.20 
&0.11 &0.10\\
$\mathrm{ K}$& 0.176$ \pm $0.002 & 1.020$ \pm $0.023 & 0.11 &0.13 
&0.10 &0.10\\
$\mathrm{Ca}$& 0.074$ \pm $0.002 & 0.565$ \pm $0.015 & 0.10 &0.11 
&0.09 &0.07\\
$\mathrm{Sc}$& 0.034$ \pm $0.002 & 0.178$ \pm $0.019 & 0.11 &0.14 
&0.08 &0.07\\
$\mathrm{Ti}$&-0.014$ \pm $0.001 & 0.185$ \pm $0.013 & 0.09 &0.09 
&0.10 &0.05\\
$\mathrm{Cr}$& 0.117$ \pm $0.000 & 0.004$ \pm $0.004 & 0.05 &0.04 
&0.06 &0.07\\
$\mathrm{Mn}$& 0.030$ \pm $0.003 &-0.346$ \pm $0.020 & 0.12 &0.15 
& 0.08 &0.09\\
$\mathrm{Co}$&-0.131$ \pm $0.002 &-0.121$ \pm $0.024 & 0.13 &0.12 
&0.13 &0.08\\
$\mathrm{Ni}$&-0.003$ \pm $0.002 &-0.048$ \pm $0.020 & 0.11 &0.13 
&0.11 &0.09\\
$\mathrm{Zn}$&-0.271$ \pm $0.002 &-0.559$ \pm $0.018 & 0.11 &0.14 
&0.08 &0.10\\
\hline
\end {tabular}
\end {center}
\end {table*}

\subsection {Light even-Z metals: Mg, Si, Ca, Ti}

\begin {figure}
\begin {center}
\resizebox  {8.cm}{4.cm} 
{\includegraphics {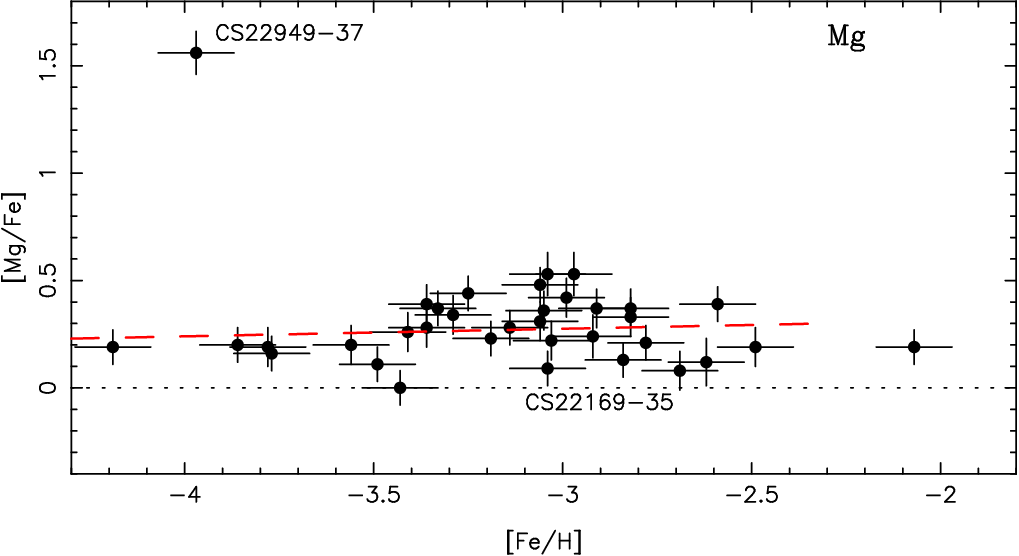} }
\resizebox  {8.cm}{4.cm} 
{\includegraphics {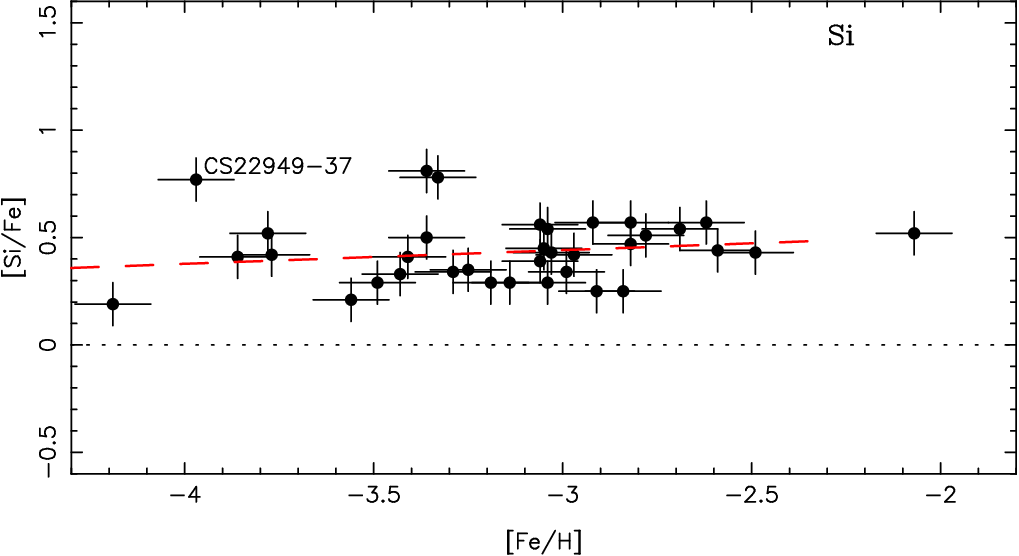} }
\resizebox  {8.cm}{4.cm} 
{\includegraphics {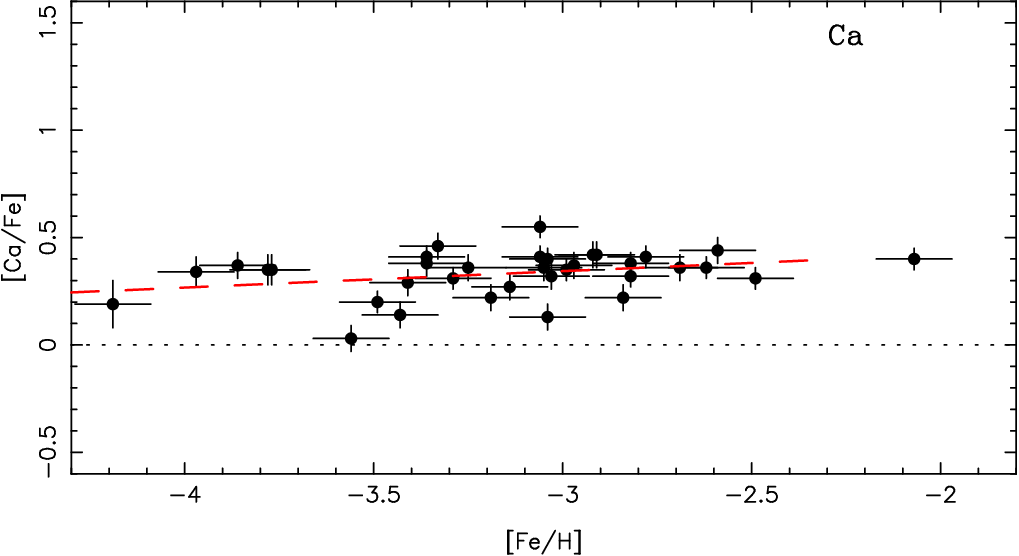} }
\resizebox  {8.cm}{4.cm} 
{\includegraphics {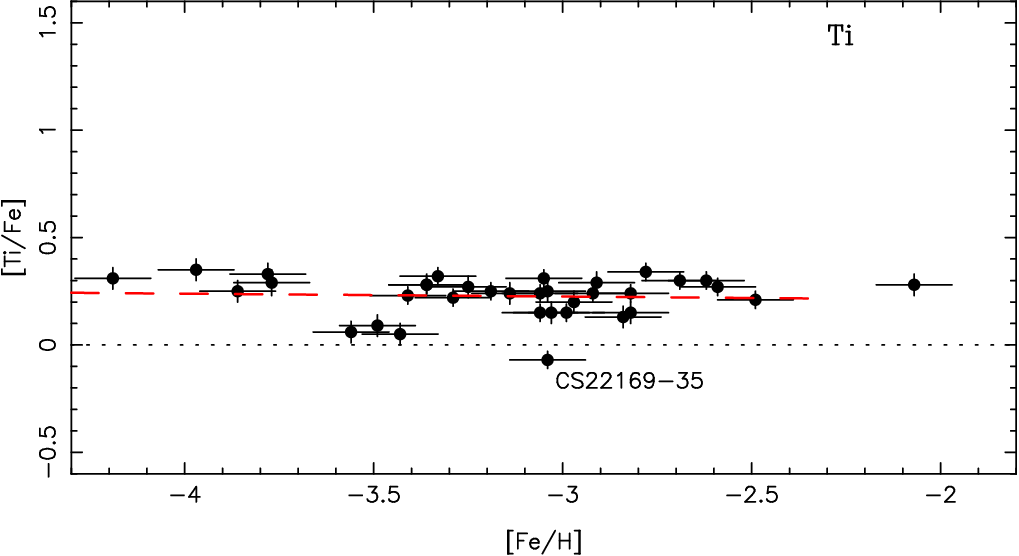} }
\caption {[Mg/Fe], [Si/Fe], [Ca/Fe] and [Ti/Fe] plotted vs. [Fe/H].
The peculiar star CS~22949--037 is not included in the computations 
of the scatter and of the regression line (dashed) for Mg. The star 
CS~22169--035 is deficient in all the light ``even'' elements.}
\label {ab-mgti}
\end {center} 
\end {figure}

In our spectra there are about 7 well-defined lines of magnesium, 15 lines of
calcium, and about 30 lines of titanium, but silicon is represented by only two
lines --  one at 390.55~nm and the other at 410.29~nm. As the first line is
severely blended by a CH line, we have chosen to use only the second one, which
unfortunately falls in the wing of the H$\delta$ line. To compute the silicon 
abundances, synthetic profiles of the line were computed taking into account the 
presence of the H$\delta$ line.
      

Following A. Chieffi (private communication), Mg is formed during hydrostatic
carbon burning in a shell and during explosive neon burning. Si and Ca are built
during incomplete explosive silicon and oxygen burning, and Ti during complete
and incomplete silicon burning. As shown in Fig. \ref{ab-mgti}, they all appear
to be enhanced relative to iron, but any slope with [Fe/H] is generally small 
(Table \ref{tab-dispel}). The even-Z ($\alpha$) elements behave similarly to O, 
but the enhancement is smaller ([Mg/Fe] = +0.27, [Si/Fe] = +0.37,
[Ca/Fe] = +0.33 and [Ti/Fe] = +0.23). The scatter around the mean value is small
($\sigma_{Mg} =0.13$ dex, $\sigma_{Si} =0.15$ dex, $\sigma_{Ca} =0.11$ dex,
$\sigma_{Ti} =0.10$ dex); however, it increases slightly as the metallicity
decreases. In Table \ref{tab-dispel} we list, for each element, the dispersion
around the mean regression line in the intervals $\mathrm{[-4.1<[Fe/H]<-3.1]}$
and $\mathrm{[-3.1<[Fe/H]<-2.1]}$, as well as the value expected from 
measurement errors only.

The nearly identical abundance ratios of these light metals at low metallicity 
suggest that there is a similarly constant ratio between the yields of
iron and of the other elements, in spite of the quite different sites
where they are produced. In the case of magnesium, the spread
around the mean value is not significantly larger than the measurement errors,
even at the lowest metallicities. An exception is the peculiar star 
CS~22949--037, which is strongly enhanced in light elements (C, O, Na, Mg, Al) 
but has a ``normal'' abundance of Si, Ca, and Ti. This star is clearly an 
outlier and has not been taken into account in the computation of the ``normal'' 
trends and dispersions of the lighter elements.

\subsection {The odd-Z metals: Na, Al, K, and Sc}

\subsubsection{NLTE effects}

In extremely metal-deficient stars, the abundances of the odd-Z elements Na, 
Al, and K are deduced from resonance lines which are very sensitive to
non-LTE effects. Hence, to determine the trends of these elements with
metallicity it is important to take these effects into account, at
least approximately.

The sodium abundance is computed from the Na D resonance lines at 589.0 nm and
589.6 nm. In some stars these lines are severely blended by interstellar lines,
and the sodium abundance cannot be measured accurately. Baum\"uller et al.
(\cite{BBG98}) have evaluated the importance of NLTE effects in metal-poor
dwarfs and subgiants. They found that the correction can reach values as high as
--0.5 dex. To account for this effect, the values of [Na/Fe] given in Tables 
\ref {tab-abund-a} to \ref{tab-abund-e} should thus be decreased by 0.5 dex.

The abundance of aluminium is based on the resonance doublet at 394.4 and 396.15 
nm. Due to the high resolution and high S/N of the spectra, both lines can
be used, and the blending of Al 394.4 nm by a CH line is easily taken into
account. The Al abundance is underestimated in LTE computations
(Baum\"uller \& Gehren (\cite {BG97}); Norris et al. \cite{NRB01}), but since
our stars are all very similar in temperature and gravity we can consider 
this correction to be similar and close to +0.65 dex for all the stars. As a
consequence, the LTE abundance given in Tables \ref {tab-abund-a} to
\ref{tab-abund-e} should be increased by about +0.65 dex. 

The K abundance has been determined from the red doublet at 766.5 and 769.9 nm. 
Ivanova \& Shimanskii (\cite {IS00}) have computed NLTE corrections
for the K lines as a function of effective temperature and gravity. In the range 
$\mathrm 4500<T_{eff}<5100$ K and $\mathrm{0.5 <log\; g <2.0}$ the NLTE 
correction reaches $\sim -0.35$ dex. Thus, the LTE abundances
given in Tables \ref {tab-abund-a} to \ref{tab-abund-e} should be decreased by
about 0.35 dex (this correction has been taken into account in Fig. 
\ref{abKSc}).
Takeda et al. (\cite{T02}) propose NLTE corrections that are slightly smaller 
($\sim -$0.25 dex), irrespective of metallicity or gravity. 

We finally note that in the range of temperature, gravity, and metallicity 
covered by our sample we can assume that these corrections are similar for all 
the stars, thus they do not alter the general abundance ratio trends, only the 
levels of the relations.
 
\subsubsection{The light elements Na and Al}

The production of Na and Al is expected to be sensitive to neutron excess
(Woosley \& Weaver \cite{WW95}), and therefore depends on the amount of 
neutron-rich nuclei 
present in the supernova before the synthesis of these two odd-Z metals. Na is
synthesized during hydrostatic carbon burning and partly in the hydrogen
envelope (Ne, Na cycle), while Al is synthesized during carbon and neon burning
and also in the hydrogen envelope (Mg, Al cycle).

\begin {figure}
\begin {center}
\resizebox  {8.cm}{4.cm} 
{\includegraphics {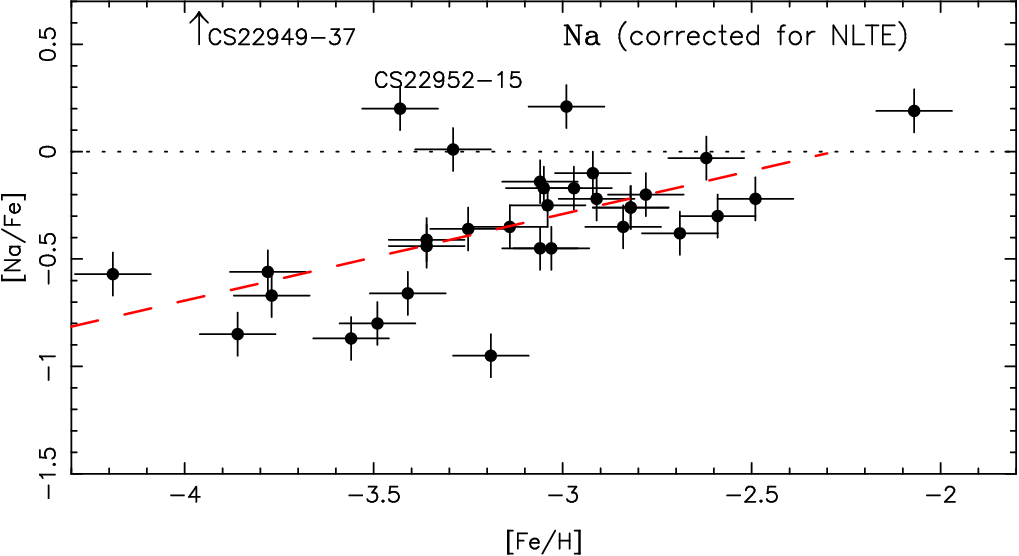} }
\resizebox  {8.cm}{4.cm} 
{\includegraphics {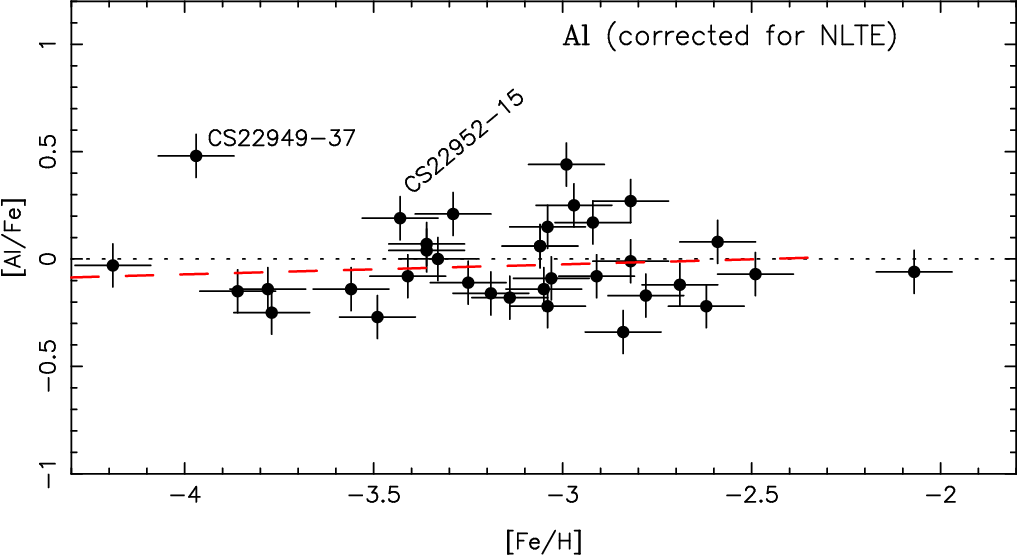} }
\caption {[Na/Fe] and [Al/Fe] plotted vs. [Fe/H]. The 
LTE abundances of these elements have been determined from 
resonance lines, but corrections for NLTE effects have been applied.
}
\label {abFe-NaAl}
\end {center} 
\end {figure}

In Fig. \ref{abFe-NaAl} we plot [Na/Fe] and [Al/Fe] vs. [Fe/H]. Both [Na/Fe]
and [Al/Fe] exhibit a rather large scatter of $\sim$ 0.2 dex (Table
\ref{tab-dispel}). On the other hand, while Na decreases significantly with
decreasing metallicity, Al remains practically constant within the range
$\mathrm{-4.0 <[Fe/H]<-3.0}$.
The striking difference in the behavior of these two elements of very
similar atomic numbers is puzzling, but there remains an alternative
interpretation of the plot of [Na/Fe] vs. [Fe/H], which we consider in Sect. 
\ref {sec-refelem}.


\subsubsection{K and Sc}

In this paper we present, for the first time,
measurements of potassium abundances 
for a large sample of very metal-poor stars.

K is produced during explosive oxygen burning, while Sc is
synthesized during explosive oxygen and neon burning. The Sc yields in the
grid of Woosley \& Weaver (\cite{WW95}) show large variations and thus appear
to be strongly influenced by the parameterisation of the explosion (Samland
\cite{Sam98}). The Sc yields also show very large variations as a function of
the mass of the progenitor in the computations of Chieffi \& Limongi
(\cite{CL02}); thus we might expect a large scatter of the ratio [Sc/Fe] vs.
[Fe/H]. 

\begin {figure}
\begin {center}
\resizebox  {8.cm}{4.cm} 
{\includegraphics {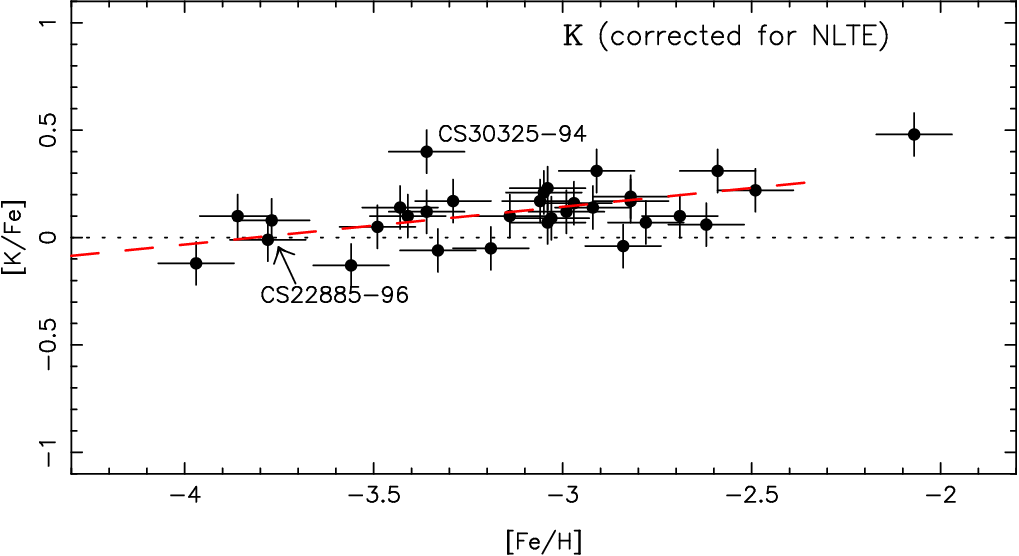} }
\resizebox  {8.cm}{4.cm} 
{\includegraphics {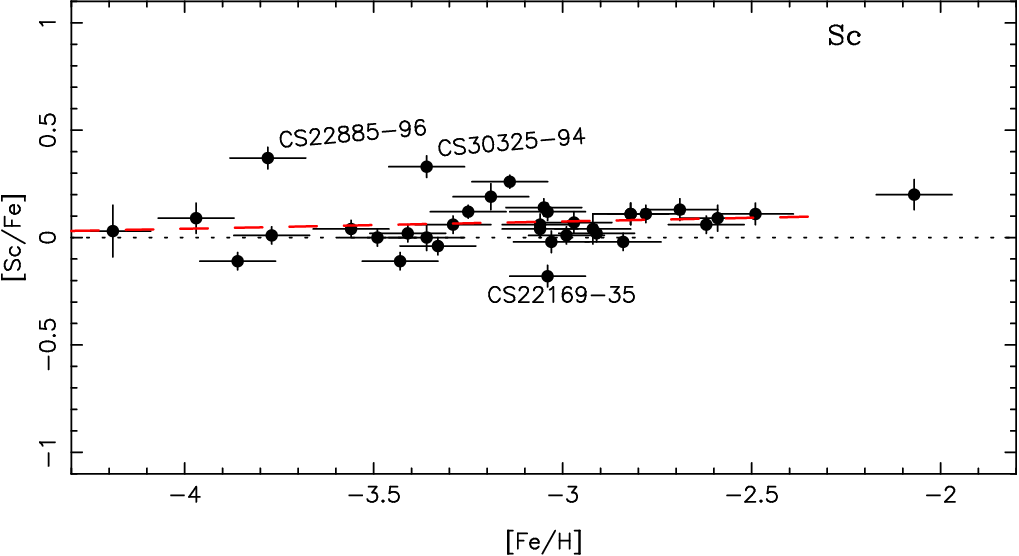} }
\caption {[K/Fe] and [Sc/Fe] plotted vs. [Fe/H]. An NLTE correction 
has been applied to our potassium abundance measurements (see text).
}
\label {abKSc}
\end {center} 
\end {figure}

In Fig. \ref{abKSc} the ratios [K/Fe] and [Sc/Fe] have been plotted vs. [Fe/H];
they seem to decrease slowly with metallicity with a moderate scatter (about
0.12 dex), although the slope is not very significant (Table \ref {tab-dispel}).
The star CS~30325--094 appears to be K- and Sc-rich, while the more metal-poor
star CS~22885--096 is rich in Sc, with a ``normal'' K abundance.

\subsection {Iron-peak elements}

Generally speaking, the iron-peak elements are built during supernova
explosions. More specifically, Cr, Mn, Fe, Co, Ni, and Zn are built during
(complete or incomplete) explosive silicon burning in two different regions
characterized by the peak temperature of the shocked material (Woosley \& Weaver
\cite{WW95}; Arnett \cite{Arn96}; Chieffi \& Limongi
\cite{CL02}; Umeda \& Nomoto \cite{UN02}). 

\subsubsection {Cr and Mn}

Six manganese lines are visible in our spectra but three of them belong to
the resonance triplet ($a^6 S - z^6 P^0$). The abundance of Mn deduced from this triplet
is systematically lower (--0.4~dex) than the abundance deduced from the other manganese
lines, and thus has not been taken into account in the mean.
 (The difference can be due to NLTE effects or to a bad estimation
of the gf values of the lines of this multiplet).  
However, for the five most metal-poor stars only the resonance triplet was detected. 
In this case  the abundance deduced from these lines
 has been systematically corrected by 0.4 dex (and are the values given in Tables
 \ref{tab-abund-a} to  \ref{tab-abund-f}).

Cr and Mn are produced mainly by incomplete explosive silicon burning (\cite{WW95};
Chieffi \& Limongi \cite{CL02}; Umeda \& Nomoto \cite{UN02}). The observed
abundances of these elements have previously been shown to decrease with
decreasing metallicity (McWilliam et al. \cite{MPS95}; Ryan et al. \cite{RNB96};
Carretta et al. \cite{CGC02}).
 
\begin {figure}
\begin {center}
\resizebox  {8.cm}{4.cm} 
{\includegraphics {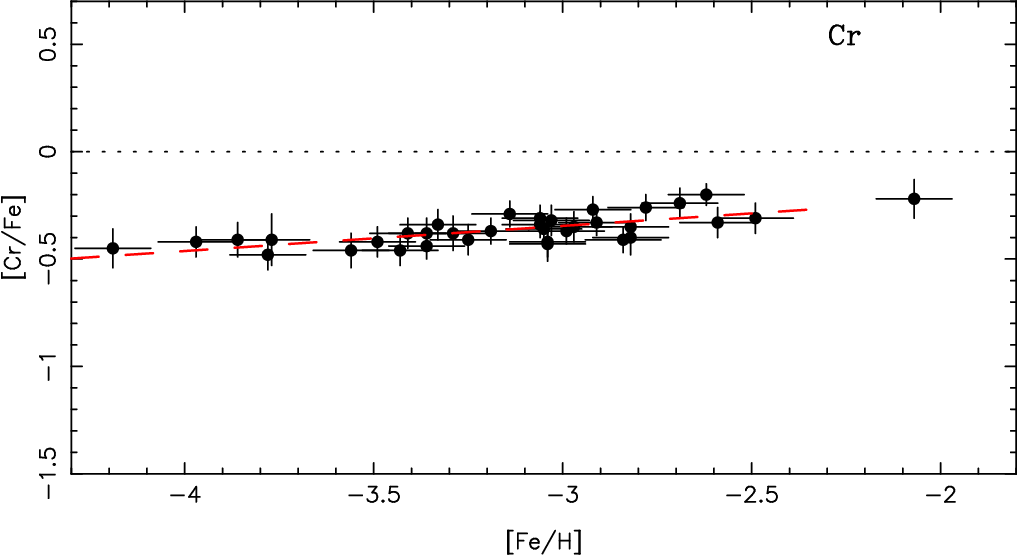} }
\resizebox  {8.cm}{4.cm} 
{\includegraphics {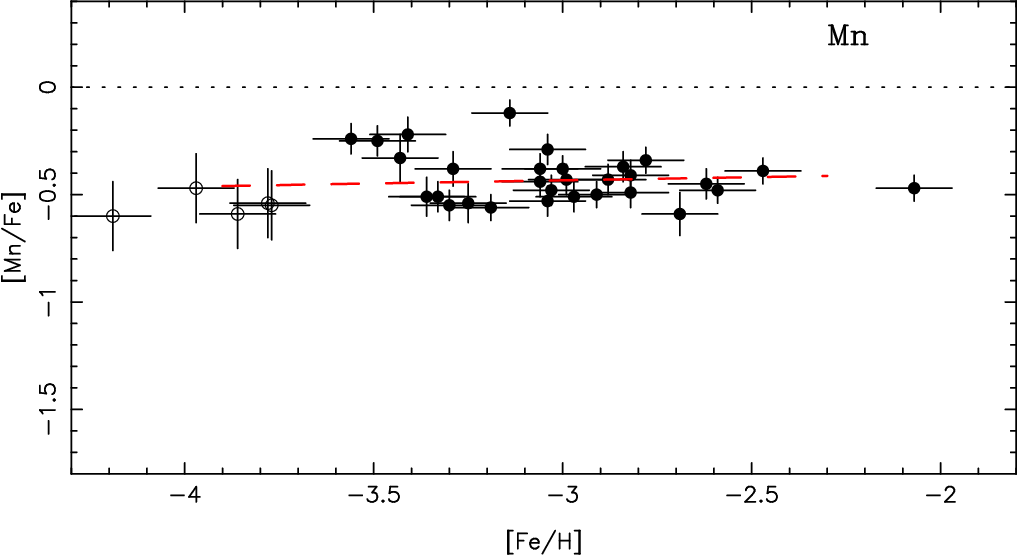} }
\caption {[Cr/Fe] and [Mn/Fe] plotted vs. [Fe/H]. For Mn the hyperfine structure
has been taken into account, and only the lines with an excitation potential
larger than 2.2 have been used. When these lines are too weak
(open symbols) the abundance has been deduced from the resonance lines
and corrected.
}
\label {abCr-Mn}
\end {center} 
\end {figure}

As shown in Fig. \ref{abCr-Mn}, the slope of [Cr/Fe] vs. [Fe/H] is
smaller than that found by Carretta et al. (\cite{CGC02}). Moreover, our
precise measurements show that [Cr/Fe] exhibits extremely small scatter ($\sigma 
= 0.05$ dex over the entire metallicity range; see Table \ref {tab-dispel}). 
This scatter is no larger than expected from measurement errors alone, 
indicating that any intrinsic scatter is extremely small and that the production 
of Fe and Cr are very closely linked. Among all elements measured in extremely 
metal poor stars, no other element follows iron so closely. We discuss this 
point further in section \ref{sec-scatter}).

Present nucleosynthesis theories do not yet provide a clear explanation for this
close link between Fe and Cr, together with the observed decrease of
[Cr/Fe] with decreasing metallicity. This is even more puzzling since the
metallicity ([Fe/H]) of a given XMP star may be considered as the ratio of the
iron yield to the volume of H gas swept up by the ejecta, which is {\em a
priori} independent of the nucleosynthesis which takes place in the exploding SN
and drives the [Cr/Fe] ratio. However, as argued by Ryan et al.
(\cite{RNB96}) and explored further by Umeda \& Nomoto (\cite{UN02}), both the
amounts of gas swept up and the supernova yields may be correlated through the
energy of the explosion, which depends in turn on the mass of the
progenitor. But the low scatter is surprising.

\begin {figure}
\begin {center}
\resizebox  {8.cm}{4.cm} 
{\includegraphics {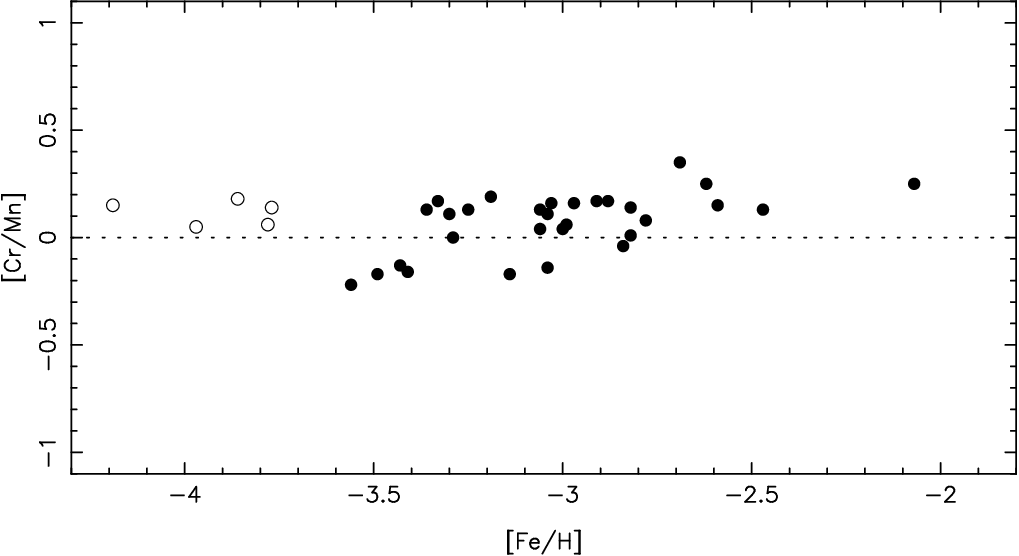} }
\caption {[Cr/Mn] plotted vs. [Fe/H]. 
 The symbols are the same as in Fig. \ref{abCr-Mn}.
The ratio Cr/Mn  is almost constant and close to the solar value. 
}
\label {abCr_Mn-Fe}
\end {center} 
\end {figure}

The relation [Cr/Mn] vs. [Fe/H] shows practically no trend with metallicity in
the range $\mathrm{-4.0<[Fe/H]<-2.5}$ (Fig. \ref{abCr_Mn-Fe}). However at low
metallicity the manganese abundance is deduced from the resonance
lines and a correction of 0.4~dex is empirically applied.
An NLTE 3D analysis of these lines would be necessary to be sure that
no significant slope is found, but it seems that the ratio Cr/Mn is
close to the solar value in the most metal-poor stars, although Mn is an odd-Z
element and Cr an even-Z element. 

\subsubsection {Co, Ni, and Zn}

\begin {figure}
\begin {center}
\resizebox  {8.cm}{4.cm} 
{\includegraphics {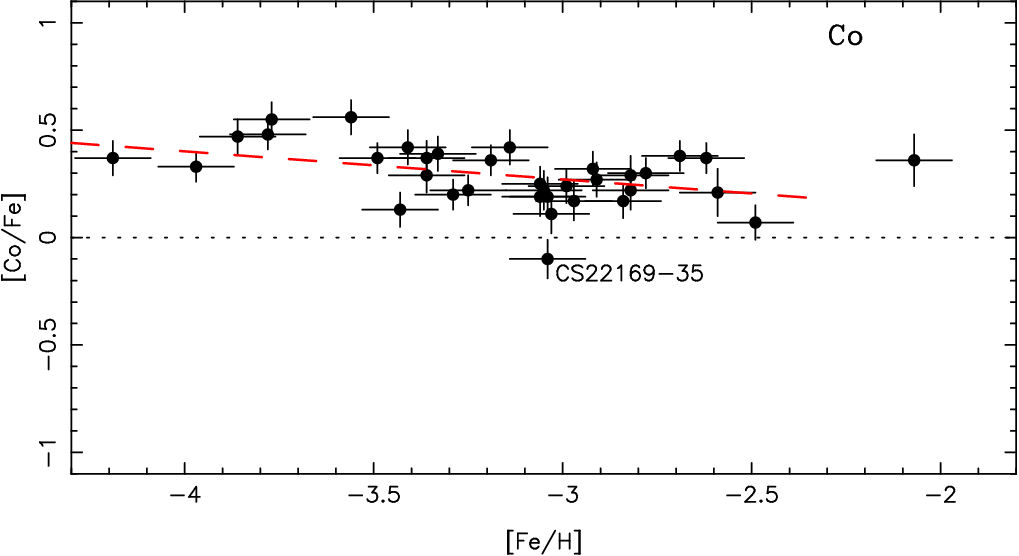} }
\resizebox  {8.cm}{4.cm} 
{\includegraphics {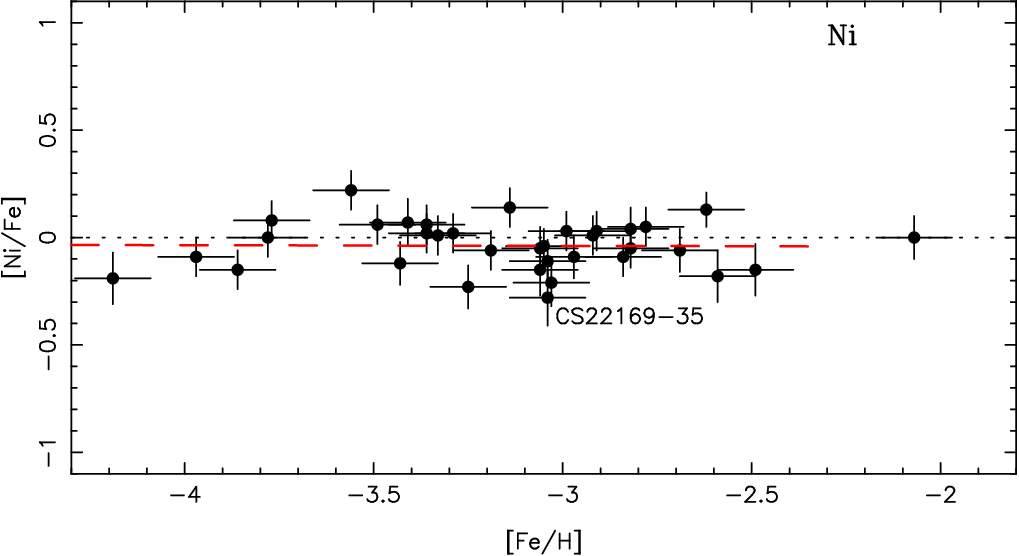} }
\resizebox  {8.cm}{4.cm} 
{\includegraphics {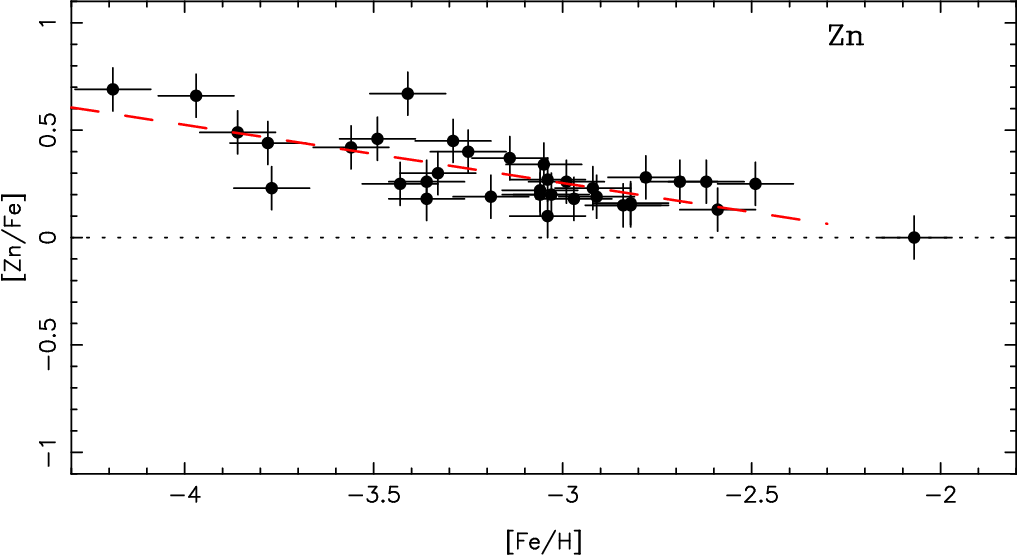} }
\caption {[Co/Fe], [Ni/Fe], and [Zn/Fe] plotted vs. [Fe/H]. These 
ratios increase ([Co/Fe], [Zn/Fe]) or remain essentially constant 
([Ni/Fe]) with decreasing metallicity.
}
\label {abCo-Zn}
\end {center} 
\end {figure}

Fe, Co, Ni, and Zn are produced mainly in complete explosive Si burning. The
abundance trends of these elements are presented in Fig. \ref{abCo-Zn}.

McWilliam et al. (\cite {MPS95}) found that [Co/Fe] increases with decreasing
[Fe/H]. We confirm this trend (Fig. \ref{abCo-Zn}), but the slope of the 
relation we obtain ($\sim 0.13$ dex per dex) is not as steep as they found. 
Also, the scatter in our data ($\approx 0.15$ dex) is significantly larger than 
expected from measurement errors alone.

Co and Ni are thought to be synthesized in the same nuclear process, but unlike 
[Co/Fe], [Ni/Fe] shows a mean value close to zero and no trend with [Fe/H]. The 
yields of Ni and Fe have a constant ratio, but the correlation is not as tight 
as that between Cr and Fe.  Three stars, CS~22189--009, CS~22885--096 and 
CS~22897--008, had been previously claimed to be Ni-rich by McWilliam et al.  
([Ni/Fe]$>+0.75$). These stars are included in our sample, but are found to 
have a normal Ni abundances.  In our computations we have rejected the line at 
423 nm for which no $gf$ value has been measured.  The ``solar $gf$ value''
computed by McWilliam et al. results in a Ni abundance from this line which
systematically disagrees with the value found from the other three
lines.

Zinc is an interesting element, as it is produced by complete silicon burning,
but it has been suggested that it could also be formed by slow or rapid neutron
capture (Heger \& Woosley \cite{HW02}; Umeda \& Nomoto \cite{UN02}). If Zn were
formed by the $s$-process, we would expect that [Zn/Fe] would decrease with
metallicity, at variance with what we observe. On the other hand, in 
CS~31082--001, a star with [Fe/H] = --3.0 and extremely rich in $r$-process 
elements, Hill et al. (\cite{HPC02}) found the Zn abundance to be normal 
relative to other stars with [Fe/H] = --3.0). We conclude that neither the 
$s$-process nor the $r$-process in their progenitors appears likely to have 
contributed a significant fraction of the Zn in these stars.

The ratio [Zn/Fe] increases with decreasing [Fe/H] more clearly than does 
[Co/Fe], in agreement with the results by Primas et al. (2000). The increase is 
quite significant and seems to be the signature of an $\alpha$-rich freeze-out 
process.

We recall that the abundances in the peculiar star CS~22949--037 were found by 
Depagne et al. (\cite{DHS02}) to correspond to the expected yields 
(Woosley \& Heger, private communication) of a rather 
massive progenitor ($\mathrm{M=35M_{\sun}}$), assuming a high mass cut 
some mixing and a 
rather large fallback (due to the large mass of the central remnant). The
[Zn/Fe] ratio of this star is, however, similar to the ``normal'' stars of the
sample and seems to be a global feature of extremely metal-poor stars.

\section{Discussion}
\begin {figure*}
\begin {center}
\resizebox  {5.8cm}{2.90cm} 
{\includegraphics {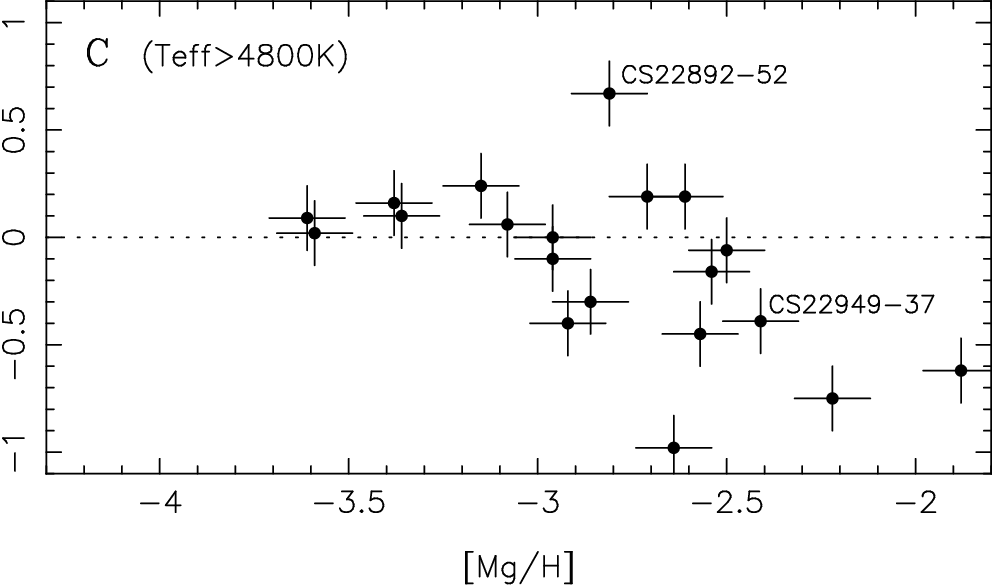} }
\resizebox  {5.8cm}{2.90cm} 
{\includegraphics {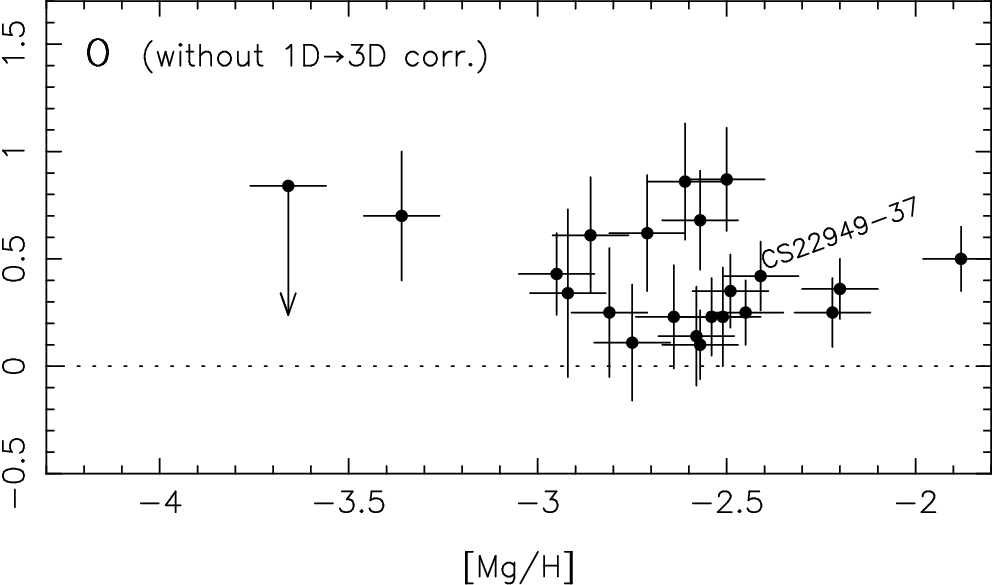} }
\resizebox  {5.8cm}{2.90cm} 
{\includegraphics {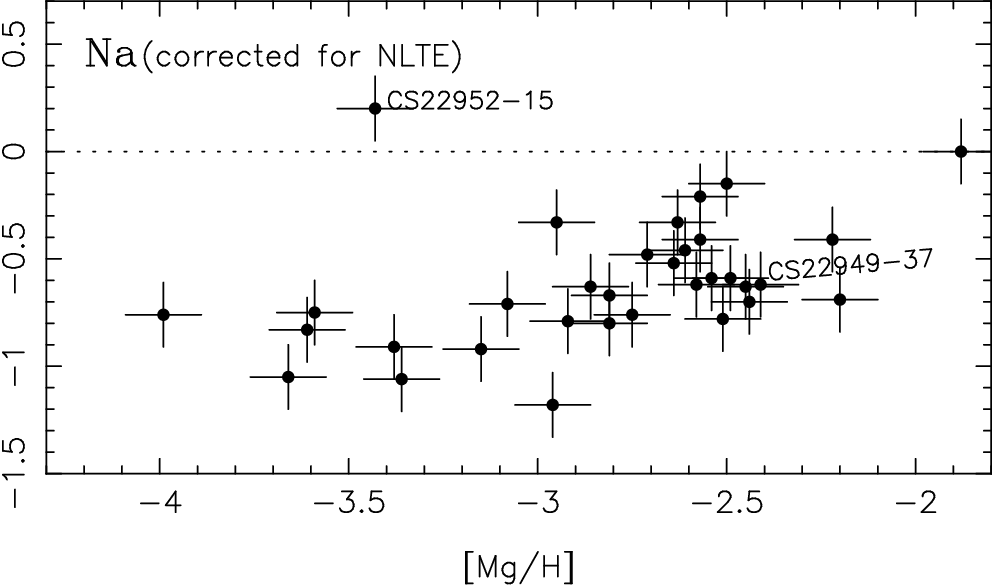} }
\resizebox  {5.8cm}{2.90cm} 
{\includegraphics {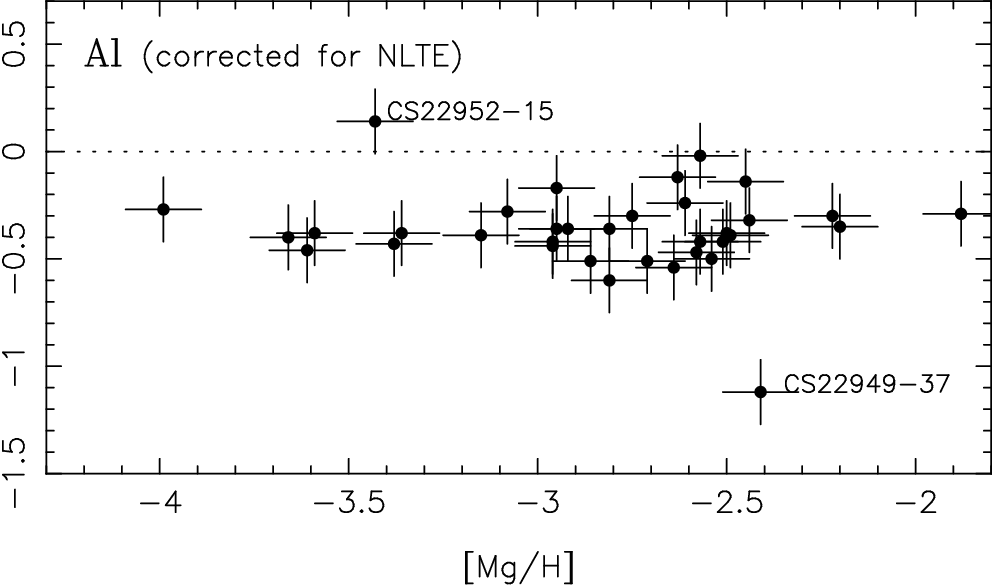} }
\resizebox  {5.8cm}{2.90cm} 
{\includegraphics {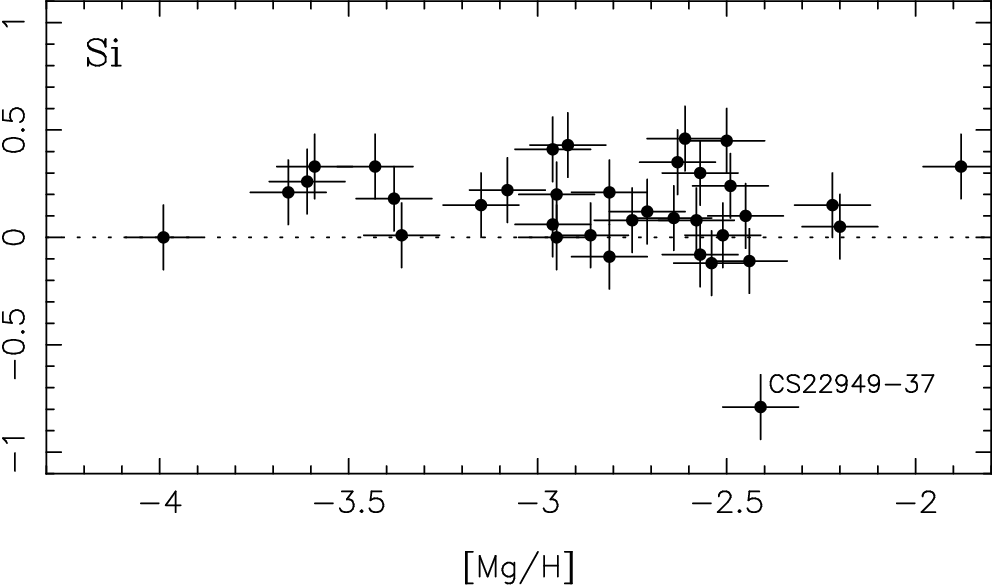} }
\resizebox  {5.8cm}{2.90cm} 
{\includegraphics {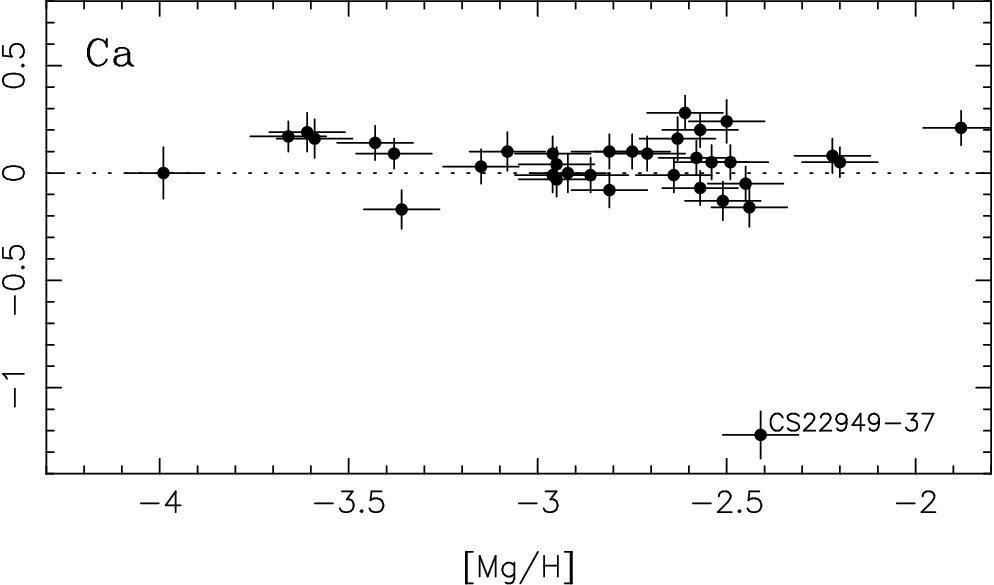} }
\resizebox  {5.8cm}{2.90cm} 
{\includegraphics {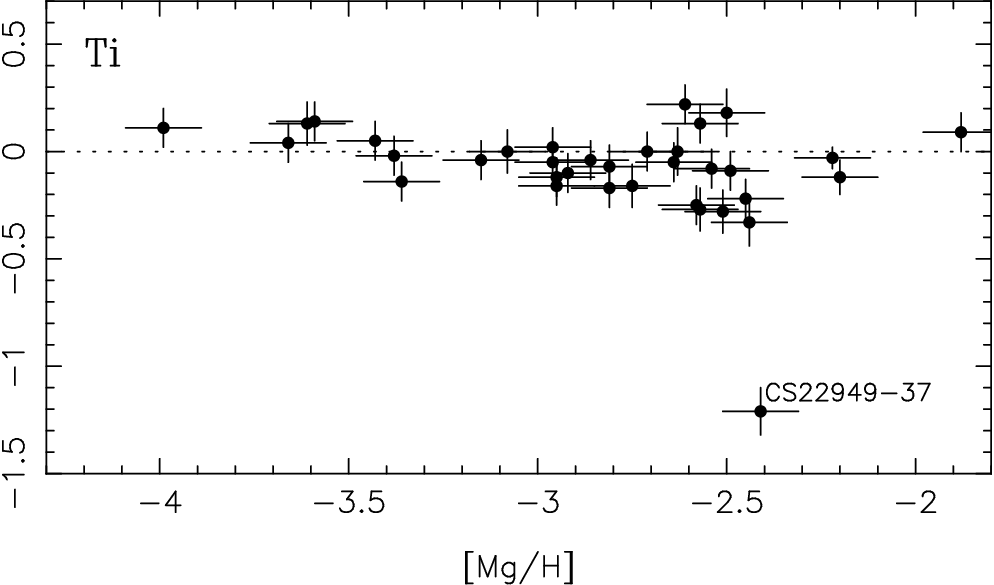} }
\resizebox  {5.8cm}{2.90cm} 
{\includegraphics {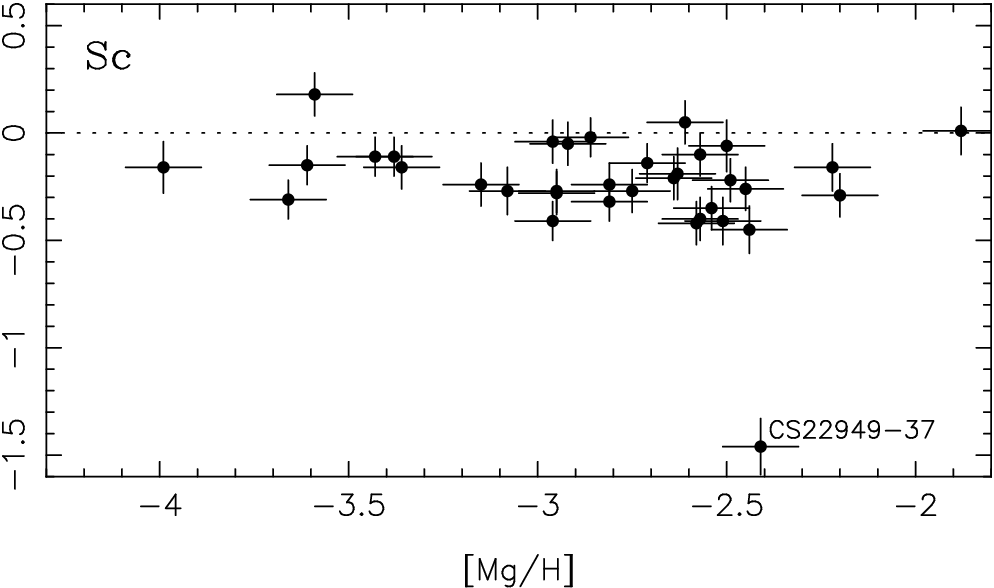} }
\resizebox  {5.8cm}{2.90cm} 
{\includegraphics {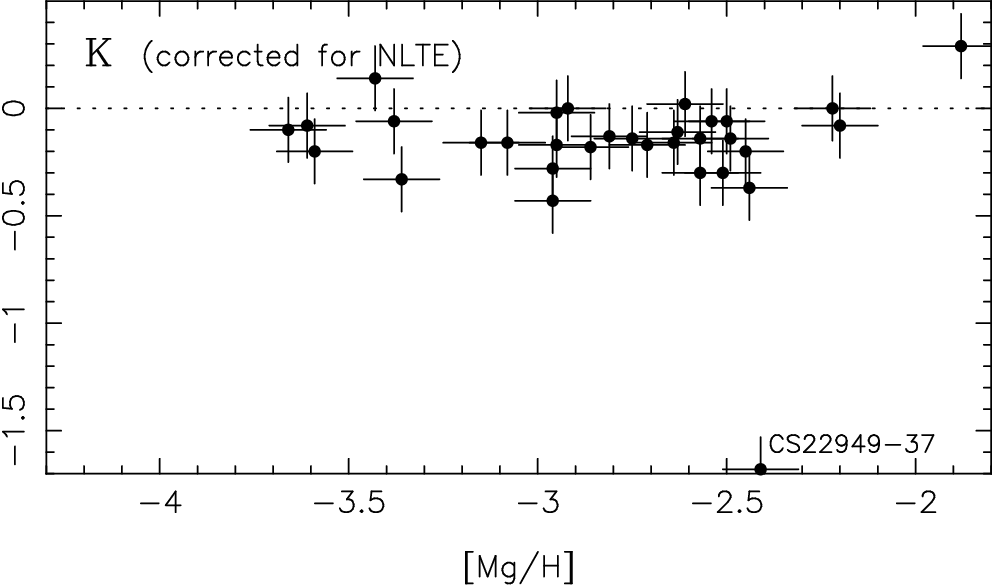} }
\resizebox  {5.8cm}{2.90cm} 
{\includegraphics {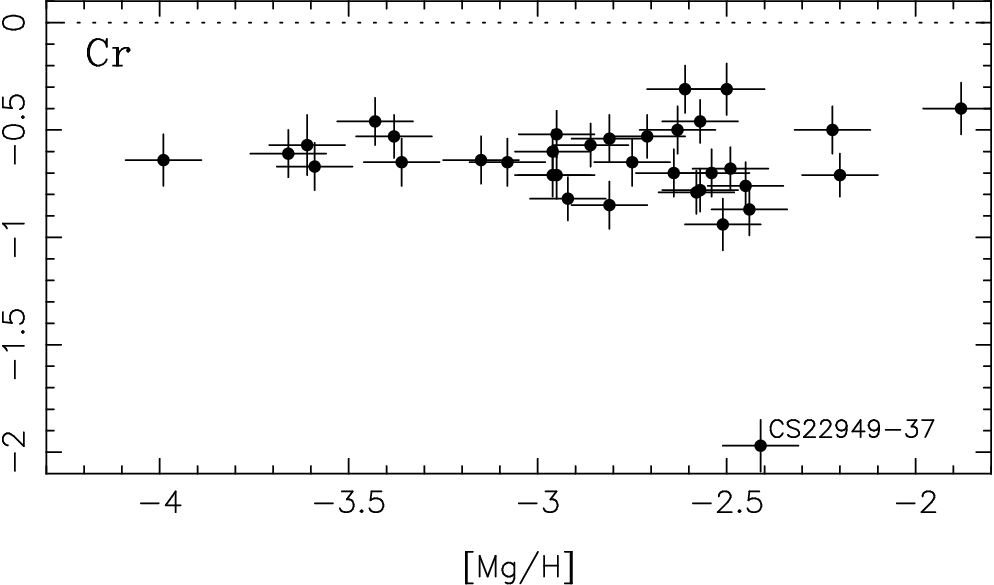} }
\resizebox  {5.8cm}{2.90cm} 
{\includegraphics {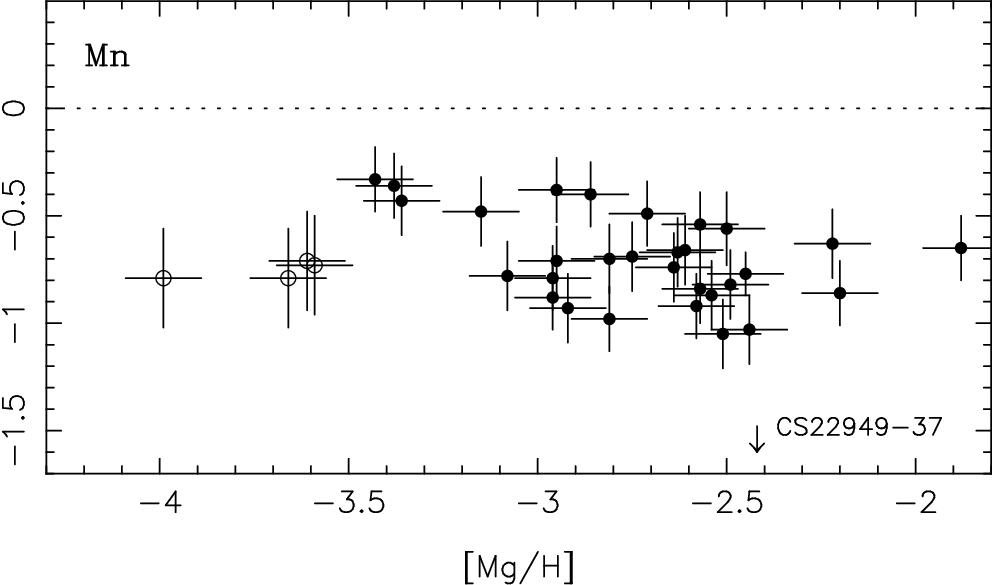} }
\resizebox  {5.8cm}{2.90cm} 
{\includegraphics {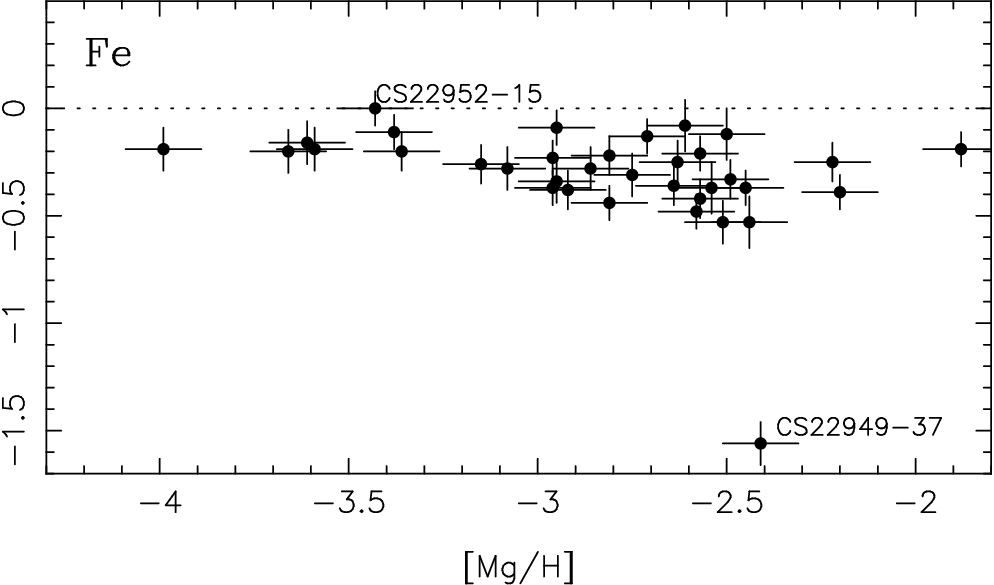} }
\resizebox  {5.8cm}{2.90cm} 
{\includegraphics {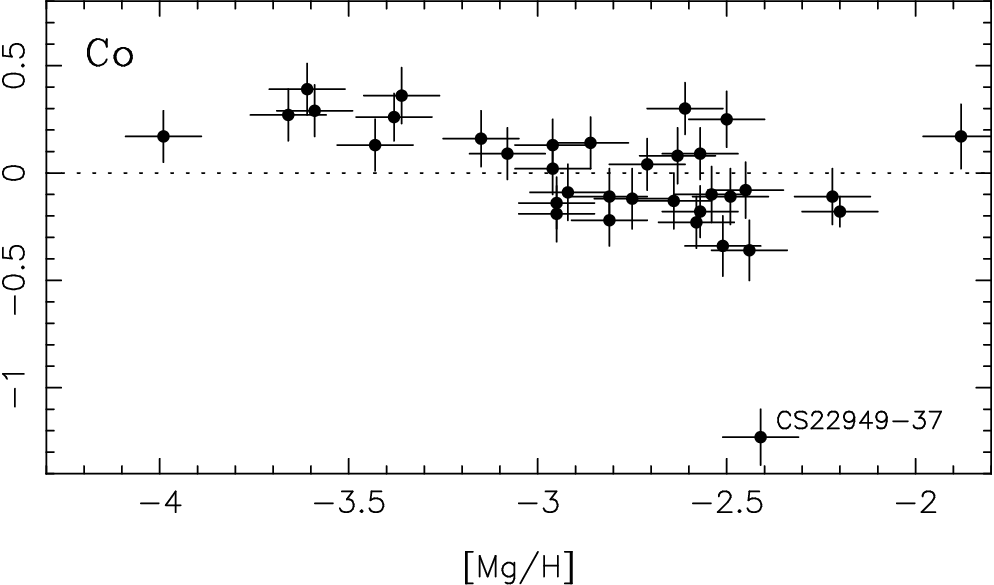} }
\resizebox  {5.8cm}{2.90cm} 
{\includegraphics {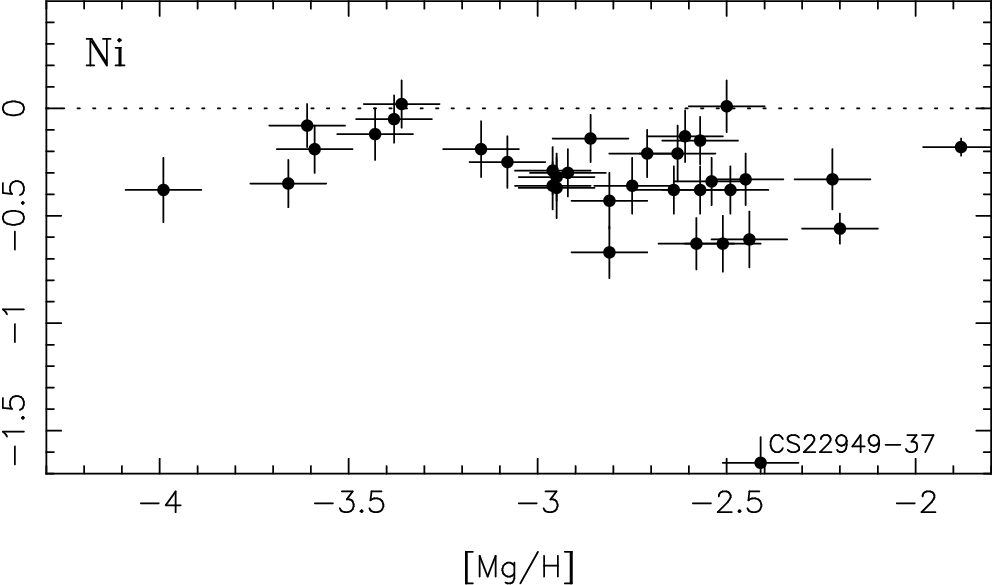} }
\resizebox  {5.8cm}{2.90cm} 
{\includegraphics {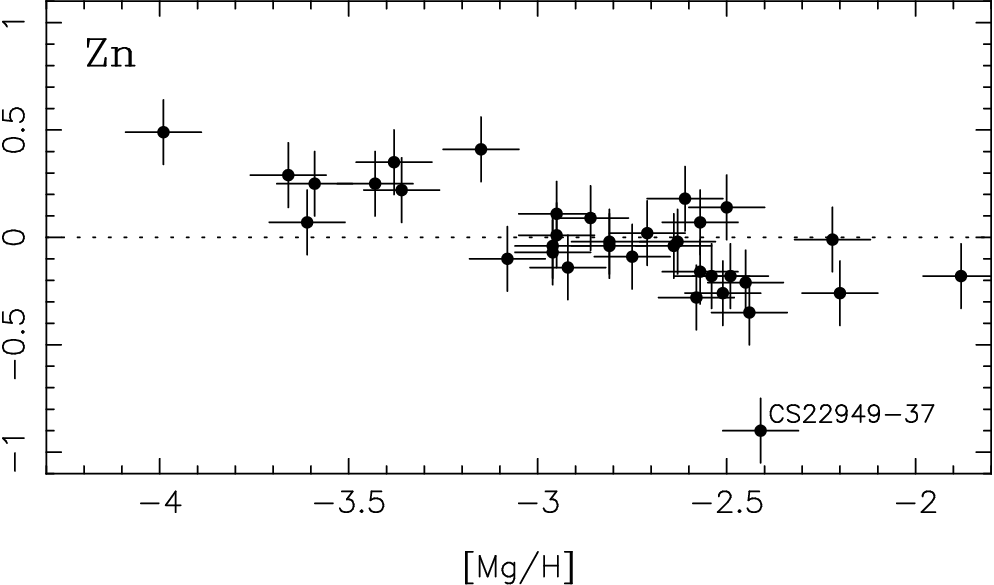} }
\caption {Abundance ratios [X/Mg] plotted vs. [Mg/H]. The scale is the
same for all the plots: the amplitude in the coordinate [X/Mg] is 2.2 dex.
For Mn the symbols are the same as in Fig. \ref{abCr-Mn}}
\label {abMg-SiZn}
\end {center} 
\end {figure*}

\subsection{Comparison with previous studies of very metal-poor stars}

The trends of the relations [X/Fe] vs. [Fe/H] reported in the present study are
generally in agreement with previous results in the literature, such as those of 
McWilliam et al. (\cite{MPS95}), Ryan et al. (\cite{RNB96}), and Norris, Ryan, 
\& Beers (\cite{NRB01}). 
However, the much smaller scatter of the ratios [X/Fe] is a notable result of 
the greatly improved spectra obtained for this study. 

Carretta et al. (\cite{CGC02}) observed three very metal-poor MS or TO stars and 
two extremely metal-poor giants ([Fe/H] $< -3.0$) with Keck spectra having a 
figure of merit F (Norris et al. \cite{NRB01}) larger than 600. The elements 
from Mg to Fe were analyzed. Within a (rather large) scatter the two analyses 
are compatible, except for a few elements, in particular Cr (their slope of 
[Cr/Fe] vs. [Fe/H] seems to be steeper than ours).

\subsection{Mg as an alternative ``reference element''}\label{sec-refelem}

Iron is a convenient ``reference element'' in high-resolution spectral analyses, 
because it has by far the largest number of usable lines and is represented by 
two ionization states. Iron may not be the best choice as a tracer of Galactic 
chemical evolution, however, since its nucleosynthesis channels are not very 
well understood, and are not necessarily even unique (e.g., explosive 
nucleosynthesis, Si burning in massive SNe II, or SNe Ia). 

From the point of view of the chemical evolution of the Galaxy oxygen would be a 
better choice (see Wheeler, Sneden \& Truran \cite{WST89}), as oxygen is the 
most abundant element after H and He, it comes from a single source, 
and its abundance should not be significantly 
affected in the explosive phase. However, its well-known 
observational difficulties would considerably degrade the accuracy of the 
derived trends: Oxygen could be measured in only 21 of our programme stars and 
the uncertainties on its abundance are large (see in particular Sect. 
\ref{sec-errors} and the error bars in Fig. \ref{ab-o}).

Mg or Ca might be good alternatives. The abundances of
these elements are accurately determined, and they are also formed mainly in
massive SNe. We choose Mg rather than Ca here (Fig. \ref{abMg-SiZn}),
because Mg is more robust in the sense that its production is dominated
by hydrostatic burning processes and it is also less affected by
explosive burning and by ``fallback'' (Woosley \& Weaver \cite{WW95}). Note that
Shigeyama \& Tsujimoto (\cite{ST98}) also recommended Mg rather than Fe as a 
useful reference element, following much the same logic. We note, however, that 
the ``plateau''-like behavior of [Mg/Fe] with increasing iron abundance in the 
range $-$4 to $-$2 implies that Mg and Fe have parallel early nucleosynthesis 
histories. Therefore, the trends of elemental ratios with [Fe/H] found in 
section \ref{CZn-Fe} should survive if [Fe/H] is replaced by [Mg/H] as a 
metallicity indicator. 

Yet, the new diagrams (see Fig. \ref{abMg-SiZn}) present a few notable 
differences. One is expected -- the scatter is never as low as in some of the 
earlier diagrams because [Mg/H] is determined less accurately than [Fe/H], 
being based on 8 lines instead of 150. The other -- new -- result is very
interesting: rather than a roughly linear variation over the full range $-4.0<$ 
[Mg/H] $<-2.5$, there is a hint that all abundance ratios are flat in the 
interval $-4.0<$ [Mg/H] $<-3.0$, with something qualitatively different 
occurring at higher metallicity. 

This pattern is in fact what is expected if the first SNe are primordial and 
have specific yields: the plateau between $-4.0$ and $-3.0$ would then reflect a 
pure zero-metallicity type of SNe, whereas at higher metallicity we may observe 
a mix of primordial and non-primordial SNe. At still higher metallicities ($\ge$ 
--2.0), the scatter is expected to decrease again because so many SN precursors 
are involved that any differences average out. 

We now consider a few elements of particular interest.

$\bullet$ Carbon\\ 
Karlsson \& Gustafsson (\cite{KG01}) have statistically simulated the chemical 
enrichment of a metal-poor system, assuming that the stellar yields are 
one-dimensional functions of the progenitor mass of the supernovae and the 
masses of the supernovae are distributed according to a Salpeter IMF. In 
particular, they computed the distribution of the abundance ratio [C/Mg]
vs. [Mg/H] in a hypothesized sample of 500 XMP stars (their Fig. 3), adopting 
the yields of either Woosley \& Weaver (\cite{WW95}) or Nomoto et al. 
(\cite{NHT97}). Our Fig. \ref{abMg-SiZn} for C is compatible with their Fig. 3a 
(yields of Woosley \& Weaver), but not with their Fig. 3b (yields of Nomoto et 
al.). 

On the other hand, both in their simulation and as observed in our present 
sample, the [C/Mg] ratio seems to decrease with increasing [Mg/H].
Following Karlsson and Gustafsson (\cite{KG01}), this effect could be the result
of different supernova masses operating at different metallicities.  SNe
producing a high [C/Mg] ratio produce only small amounts of Mg; on the contrary, 
SNe producing a low [C/Mg] ratio also produce substantial Mg.  

\begin {figure}
\begin {center}
\resizebox  {6.cm}{6.cm} 
{\includegraphics {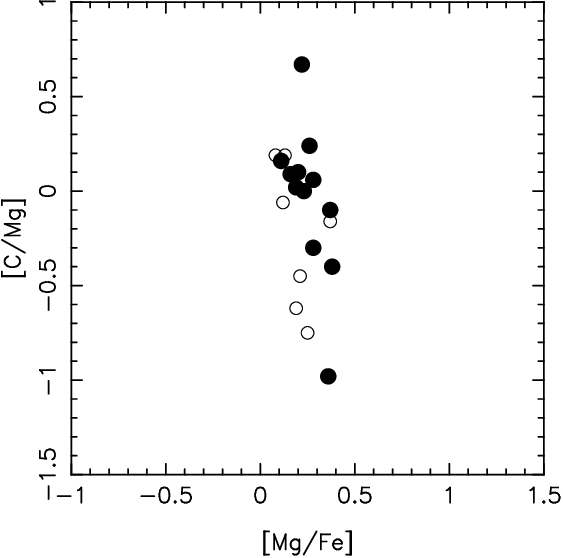}}
\caption {[C/Mg] vs. [Mg/Fe] for all stars of our sample with temperatures 
higher than 4800 K (and thus presumably not mixed). Filled circles represent 
the most metal-poor stars ([Fe/H] $\leq -3.0$), open circles the stars with 
[Fe/H]$> -3.0$.  The variation of [C/Mg] is larger than expected from the 
computations of Karlsson \& Gustafsson (\cite{KG01}), suggesting that
the scatter in [C/Mg] is not explained by the theoretical yields.
}
\label {abmg-C-Mg}
\end {center}  
\end {figure}

Karlsson \& Gustafsson show that the patterns they predict become barely
visible (or invisible) in an observational sample smaller than N $\approx
500$ and affected by uncertainties of the order of 0.1 dex. We have examined how
their [C/Mg] vs. [Mg/Fe] diagram would appear for our sample (Fig. 
\ref{abmg-C-Mg}). Not only is no fine structure visible, but our diagram is 
considerably more extended in the vertical direction, strongly suggesting that 
the scatter in [C/Mg] is not explained by the theoretical yields.

$\bullet$ Sodium and Aluminium\\ [0cm]
[Na/Fe] and [Al/Fe] exhibit very different behaviours as functions of [Fe/H] 
(Fig. \ref{abFe-NaAl}). However, when Mg is used as a reference element (Fig. 
\ref{abMg-SiZn}), the behavior of these elements appears rather similar, and a 
plateau at the lowest metallicities appears for Na as well as for Al: below 
$\mathrm{[Mg/H] = -3.0}$, [Na/Mg] remains constant at about [Na/Mg] = 0.9 (in 
fact this plateau appears also in Fig. \ref{abFe-NaAl}:  [Na/Fe] $\approx 0.7$ 
for [Fe/H]$<-3.4$). However, the rise in [Na/Mg] at higher metallicity is not 
seen for Al.

The discrepant position of CS~22952--015 in the diagrams of [Na/Mg] and [Al/Mg] 
vs. [Mg/H] will be discussed in Sect. \ref{pec-stars}.

$\bullet$ Iron peak elements\\ 
There is no clear slope of [Cr/Mg] vs. [Mg/H], as was seen 
for [Cr/Fe] vs. [Fe/H]. This raises the suspicion that the slope in the latter
diagram may be an artifact due to different NLTE corrections for the two 
elements as a function of metallicity (Th\'evenin \& Idiart \cite{TI99}). These 
corrections have not been applied, as they are not known for Cr and have not 
been published line by line for Fe. We return to this question below.

No significant slope is found for [Mn/Mg], (in agreement with absence of slope for [Mn/Fe]). Similarly, for [Mg/H]$< -3.0$ no clearly significant slope is found for the other elements (Fe, Co, Ni, Zn) .

\subsection{Cosmic vs. observational scatter in the abundance ratios}
\label{sec-scatter}
A key motivation for the present programme was to explore to what extent the 
scatter in the observed abundance ratios is due to observational error, and to 
what extent it reflects physical conditions in the early Galaxy when these stars 
were formed. McWilliam et al. (\cite{MPS95}) and McWilliam \& Searle (\cite{MS99})  
already noted that the scatter in 
some of their diagrams of [X/Fe] vs. [Fe/H] could be entirely accounted for by 
observational errors. The issue was summarized by Ryan, Norris, \& Beers 
(\cite{RNB96}) as follows. ``The abundance patterns, especially those of Cr, Mn, 
or Co, raise the following question: why should all halo supernova ejecta around 
this epoch that possess a particular [Cr/Fe] ratio (or [Mn/Fe] ratio or [Co/Fe] 
ratio) {\it subsequently} form into stars of the same [Fe/H]? Put differently, 
how do the ejecta know how much interstellar hydrogen to combine with?''.

Our observations were designed to achieve twice the spectral resolution and 3--4 
times the S/N ratio of the earlier data in order to test this very point.
With our much lower observational errors, we can conclude that while the 
scatter for C, Na, Mg, Al, and Si is probably real, the very small scatter of
Ca, Cr, and Ni {\it still} do not leave room for the existence of an intrinsic 
scatter!

Consider the case of chromium, which has the lowest observed scatter
(r.m.s. 0.05 dex).  
The problem mentioned earlier is very acute and derives from the {\it
simultaneous} absence of scatter and presence of of a slope of
[Cr/Fe] versus metallicity.  Although one can
argue that the amount of  hydrogen swept up by the
ejecta is mainly determined by the energy of the explosion of the SN
(Cioffi et al. \cite{CMB88}), thus relating the abundance ratios
produced by the SN to the final [Fe/H] of the enriched gas, it is
still difficult to believe that there is so little room for 
noise in the mixing process.
However, if the slope is in fact spurious (e.g., due to neglected differential 
NLTE corrections between Cr and Fe), as suggested by the diagram of [Cr/Mg] vs. 
[Mg/H] (Fig. \ref{abMg-SiZn}), the problem vanishes. One would simply conclude
 that Cr and Fe are produced together, 
independent of the metallicity of the SN progenitor, and  the amount
of mixing cannot be localized anymore along the metallicity axis. 
Until detailed NLTE computations for Cr and Fe become available we cannot decide 
if this interpretation is correct.

For all the elements discussed here, our results show that the scatter of their 
production ratios is very small, far below the values derived earlier. This 
implies that we are observing either the ejecta of fairly large bursts of 
massive stars, so the sampling of the IMF is reasonably good, or the result of 
several events promptly mixed by strong turbulence.

\subsection{The nature of the first supernovae}

Theoretical work  (Bromm et al. \cite{BFC01}  and references therein) predict that
the first stellar generation is made of very massive stars,  with masses
above 100 M$_{\odot}$, because zero-metal matter lacks adequate cooling
mechanisms for fragmenting down to classical supernova-progenitor masses.
Results of WMAP (Kogut et al. \cite{KSB03}) on an early reionization of the 
Universe have triggered further claims (Cen \cite{Cen03}) of a very massive stellar
generation.This has  very important implications for early 'stellar'
nucleosynthesis. According to current models, such stars either end up  
as pair-instability supernovae, or as collapsed black holes, in the latter case
with no contribution to the metal enrichment of the ISM
(Heger \& Woosley \cite{HW02}; Umeda \& Nomoto \cite{UN02}).
A comparison of the yields of Heger \& Woosley with our results show a 
clear disagreement, in particular the predicted strong odd-even effect,
not seen in our observations, and  a strong decline
of Zn with metallicity also not observed.
Nakamura et al. (\cite{NUI01}) and Nomoto et al. (\cite{NHT97}) have
computed yields of SNe with a progenitor mass of 25 M$_{\odot}$
of very high energy (also called hypernovae). 
Their predicted yields have some
positive features at very low metallicity, such as 
the high [Zn/Fe] and [Co/Fe], and a low [Mn/Fe], as observed.
However they also predict a lack of [O/Fe] enhancement, in clear
disagreement with our observations.

Our conclusion is that classical SNe are still the best candidates
to explain our observational results.

\subsection{Peculiar objects} \label{pec-stars}
\subsubsection{CS~22949--037}

The highly peculiar abundances of CS~22949--037 were studied in detail by 
Depagne et al. (\cite{DHS02}). They may be explained by a single progenitor or 
by an enrichment event dominated by massive SNe II with substantial fallback; 
this applies also to the similar star CS~29498--043 analysed by Aoki et al. 
(\cite{ANR02a}, \cite{ANR02b}).

Tsujimoto \& Shigeyama (\cite{TS03}) propose another interpretation of
CS~22949--037 and CS~29498--043. The high [Mg/Fe] ratio could be due to a 
low-energy explosion, strong enough to eject the layers containing the light
elements, but ejecting little iron and other iron-peak elements. In this
interpretation, the lack of Fe relative to the lighter elements O and Mg is
also associated with a large fallback on the remnant, but due to a low explosion 
energy rather than a large mass of the collapsed core as in the model adopted by 
Depagne et al. (\cite{DHS02}). However, the normal Cr/Mn/Fe/C/Ni ratios observed 
in this star would rather suggest a normal explosion
energy, the larger fallback being due to a larger mass of the single or
multiple progenitors. 

Further exploration of the precise abundance patterns of these two putative 
low-energy supernova descendents should be quite interesting. This 
interpretation would link the exceptional cases of the most extreme metal-poor 
stars with rather low-mass progenitors. It would, however, conflict with
the usual interpretation which associate the more massive progenitors with both 
the earliest explosions and the largest volume of hydrogen swept up, resulting 
in low metallicity in the ISM. The problem clearly requires further 
investigation.

\subsubsection{CS~22952--015}

CS~22952--015 is known to be Mg deficient (McWilliam et al., \cite{MPS95}). In 
the abundance ratio plots vs. [Mg/H] (Fig. \ref{abMg-SiZn}), it does indeed 
appear Fe-rich, but its most notable characteristic is the large values of 
[Na/Mg] and [Al/Mg]. An interesting point is that this effect is not seen as 
clearly in the diagrams of [Na/Fe] and [Al/Fe] vs. [Fe/H] (Fig. 
\ref{abFe-NaAl}), because all three elements Na, Al, and Fe are enhanced 
relative to Mg in CS~22952--015.

\subsubsection{CS~22169--035}

CS~22169--035 appears to be particularly deficient in Ti (Fig. \ref{ab-mgti}), 
but in fact all the ratios [Mg/Fe], [Si/Fe], [Ca/Fe], [Co/Fe], [Ni/Fe], and 
[Zn/Fe] are low as well. When Mg is used as the reference element, this star has 
a normal position in the diagrams, and the abundance anomalies are most simply 
characterised as a deficiency of Fe.

\subsection{Yields of the first supernovae} 

With Mg chosen as the reference element (Fig. \ref{abMg-SiZn}), the most 
metal-poor stars in our sample ([X/H]$<-2.9$) define a plateau at abundance 
ratios [X/Mg] corresponding to the yields of the first supernovae, thus 
providing constraints on these yields. The mean value of [X/Mg] of each plateau 
is given in Table \ref{yieldsuper} and represent our best estimate of the yields 
from the first metal producers in the Galaxy.

\begin {table}[t]
\caption {Mean [X/Mg] ratios for stars with [Mg/H]$<-2.9$, corresponding to the 
yields of the first supernovae. The r.m.s. is the scatter around the mean.}
\label{yieldsuper}
\begin {center}
\begin {tabular}{lrrrr}
\hline
          &      &$r.m.s.$&  n    \\     
$[$Na/Mg] & $-$0.84 & 0.22 & 11   \\
$[$Al/Mg] & $-$0.33 & 0.15 & 13   \\
$[$Si/Mg] & $+$0.21 & 0.14 & 14   \\
$[$K/Mg]  & $-$0.14 & 0.14 & 13   \\
$[$Ca/Mg] & $+$0.06 & 0.09 & 14   \\
$[$Sc/Mg] & $-$0.17 & 0.14 & 14   \\
$[$Ti/Mg] & $-$0.01 & 0.09 & 14   \\
$[$Cr/Mg] & $-$0.63 & 0.09 & 14   \\
$[$Mn/Mg] & $-$0.65 & 0.20 & 14   \\
$[$Fe/Mg] & $-$0.21 & 0.10 & 14   \\
$[$Co/Mg] & $+$0.13 & 0.17 & 14   \\
$[$Ni/Mg] & $-$0.23 & 0.13 & 14   \\
$[$Zn/Mg] & $+$0.15 & 0.19 & 14   \\
\hline
\end {tabular}
\end {center}
\end {table}

\begin {figure}
\begin {center}
\resizebox  {7.cm}{3.5cm} 
{\includegraphics {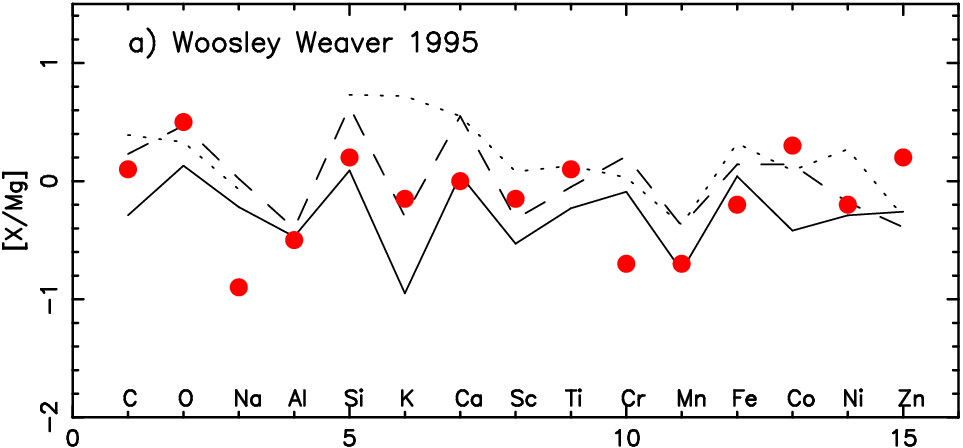} }
\resizebox  {7.cm}{3.5cm} 
{\includegraphics {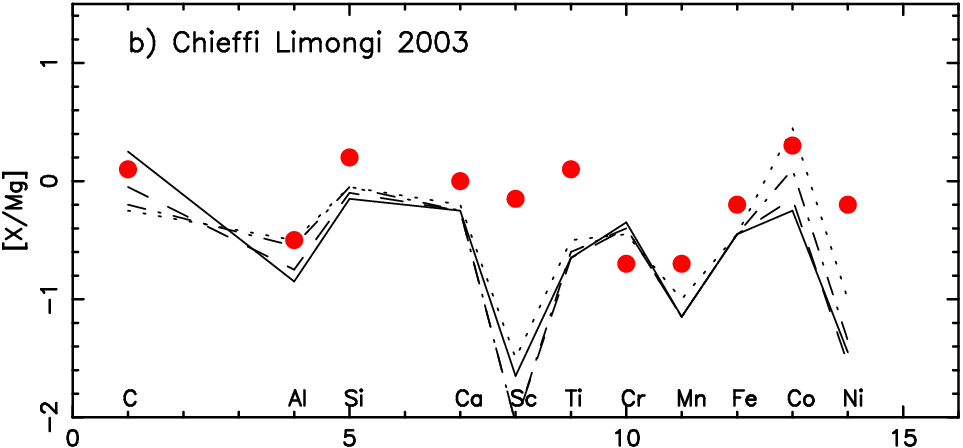} }
\caption {Mean [X/Mg] values for [Mg/H]$<-3$ (Fig. \ref{ab-mgti} and Table 
\ref{yieldsuper}) compared to the model yields by: 
{\bf a)} Woosley and Weaver for 15 (dotted line), 25 (dashed), and 35 
$M_{\odot}$ SNe (full), and 
{\bf b)} Limongi and Chieffi for 15 (dotted line), 20 (dotted-dashed), 35 
(dashed), and 50 $M_{\odot}$ (full).
}
\label{compayields}
\end {center}  
\end {figure}

In Fig. \ref{compayields}a, these mean values are compared to the values
predicted by Woosley \& Weaver (\cite{WW95}) for their zero-metal supernova
models 15A, 25B and 35C (progenitor masses 15, 25 and 35 $M_{\odot}$). The 
Woosley \& Weaver values of [X/Mg] for the elements from C to Ca have been
enhanced in order to bring their predicted value of [Fe/Mg] into agreement with
our observations. 

In Fig. \ref{compayields}b we compare with the zero-metal models of Chieffi \& 
Limongi (\cite{CL03}) for 15, 20, 35, 50 $M_{\odot}$ stars. The predicted [X/Mg] 
values were already adjusted by Chieffi \& Limongi to obtain [Mg/Fe]$\approx 
+0.45$). These comparisons suggest that some adjustment of the models is indeed 
required.

The connection between the abundances observed in these very old stars and those 
observed in intergalactic clouds -- in particular the damped Ly$_\alpha$ 
systems (DLA) -- will be further discussed in subsequent papers.

\section {Conclusions}

We have studied the abundances of 17 elements from C to Zn in a sample of 35
halo giant stars in the metallicity range --4.0 $<$ [Fe/H] $<$ --2.7. Our 
VLT/UVES spectra have resolving power $R = 47,000$ and S/N ratio per pixel 
between 100 and 200 -- far better than former data obtained with 4-m class 
telescopes. The lowest possible metallicity range was chosen because, according 
to previous theoretical work, this is where one expects to see the imprint of SN 
ejecta from either single SNe or single bursts of star formation.

$\bullet$ We have shown that in very metal-poor giants the continuous opacity in the UV is dominated by the Rayleigh scattering, and it is therefore crucial in this region to properly account for continuum scattering, as it has been done in 
this work. (A still better approach would be to take also scattering into account
in the line formation).

$\bullet$ Our first and probaby most important conclusion is
the existence of a surprisingly well-defined pattern of abundance ratios, with a
cosmic scatter often at (or below) the level of detection
(with the exception of carbon). Given the exceptional
quality of our observations, our upper limit to the cosmic scatter is 3-4 times 
lower than former determinations.
  
$\bullet$ Trends with metallicity, visible as mild slopes in diagrams of [X/Fe] 
vs. [Fe/H] (X being any element between Na and Zn), tend to disappear 
when [X/Mg] vs. [Mg/H] are chosen as the 
diagnostic diagrams (at least when [Mg/H] $ \leq -3.0$). 
 At the lowest [Mg/H], practically no slope is present, 
suggesting that the level of ``primordial yields'' may have been reached.

$\bullet$ In the diagram [Cr/Fe] vs. [Fe/H] the intrinsic scatter is so small
as to raise the suspicion that the mild slope might be due to residual, 
metallicity-dependent, differential NLTE corrections to our abundances for Cr 
and Fe, rather than to a variation of the [Cr/Fe] ratio in the yields. 
The lack of any slope when Mg is taken as the reference element supports this 
hypothesis. Moreover, if there were real changes in the [Cr/Fe] ratio in the yields, we would have to explain how the ejecta could be diluted in the environment without producing significant scatter in [Cr/Fe] at given [Fe/H], a rather unlikely scenario.
Our result for [Cr/Fe] makes clear that progress on the interpretation
of our observational data requires further progress in stellar atmosphere
theory. NLTE computations are available for only a few elements, and 3-D effects
have not yet been computed for giants -- a special concern for our sample, which 
consists exclusively of giants. 

$\bullet$ The low scatter in our observed abundance ratios suggests that they 
are the results of enrichment events that {\it are not} single SNe (we should 
then see scatter due to the different masses of each SN), but rather of single 
``burst'' events.
 
$\bullet$ Our results are clearly incompatible with the predicted yields of
pair-instability supernovae/hypernovae (Umeda \& Nomoto \cite{UN02}, Heger \& Woosley \cite{HW02}). Neither the expected strong odd-even
effect nor the predicted Zn deficiency is observed. Very high-energy SNe, with
explosive energies of the order of $10^{53}$ ergs, are not good candidates
either, as they under-produce oxygen (Nakamura et al. \cite{NUN99}).  Only when 
mixing and fallback are added is a better fit obtained (Umeda \& Nomoto 
\cite{UN02}).

$\bullet$ A detailed comparison of our observations with theoretical supernova 
yields is deferred to a forthcoming paper. However, a first comparison has been 
made with standard sources of theoretical yields of SNe with progenitor masses 
in the usual range from 12 to 70 $M_{\sun}$. While not perfect, they still give 
the best available fit, at least when compared to more exotic processes. This 
has implications for the kinds of massive stars that are thought to have 
reionized the Universe as early as 200 million years after the Big Bang,
according to the recent results from WMAP.
    
Our abundance results for elements heavier than Zn (e.g., Sr, Ba, etc.) and 
detailed comparisons with Galactic chemical evolution models will be discussed 
in forthcoming papers.

\begin {acknowledgements}
This paper makes use of data from the DENIS survey and from the 2MASS All Sky
Survey (a joint project of the University of Massachusetts and the Infrared
Processing and Analysis Center/California Institute of Technology, funded by
the National Aeronautics and Space Administration and the National Science
Foundation). We thank G. Simon for the communication of DENIS photometry in
advance of publication, A. Chieffi for very useful comments, and the ESO staff 
for assistance during all the runs of our Large Programme. TCB acknowledges
partial support from grants AST 00-98508 and AST 00-98549 awarded by the U.S.
National Science Foundation. BA \& JA thank the Carlsberg Foundation and the
Swedish and Danish Natural Science Research Council for partial financial
support of this work.
\end {acknowledgements}


\begin {table*}[t]
\caption {Abundances of the elements in the selected stars. Na Al and 
K are NOT corrected for non-LTE effects and O is computed with 
1D models}
\label {tab-abund-a}
\begin {center}
\begin {tabular}{|cccccc|}
\multicolumn {4}{c}{HD~2796}      [Fe/H]=--2.47\\
\hline
     & $log \varepsilon$& [M/H]&  [M/Fe] & $\sigma$ & N  \\
C    &     5.55  &  -2.97   &  -0.51  &      -   &     - \\
N    &       -   &    -     &   -     &      -   &     - \\
O    &     6.80  &  -1.97   &  0.50   &      -   &     - \\
Na   &     4.20  &  -2.13   &  0.34   &      -   &     2 \\
Mg   &     5.36  &  -2.22   &  0.25   &    0.14  &     7 \\
Al   &     3.35  &  -3.12   & -0.66   &      -   &     2 \\
Si   &     5.48  &  -2.07   &  0.40   &      -   &     1 \\
K    &     3.25  &  -1.87   &  0.60   &      -   &     2 \\
Ca   &     4.22  &  -2.14   &  0.32   &    0.10  &    16 \\
Sc   &     0.79  &  -2.38   &  0.09   &    0.11  &     7 \\
Ti I &     2.75  &  -2.27   &  0.20   &    0.06  &    12 \\
Ti II&     2.79  &  -2.23   &  0.24   &    0.09  &    28 \\
Cr   &     2.95  &  -2.72   & -0.26   &    0.13  &     7 \\
Mn   &     2.54  &  -2.85   & -0.39   &    0.01  &     3 \\
Fe I &     5.05  &  -2.45   &  0.01   &    0.13  &   114 \\
Fe II&     5.02  &  -2.48   & -0.02   &    0.13  &    17 \\
Co   &     2.59  &  -2.33   &  0.14   &    0.08  &     2 \\
Ni   &     3.70  &  -2.55   & -0.09   &    0.14  &     3 \\
Zn   &     2.37  &  -2.23   &  0.24   &      -   &     1 \\
\hline
\end {tabular}
\begin {tabular}{|cccccc|}
\multicolumn {4}{c}{HD~122563}     [Fe/H]=--2.82\\
\hline
     & $log \varepsilon$& [M/H]&  [M/Fe] & $\sigma$ & N\\
C    &     5.30  &   -3.22  &   -0.41  &      -   &     - \\
N    &       -   &     -    &      -   &      -   &     - \\
O    &     6.57  &   -2.20  &    0.62  &      -   &     - \\
Na   &     3.75  &   -2.58  &    0.23  &      -   &     2 \\
Mg   &     5.13  &   -2.45  &    0.36  &    0.12  &     8 \\
Al   &     3.28  &   -3.19  &   -0.38  &      -   &     2 \\
Si   &     5.20  &   -2.35  &    0.47  &      -   &     1 \\
K    &     2.82  &   -2.30  &    0.51  &      -   &     1 \\
Ca   &     3.86  &   -2.50  &    0.31  &    0.10  &    16 \\
Sc   &     0.46  &   -2.71  &    0.11  &    0.10  &     7 \\
Ti I &     2.32  &   -2.70  &    0.12  &    0.10  &    14 \\
Ti II&     2.38  &   -2.64  &    0.18  &    0.10  &    28 \\
Cr   &     2.46  &   -3.21  &   -0.40  &    0.17  &     7 \\
Mn   &     2.17  &   -3.22  &   -0.41  &    0.06  &     3 \\
Fe I &     4.72  &   -2.78  &    0.03  &    0.17  &   142 \\
Fe II&     4.65  &   -2.85  &   -0.04  &    0.13  &    18 \\
Co   &     2.39  &   -2.53  &    0.29  &    0.13  &     4 \\
Ni   &     3.47  &   -2.78  &    0.03  &    0.09  &     3 \\
Zn   &     1.94  &   -2.66  &    0.16  &      -   &     2 \\
\hline
\end {tabular}
\begin {tabular}{|cccccc|}
\multicolumn {4}{c}{HD~186478}    [Fe/H]=--2.59\\
\hline
     & $log \varepsilon$& [M/H]&  [M/Fe] & $\sigma$ & N\\
C    &     5.60  &   -2.92  &   -0.34  &      -   &     - \\
N    &     5.75  &   -2.17  &    0.42  &      -   &     - \\
O    &     6.93  &   -1.84  &    0.75  &      -   &     - \\
Na   &     3.94  &   -2.39  &    0.20  &      -   &     2 \\
Mg   &     5.38  &   -2.20  &    0.39  &    0.08  &     7 \\
Al   &     3.32  &   -3.15  &   -0.57  &      -   &     2 \\
Si   &     5.40  &   -2.15  &    0.44  &      -   &     1 \\
K    &     3.19  &   -1.93  &    0.66  &      -   &     2 \\
Ca   &     4.21  &   -2.15  &    0.44  &    0.10  &    16 \\
Sc   &     0.68  &   -2.49  &    0.10  &    0.12  &     7 \\
Ti I &     2.68  &   -2.34  &    0.25  &    0.07  &    13 \\
Ti II&     2.72  &   -2.30  &    0.29  &    0.10  &    26 \\
Cr   &     2.76  &   -2.91  &   -0.33  &    0.14  &     7 \\
Mn   &     2.33  &   -3.06  &   -0.48  &    0.04  &     3 \\
Fe I &     4.93  &   -2.57  &    0.01  &    0.18  &   140 \\
Fe II&     4.90  &   -2.60  &   -0.01  &    0.14  &    18 \\
Co   &     2.54  &   -2.38  &    0.21  &    0.19  &     4 \\
Ni   &     3.49  &   -2.76  &   -0.18  &    0.15  &     3 \\
Zn   &     2.14  &   -2.46  &    0.13  &      -   &     1 \\
\hline
\end {tabular}
\begin {tabular}{|cccccc|}
\multicolumn {4}{c}{BD+17:3248}    [Fe/H]=--2.07\\
\hline
     & $log \varepsilon$& [M/H]&  [M/Fe] & $\sigma$ & N\\
C    &     6.02  &  -2.50  &   -0.44  &      -   &     - \\
N    &     6.42  &  -1.50  &    0.57  &      -   &     - \\
O    &     7.39  &  -1.38  &    0.69  &      -   &     - \\
Na   &     4.95  &  -1.38  &    0.69  &      -   &     2 \\
Mg   &     5.70  &  -1.88  &    0.19  &    0.10  &     7 \\
Al   &     3.70  &  -2.77  &   -0.71  &      -   &     2 \\
Si   &     6.00  &  -1.55  &    0.52  &      -   &     2 \\
K    &     3.88  &  -1.24  &    0.82  &      -   &     2 \\
Ca   &     4.69  &  -1.67  &    0.40  &    0.08  &    14 \\
Sc   &     1.30  &  -1.87  &    0.20  &    0.16  &     6 \\
Ti I &     3.18  &  -1.84  &    0.23  &    0.08  &    12 \\
Ti II&     3.28  &  -1.74  &    0.33  &    0.17  &    29 \\
Cr   &     3.39  &  -2.28  &   -0.22  &    0.20  &     6 \\
Mn   &     2.86  &  -2.53  &   -0.47  &    0.02  &     3 \\
Fe I &     5.46  &  -2.04  &    0.02  &    0.15  &   141 \\
Fe II&     5.41  &  -2.09  &   -0.03  &    0.09  &    16 \\
Co   &     3.21  &  -1.71  &    0.36  &      -   &     2 \\
Ni   &     4.19  &  -2.06  &    0.00  &    0.08  &     3 \\
Zn   &     2.54  &  -2.06  &    0.00  &      -   &     1 \\
\hline
\end {tabular}
\begin {tabular}{|cccccc|}
\multicolumn {4}{c}{BD-18:5550}    [Fe/H]=--3.06\\
\hline
     & $log \varepsilon$& [M/H]&  [M/Fe] & $\sigma$ & N\\
C    &     5.44  &  -3.08  &   -0.02  &      -   &     - \\
N    &       -   &    -    &      -   &      -   &     - \\
O    &     6.13  &  -2.64  &    0.42  &      -   &     - \\
Na   &     3.32  &  -3.01  &    0.05  &      -   &     2 \\
Mg   &     4.83  &  -2.75  &    0.31  &    0.15  &     8 \\
Al   &     2.82  &  -3.65  &   -0.59  &      -   &     2 \\
Si   &     4.88  &  -2.67  &    0.39  &      -   &     1 \\
K    &     2.58  &  -2.54  &    0.52  &      -   &     2 \\
Ca   &     3.71  &  -2.65  &    0.41  &    0.10  &    16 \\
Sc   &     0.15  &  -3.02  &    0.04  &    0.09  &     7 \\
Ti I &     2.12  &  -2.90  &    0.16  &    0.04  &    13 \\
Ti II&     2.10  &  -2.92  &    0.14  &    0.09  &    30 \\
Cr   &     2.27  &  -3.40  &   -0.34  &    0.10  &     7 \\
Mn   &     1.95  &  -3.44  &   -0.38  &    0.05  &     3 \\
Fe I &     4.44  &  -3.06  &    0.00  &    0.12  &   148 \\
Fe II&     4.44  &  -3.06  &    0.00  &    0.10  &    17 \\
Co   &     2.05  &  -2.87  &    0.19  &    0.12  &     3 \\
Ni   &     3.14  &  -3.11  &   -0.05  &    0.10  &     3 \\
Zn   &     1.76  &  -2.84  &    0.22  &      -   &     1 \\
\hline
\end {tabular}
\begin {tabular}{|cccccc|}
\multicolumn {4}{c}{CD-38:245}    [Fe/H]=--4.19\\
\hline
     & $log \varepsilon$& [M/H]&  [M/Fe] & $\sigma$ & N\\
C    &  $<$4.00  &$<$-4.52  & $<$-0.33 &      -   &     - \\
N    &       -   &     -    &      -   &      -   &     - \\
O    &       -   &     -    &      -   &      -   &     - \\
Na   &     2.08  &   -4.25  &   -0.06  &      -   &     2 \\
Mg   &     3.59  &   -3.99  &    0.20  &    0.08  &     6 \\
Al   &     1.61  &   -4.86  &   -0.67  &      -   &     2 \\
Si   &     3.56  &   -3.99  &    0.20  &      -   &     1 \\
K    &       -   &      -   &      -   &      -   &     - \\
Ca   &     2.37  &   -3.99  &    0.20  &    0.17  &     8 \\
Sc   &    -0.98  &   -4.15  &    0.04  &      -   &     2 \\
Ti I &     1.20  &   -3.82  &    0.37  &    0.04  &     5 \\
Ti II&     1.08  &   -3.94  &    0.25  &    0.11  &    24 \\
Cr   &     1.04  &   -4.63  &   -0.44  &    0.12  &     5 \\
Mn   &     0.61  &   -4.78  &   -0.60  &    0.03  &     3 \\
Fe I &     3.30  &   -4.20  &   -0.01  &    0.20  &    95 \\
Fe II&     3.33  &   -4.17  &    0.02  &    0.13  &     7 \\
Co   &     1.10  &   -3.82  &    0.37  &    0.08  &     3 \\
Ni   &     1.88  &   -4.37  &   -0.19  &      -   &     2 \\
Zn   &     1.10  &   -3.50  &    0.69  &      -   &     1 \\
\hline
\end {tabular}
\end {center}
\end{table*}

\begin {table*}[t]
\caption {Abundances of the elements in the selected stars. Na Al and 
K are NOT corrected for non-LTE effects and O is computed with 
1D models}
\label {tab-abund-b}
\begin {center}
\begin {tabular}{|cccccc|}
\multicolumn {4}{c}{BS~16467--062}   [Fe/H]=--3.77\\
\hline
     & $log \varepsilon$& [M/H]&  [M/Fe] & $\sigma$ & N\\
C    &     5.00  &   -3.52  &    0.25  &      -   &     - \\
N    &       -   &     -    &      -   &      -   &     - \\
O    &       -   &     -    &      -   &      -   &     - \\
Na   &     2.39  &   -3.94  &   -0.17  &      -   &     2 \\
Mg   &     3.97  &   -3.61  &    0.16  &    0.09  &     7 \\
Al   &     1.80  &   -4.67  &   -0.90  &      -   &     2 \\
Si   &     4.20  &   -3.35  &    0.42  &      -   &     1 \\
K    &     1.78  &   -3.34  &    0.43  &      -   &     2 \\
Ca   &     2.94  &   -3.42  &    0.35  &    0.19  &    12 \\
Sc   &    -0.59  &   -3.76  &    0.01  &    0.06  &     4 \\
Ti I &     1.65  &   -3.37  &    0.40  &    0.17  &    11 \\
Ti II&     1.43  &   -3.59  &    0.18  &    0.18  &    23 \\
Cr   &     1.49  &   -4.18  &   -0.41  &    0.29  &     5 \\
Mn   &     1.07  &   -4.32  &   -0.55  &    0.03  &     3 \\
Fe I &     3.67  &   -3.83  &   -0.06  &    0.14  &   130 \\
Fe II&     3.79  &   -3.71  &    0.06  &    0.12  &     4 \\
Co   &     1.70  &   -3.22  &    0.55  &    0.10  &     4 \\
Ni   &     2.56  &   -3.69  &    0.08  &    0.03  &     3 \\
Zn   &     1.06  &   -3.54  &    0.23  &      -   &     1 \\
\hline
\end {tabular}
\begin {tabular}{|cccccc|}
\multicolumn {4}{c}{BS~16477--003}   [Fe/H]=--3.36\\
\hline
     & $log \varepsilon$& [M/H]&  [M/Fe] & $\sigma$ & N\\
C    &     5.50  &   -3.02  &    0.34  &      -   &     - \\
N    &       -   &     -    &      -   &      -   &     - \\
O    &       -   &     -    &      -   &      -   &     - \\
Na   &     3.04  &   -3.29  &    0.07  &      -   &     2 \\
Mg   &     4.50  &   -3.08  &    0.28  &    0.15  &     8 \\
Al   &     2.51  &   -3.96  &   -0.61  &      -   &     2 \\
Si   &     4.69  &   -2.86  &    0.50  &      -   &     1 \\
K    &     2.23  &   -2.89  &    0.47  &      -   &     1 \\
Ca   &     3.38  &   -3.98  &    0.38  &    0.15  &    17 \\
Sc   &    -0.18  &   -3.35  &    0.01  &    0.14  &     7 \\
Ti I &     1.98  &   -3.04  &    0.32  &    0.10  &    14 \\
Ti II&     1.90  &   -3.12  &    0.24  &    0.12  &    29 \\
Cr   &     1.94  &   -3.73  &   -0.38  &    0.12  &     7 \\
Mn   &     1.53  &   -3.86  &   -0.51  &    0.11  &     3 \\
Fe I &     4.14  &   -3.36  &    0.00  &    0.12  &   141 \\
Fe II&     4.15  &   -3.35  &    0.01  &    0.05  &    16 \\
Co   &     1.93  &   -2.99  &    0.37  &    0.09  &     4 \\
Ni   &     2.92  &   -3.33  &    0.03  &    0.07  &     3 \\
Zn   &     1.42  &   -3.18  &    0.18  &      -   &     1 \\
\hline
\end {tabular}
\begin {tabular}{|cccccc|}
\multicolumn {4}{c}{BS~17569--049}    [Fe/H]=--2.88\\
\hline
     & $log \varepsilon$& [M/H]&  [M/Fe] & $\sigma$ & N\\
C    &     5.45  &   -3.07  &   -0.22  &      -   &     - \\
N    &     5.80  &   -2.12  &    0.73  &      -   &     - \\
O    &       -   &     -    &      -   &      -   &     - \\
Na   &     3.87  &   -2.46  &    0.42  &      -   &     2 \\
Mg   &     4.95  &   -2.63  &    0.25  &    0.20  &     8 \\
Al   &     3.12  &   -3.35  &   -0.48  &      -   &     2 \\
Si   &     5.27  &   -2.28  &    0.60  &      -   &     1 \\
K    &     2.73  &   -2.39  &    0.49  &      -   &     2 \\
Ca   &     3.89  &   -2.47  &    0.41  &    0.13  &    16 \\
Sc   &     0.35  &   -2.82  &    0.06  &    0.16  &     7 \\
Ti I &     2.42  &   -2.60  &    0.28  &    0.06  &    13 \\
Ti II&     2.37  &   -2.65  &    0.23  &    0.12  &    30 \\
Cr   &     2.54  &   -3.13  &   -0.26  &    0.08  &     7 \\
Mn   &     2.09  &   -3.30  &   -0.43  &    0.05  &     3 \\
Fe I &     4.65  &   -2.85  &    0.03  &    0.17  &   147 \\
Fe II&     4.60  &   -2.90  &   -0.03  &    0.10  &    18 \\
Co   &     2.37  &   -2.55  &    0.33  &    0.09  &     4 \\
Ni   &     3.41  &   -2.84  &    0.04  &    0.07  &     3 \\
Zn   &     1.95  &   -2.65  &    0.23  &      -   &     1 \\
\hline
\end {tabular}
\begin {tabular}{|cccccc|}
\multicolumn {4}{c}{CS~22169--035}    [Fe/H]=--3.04\\
\hline
     & $log \varepsilon$& [M/H]&  [M/Fe] & $\sigma$ & N\\
C    &     5.20  &   -3.32  &    0.28  &      -   &     - \\
N    &       -   &     -    &      -   &      -   &     - \\
O    &       -   &     -    &      -   &      -   &     - \\
Na   &       -   &     -    &      -   &      -   &     0 \\
Mg   &     4.63  &   -2.95  &    0.09  &    0.11  &     8 \\
Al   &     2.56  &   -3.91  &   -0.87  &      -   &     2 \\
Si   &     4.80  &   -2.75  &    0.29  &      -   &     1 \\
K    &     2.50  &   -2.62  &    0.42  &      -   &     1 \\
Ca   &     3.45  &   -2.91  &    0.13  &    0.10  &    16 \\
Sc   &    -0.05  &   -3.22  &   -0.18  &    0.11  &     7 \\
Ti I &     1.94  &   -3.08  &   -0.04  &    0.02  &    11 \\
Ti II&     1.88  &   -3.14  &   -0.10  &    0.14  &    31 \\
Cr   &     2.20  &   -3.47  &   -0.43  &    0.18  &     7 \\
Mn   &     2.06  &   -3.33  &   -0.29  &    0.05  &     3 \\
Fe I &     4.46  &   -3.04  &    0.00  &    0.19  &   149 \\
Fe II&     4.46  &   -3.04  &    0.00  &    0.13  &    19 \\
Co   &     1.78  &   -3.14  &   -0.10  &    0.13  &     4 \\
Ni   &     2.93  &   -3.32  &   -0.28  &    0.17  &     3 \\
Zn   &     1.66  &   -2.94  &    0.10  &      -   &     1 \\
\hline
\end {tabular}
\begin {tabular}{|cccccc|}
\multicolumn {4}{c}{CS~22172--002}    [Fe/H]=--3.86\\
\hline
     & $log \varepsilon$& [M/H]&  [M/Fe] & $\sigma$ & N\\
C    &     4.63  &   -3.89  &   -0.03  &      -   &     - \\
N    &       -   &     -    &      -   &      -   &     - \\
O    &  $<$5.95  &$<$-2.82  & $<$1.04  &      -   &     - \\
Na   &     2.12  &   -4.21  &   -0.35  &      -   &     2 \\
Mg   &     3.92  &   -3.66  &    0.20  &    0.08  &     8 \\
Al   &     1.81  &   -4.66  &   -0.80  &      -   &     2 \\
Si   &     4.10  &   -3.45  &    0.41  &      -   &     1 \\
K    &     1.71  &   -3.41  &    0.45  &      -   &     2 \\
Ca   &     2.87  &   -3.49  &    0.37  &    0.10  &    11 \\
Sc   &    -0.80  &   -3.97  &   -0.11  &    0.06  &     6 \\
Ti I &     1.51  &   -3.51  &    0.35  &    0.09  &    14 \\
Ti II&     1.30  &   -3.72  &    0.14  &    0.12  &    28 \\
Cr   &     1.40  &   -4.27  &   -0.41  &    0.17  &     6 \\
Mn   &     0.94  &   -4.45  &   -0.59  &    0.02  &     3 \\
Fe I &     3.64  &   -3.86  &    0.00  &    0.17  &   139 \\
Fe II&     3.64  &   -3.86  &    0.00  &    0.10  &    12 \\
Co   &     1.53  &   -3.39  &    0.47  &    0.11  &     4 \\
Ni   &     2.24  &   -4.01  &   -0.15  &    0.05  &     3 \\
Zn   &     1.23  &   -3.37  &    0.49  &      -   &     1 \\
\hline
\end {tabular}
\begin {tabular}{|cccccc|}
\multicolumn {4}{c}{CS~22186--025}    [Fe/H]=--3.00\\
\hline
     & $log \varepsilon$& [M/H]&  [M/Fe] & $\sigma$ & N\\
C    &     4.90  &   -3.62  &   -0.62  &      -   &     - \\
N    &       -   &     -    &      -   &      -   &     - \\
O    &     6.36  &   -2.41  &    0.59  &      -   &     - \\
Na   &     3.67  &   -2.66  &    0.34  &      -   &     2 \\
Mg   &     4.94  &   -2.64  &    0.36  &    0.14  &     8 \\
Al   &     2.69  &   -3.78  &   -0.78  &      -   &     2 \\
Si   &     5.00  &   -2.55  &    0.45  &      -   &     1 \\
K    &     2.67  &   -2.45  &    0.55  &      -   &     1 \\
Ca   &     3.71  &   -2.65  &    0.35  &    0.11  &    16 \\
Sc   &     0.32  &   -2.85  &    0.15  &    0.06  &     7 \\
Ti I &     2.34  &   -2.68  &    0.32  &    0.09  &    14 \\
Ti II&     2.32  &   -2.70  &    0.30  &    0.06  &    26 \\
Cr   &     2.33  &   -3.34  &   -0.34  &    0.12  &     7 \\
Mn   &     2.01  &   -3.38  &   -0.38  &    0.01  &     3 \\
Fe I &     4.50  &   -3.00  &    0.00  &    0.14  &   144 \\
Fe II&     4.50  &   -3.00  &    0.00  &    0.09  &    16 \\
Co   &     2.15  &   -2.77  &    0.23  &    0.12  &     4 \\
Ni   &     3.23  &   -3.02  &   -0.02  &    0.02  &     3 \\
Zn   &     1.92  &   -2.68  &    0.32  &      -   &     1 \\
\hline
\end {tabular}
\end {center}
\end{table*}

\begin {table*}[t]
\caption {Abundances of the elements in the selected stars. Na, Al, and 
K are {\it not} corrected for non-LTE effects and O is computed with 
1-D models}
\label {tab-abund-c}
\begin {center}
\begin {tabular}{|cccccc|}
\multicolumn {4}{c}{CS~22189--009}    [Fe/H]=--3.49\\
\hline
     & $log \varepsilon$& [M/H]&  [M/Fe] & $\sigma$ & N\\
C    &     5.30  &   -3.22  &    0.27  &      -   &     - \\
N    &      -    &     -    &     -    &      -   &     - \\
O    &      -    &     -    &     -    &      -   &     - \\
Na   &     2.54  &   -3.79  &   -0.30  &      -   &     2 \\
Mg   &     4.20  &   -3.38  &    0.11  &    0.06  &     7 \\
Al   &     2.06  &   -4.41  &   -0.92  &      -   &     2 \\
Si   &     4.35  &   -3.20  &    0.29  &      -   &     1 \\
K    &     2.03  &   -3.09  &    0.40  &      -   &     1 \\
Ca   &     3.07  &   -3.29  &    0.20  &    0.10  &    15 \\
Sc   &    -0.32  &   -3.49  &    0.00  &    0.08  &     7 \\
Ti I &     1.66  &   -3.36  &    0.13  &    0.16  &    14 \\
Ti II&     1.57  &   -3.45  &    0.04  &    0.11  &    30 \\
Cr   &     1.76  &   -3.91  &   -0.42  &    0.13  &     7 \\
Mn   &     1.65  &   -3.74  &   -0.25  &    0.05  &     3 \\
Fe I &     4.01  &   -3.49  &    0.00  &    0.15  &   150 \\
Fe II&     4.01  &   -3.49  &    0.00  &    0.11  &    11 \\
Co   &     1.80  &   -3.12  &    0.37  &    0.06  &     4 \\
Ni   &     2.82  &   -3.43  &    0.06  &    0.06  &     3 \\
Zn   &     1.57  &   -3.03  &    0.46  &      -   &     1 \\
\hline
\end {tabular}
\begin {tabular}{|cccccc|}
\multicolumn {4}{c}{CS~22873--055}    [Fe/H]=--2.99\\
\hline
     & $log \varepsilon$& [M/H]&  [M/Fe] & $\sigma$ & N\\
C    &     4.56  &   -3.96  &   -0.98  &      -   &     - \\
N    &      -    &     -    &     -    &      -   &     - \\
O    &     6.30  &   -2.47  &    0.52  &      -   &     - \\
Na   &     4.05  &   -2.28  &    0.71  &      -   &     2 \\
Mg   &     5.01  &   -2.57  &    0.42  &    0.14  &     7 \\
Al   &     3.28  &   -3.19  &   -0.21  &      -   &     2 \\
Si   &     4.90  &   -2.65  &    0.34  &      -   &     2 \\
K    &     2.60  &   -2.52  &    0.47  &      -   &     1 \\
Ca   &     3.72  &   -2.64  &    0.35  &    0.09  &    16 \\
Sc   &     0.20  &   -2.97  &    0.02  &    0.05  &     7 \\
Ti I &     2.19  &   -2.83  &    0.16  &    0.05  &    13 \\
Ti II&     2.17  &   -2.85  &    0.14  &    0.11  &    29 \\
Cr   &     2.32  &   -3.35  &   -0.37  &    0.09  &     7 \\
Mn   &     1.98  &   -3.41  &   -0.43  &    0.05  &     3 \\
Fe I &     4.52  &   -2.98  &    0.00  &    0.14  &   145 \\
Fe II&     4.51  &   -2.99  &    0.00  &    0.11  &    17 \\
Co   &     2.17  &   -2.75  &    0.24  &    0.09  &     4 \\
Ni   &     3.30  &   -2.95  &    0.04  &    0.05  &     3 \\
Zn   &     1.87  &   -2.73  &    0.26  &     -    &     1 \\
\hline
\end {tabular}
\begin {tabular}{|cccccc|}
\multicolumn {4}{c}{CS~22873--166}   [Fe/H]=--2.97\\
\hline
     & $log \varepsilon$& [M/H]&  [M/Fe] & $\sigma$ & N\\
C    &     5.40  &   -3.12  &   -0.15  &      -   &     - \\
N    &     6.00  &   -1.92  &    1.05  &      -   &     - \\
O    &       -   &     -    &      -   &      -   &     - \\
Na   &     3.69  &   -2.64  &    0.33  &      -   &     2 \\
Mg   &     5.14  &   -2.44  &    0.52  &    0.18  &     7 \\
Al   &     3.11  &   -3.36  &   -0.40  &      -   &     2 \\
Si   &     5.00  &   -2.55  &    0.42  &      -   &     1 \\
K    &     2.66  &   -2.46  &    0.51  &      -   &     1 \\
Ca   &     3.76  &   -2.60  &    0.36  &    0.10  &    16 \\
Sc   &     0.28  &   -2.89  &    0.08  &    0.10  &     7 \\
Ti I &     2.25  &   -2.77  &    0.20  &    0.09  &    14 \\
Ti II&     2.25  &   -2.77  &    0.20  &    0.13  &    30 \\
Cr   &     2.36  &   -3.47  &   -0.51  &    0.12  &     7 \\
Mn   &     1.92  &   -3.28  &   -0.32  &    0.06  &     3 \\
Fe I &     4.58  &   -2.92  &    0.04  &    0.19  &   143 \\
Fe II&     4.49  &   -3.01  &   -0.04  &    0.11  &    18 \\
Co   &     2.12  &   -2.80  &    0.17  &    0.13  &     4 \\
Ni   &     3.20  &   -3.05  &   -0.09  &    0.10  &     3 \\
Zn   &     1.81  &   -2.79  &    0.18  &      -   &     1 \\
\hline
\end {tabular}
\begin {tabular}{|cccccc|}
\multicolumn {4}{c}{CS~22878--101}   [Fe/H]=--3.25\\
\hline
     & $log \varepsilon$& [M/H]&  [M/Fe] & $\sigma$ & N\\
C    &     5.00  &     -    &      -   &      -   &     - \\
N    &       -   &     -    &      -   &      -   &     - \\
O    &       -   &     -    &      -   &      -   &     - \\
Na   &     3.22  &   -3.11  &    0.14  &      -   &     2 \\
Mg   &     4.77  &   -2.81  &    0.44  &    0.11  &     7 \\
Al   &     2.46  &   -4.01  &   -0.76  &      -   &     2 \\
Si   &     4.65  &   -2.90  &    0.35  &      -   &     1 \\
K    &           &     -    &          &      -   &     0 \\
Ca   &     3.47  &   -2.89  &    0.36  &    0.16  &    17 \\
Sc   &     0.04  &   -3.13  &    0.12  &    0.03  &     7 \\
Ti I &     2.06  &   -2.96  &    0.29  &    0.14  &    14 \\
Ti II&     2.02  &   -3.00  &    0.25  &    0.09  &    28 \\
Cr   &     2.01  &   -3.66  &   -0.41  &    0.13  &     7 \\
Mn   &     1.60  &   -3.79  &   -0.54  &    0.12  &     3 \\
Fe I &     4.21  &   -3.29  &   -0.04  &    0.12  &   144 \\
Fe II&     4.29  &   -3.21  &    0.04  &    0.14  &    17 \\
Co   &     1.89  &   -3.03  &    0.22  &    0.07  &     4 \\
Ni   &     2.77  &   -3.48  &   -0.23  &    0.10  &     3 \\
Zn   &     1.75  &   -2.85  &    0.40  &      -   &     1 \\
\hline
\end {tabular}
\begin {tabular}{|cccccc|}
\multicolumn {4}{c}{CS~22885--096}    [Fe/H]=--3.78\\
\hline
     & $log \varepsilon$& [M/H]&  [M/Fe] & $\sigma$ & N\\
C    &     4.95  &   -3.57  &    0.24  &      -   &     - \\
N    &      -    &     -    &     -    &      -   &     - \\
O    &      -    &     -    &     -    &      -   &     - \\
Na   &     2.49  &   -3.84  &   -0.06  &      -   &     2 \\
Mg   &     3.99  &   -3.59  &    0.19  &    0.12  &     7 \\
Al   &     1.90  &   -4.57  &   -0.79  &      -   &     2 \\
Si   &     4.29  &   -3.26  &    0.52  &      -   &     1 \\
K    &     1.68  &   -3.44  &    0.34  &      -   &     2 \\
Ca   &     2.93  &   -3.43  &    0.35  &    0.15  &    12 \\
Sc   &    -0.24  &   -3.41  &    0.37  &    0.08  &     5 \\
Ti I &     1.58  &   -3.44  &    0.34  &    0.07  &     9 \\
Ti II&     1.56  &   -3.46  &    0.32  &    0.11  &    24 \\
Cr   &     1.41  &   -4.26  &   -0.48  &    0.13  &     7 \\
Mn   &     1.07  &   -4.32  &   -0.54  &    0.02  &     3 \\
Fe I &     3.69  &   -3.81  &   -0.03  &    0.17  &   120 \\
Fe II&     3.75  &   -3.75  &    0.03  &    0.14  &     5 \\
Co   &     1.62  &   -3.30  &    0.48  &    0.05  &     4 \\
Ni   &     2.47  &   -3.78  &    0.00  &    0.06   &    3 \\
Zn   &     1.26  &   -3.34  &    0.44  &      -   &     1 \\
\hline
\end {tabular}
\begin {tabular}{|cccccc|}
\multicolumn {4}{c}{CS~22891--209}   [Fe/H]=- -3.29\\
\hline
     & $log \varepsilon$& [M/H]&  [M/Fe] & $\sigma$ & N\\
C    &     4.71  &   -3.81  &   -0.52  &      -   &     - \\
N    &      -    &     -    &     -    &      -   &     - \\
O    &     6.25  &   -2.52  &    0.77  &      -   &     - \\
Na   &     3.55  &   -2.78  &    0.51  &      -   &     2 \\
Mg   &     4.63  &   -2.95  &    0.34  &    0.14  &     8 \\
Al   &     2.75  &   -3.72  &   -0.44  &      -   &     2 \\
Si   &     4.60  &   -2.95  &    0.34  &      -   &     1 \\
K    &     2.35  &   -2.77  &    0.52  &      -   &     2 \\
Ca   &     3.38  &   -2.98  &    0.31  &    0.09  &    16 \\
Sc   &    -0.06  &   -3.23  &    0.06  &    0.05  &     7 \\
Ti I &     1.97  &   -3.05  &    0.24  &    0.04  &    12 \\
Ti II&     1.94  &   -3.08  &    0.21  &    0.12  &    30 \\
Cr   &     2.01  &   -3.66  &   -0.38  &    0.16  &     7 \\
Mn   &     1.73  &   -3.66  &   -0.38  &    0.08  &     3 \\
Fe I &     4.21  &   -3.29  &    0.00  &    0.14  &   145 \\
Fe II&     4.22  &   -3.28  &    0.00  &    0.11  &    18 \\
Co   &     1.83  &   -3.09  &    0.20  &    0.07  &     4 \\
Ni   &     2.98  &   -3.27  &    0.02  &    0.04  &     3 \\
Zn   &     1.76  &   -2.84  &    0.45  &      -   &     1 \\
\hline
\end {tabular}
\end {center}
\end{table*}

\begin {table*}[t]
\caption {Abundances of the elements in the selected stars. Na, Al, and 
K are {\it not} corrected for non-LTE effects and O is computed with 
1-D models}
\label {tab-abund-d}
\begin {center}
\begin {tabular}{|cccccc|}
\multicolumn {4}{c}{CS~22892--052}    [Fe/H]=--3.03\\
\hline
     & $log \varepsilon$& [M/H]&  [M/Fe] & $\sigma$ & N\\
C    &     6.38  &   -2.14  &    0.89  &      -   &     - \\
N    &     5.42  &   -2.50  &    0.53  &      -   &     - \\
O    &     6.21  &   -2.56  &    0.47  &      -   &     - \\
Na   &     3.35  &   -2.98  &    0.05  &      -   &     2 \\
Mg   &     4.77  &   -2.81  &    0.22  &    0.13  &     7 \\
Al   &     2.70  &   -3.77  &   -0.74  &      -   &     1 \\
Si   &     4.95  &   -2.60  &    0.43  &      -   &     1 \\
K    &     2.53  &   -2.59  &    0.44  &      -   &     1 \\
Ca   &     3.65  &   -2.71  &    0.32  &    0.11  &    14 \\
Sc   &     0.12  &   -3.05  &   -0.02  &    0.10  &     6 \\
Ti I &     2.15  &   -2.87  &    0.16  &    0.12  &    14 \\
Ti II&     2.12  &   -2.90  &    0.13  &    0.13  &    28 \\
Cr   &     2.32  &   -3.35  &   -0.32  &    0.14  &     7 \\
Mn   &     1.88  &   -3.51  &   -0.48  &    0.07  &     3 \\
Fe I &     4.46  &   -3.04  &   -0.01  &    0.14  &   141 \\
Fe II&     4.48  &   -3.02  &    0.01  &    0.07  &    18 \\
Co   &     2.00  &   -2.92  &    0.11  &    0.14  &     4 \\
Ni   &     3.01  &   -3.24  &   -0.21  &    0.11  &     3 \\
Zn   &     1.77  &   -2.83  &    0.20  &      -   &     1 \\
\hline
\end {tabular}
\begin {tabular}{|cccccc|}
\multicolumn {4}{c}{CS~22896--154}   [Fe/H]=--2.69\\
\hline
     & $log \varepsilon$& [M/H]&  [M/Fe] & $\sigma$ & N\\
C    &     6.10  &   -2.42  &    0.27  &      -   &      - \\
N    &      -    &     -    &     -    &      -   &      - \\
O    &     7.02  &   -1.75  &    0.94  &      -   &      - \\
Na   &     3.76  &   -2.57  &    0.12  &      -   &      2 \\
Mg   &     4.97  &   -2.61  &    0.08  &    0.12  &      7 \\
Al   &     3.02  &   -3.45  &   -0.76  &      -   &      2 \\
Si   &     5.40  &   -2.15  &    0.54  &      -   &      1 \\
K    &     2.88  &   -2.24  &    0.45  &      -   &      2 \\
Ca   &     4.03  &   -2.33  &    0.36  &    0.14  &     16 \\
Sc   &     0.61  &   -2.56  &    0.13  &    0.08  &      6 \\
Ti I &     2.60  &   -2.42  &    0.27  &    0.06  &     13 \\
Ti II&     2.66  &   -2.36  &    0.33  &    0.12  &     24 \\
Cr   &     2.75  &   -2.92  &   -0.23  &    0.12  &      7 \\
Mn   &     2.12  &   -3.27  &   -0.59  &    0.13  &      3 \\
Fe I &     4.82  &   -2.68  &    0.01  &    0.19  &    141 \\
Fe II&     4.81  &   -2.69  &    0.00  &    0.14  &     16 \\
Co   &     2.61  &   -2.31  &    0.38  &    0.04  &      3 \\
Ni   &     3.51  &   -2.74  &   -0.05  &    0.09  &      3 \\
Zn   &     2.17  &   -2.43  &    0.26  &      -   &      1 \\
\hline
\end {tabular}
\begin {tabular}{|cccccc|}
\multicolumn {4}{c}{CS~22897--008}    [Fe/H]=--3.41\\
\hline
     & $log \varepsilon$& [M/H]&  [M/Fe] & $\sigma$ & N\\
C    &     5.61  &   -2.91  &    0.50  &      -   &     - \\
N    &       -   &     -    &      -   &      -   &     - \\
O    &       -   &     -    &      -   &      -   &     - \\
Na   &     2.76  &   -3.57  &   -0.16  &      -   &     2 \\
Mg   &     4.43  &   -3.15  &    0.26  &    0.12  &     7 \\
Al   &     2.33  &   -4.14  &   -0.73  &      -   &     2 \\
Si   &     4.55  &   -3.00  &    0.41  &      -   &     1 \\
K    &     2.16  &   -2.96  &    0.45  &      -   &     2 \\
Ca   &     3.24  &   -3.12  &    0.29  &    0.11  &    15 \\
Sc   &    -0.22  &   -3.39  &    0.02  &    0.05  &     6 \\
Ti I &     1.88  &   -3.14  &    0.27  &    0.05  &    13 \\
Ti II&     1.79  &   -3.23  &    0.18  &    0.11  &    26 \\
Cr   &     1.88  &   -3.79  &   -0.38  &    0.15  &     7 \\
Mn   &     1.76  &   -3.63  &   -0.22  &    0.08  &     3 \\
Fe I &     4.08  &   -3.42  &   -0.01  &    0.15  &   140 \\
Fe II&     4.10  &   -3.40  &    0.01  &    0.15  &    17 \\
Co   &     1.93  &   -2.99  &    0.42  &    0.11  &     4 \\
Ni   &     2.91  &   -3.34  &    0.07  &    0.12  &     3 \\
Zn   &     1.86  &   -2.74  &    0.67  &      -   &     1 \\
\hline
\end {tabular}
\begin {tabular}{|cccccc|}
\multicolumn {4}{c}{CS~22948--066}    [Fe/H]=--3.14\\
\hline
     & $log \varepsilon$& [M/H]&  [M/Fe] & $\sigma$ & N\\
C    &     5.36  &   -3.16  &   -0.02  &      -   &    - \\
N    &      -    &     -    &     -    &      -   &    - \\
O    &     6.52  &   -2.25  &    0.89  &      -   &    - \\
Na   &     3.34  &   -2.99  &    0.15  &      -   &    2 \\
Mg   &     4.72  &   -2.86  &    0.28  &    0.08  &    7 \\
Al   &     2.50  &   -3.97  &   -0.83  &      -   &    2 \\
Si   &     4.70  &   -2.85  &    0.29  &      -   &    1 \\
K    &     2.43  &   -2.69  &    0.45  &      -   &    1 \\
Ca   &     3.49  &   -2.87  &    0.27  &    0.13  &   17 \\
Sc   &     0.29  &   -2.88  &    0.26  &    0.03  &    7 \\
Ti I &     2.16  &   -2.86  &    0.28  &    0.11  &   14 \\
Ti II&     2.08  &   -2.94  &    0.20  &    0.08  &   28 \\
Cr   &     2.24  &   -3.43  &   -0.29  &    0.11  &    7 \\
Mn   &     2.13  &   -3.26  &   -0.12  &    0.03  &    3 \\
Fe I &     4.35  &   -3.15  &   -0.01  &    0.11  &  145 \\
Fe II&     4.37  &   -3.13  &    0.01  &    0.06  &   15 \\
Co   &     2.20  &   -2.72  &    0.42  &    0.09  &    4 \\
Ni   &     3.25  &   -3.00  &    0.14  &    0.05  &    3 \\
Zn   &     1.83  &   -2.77  &    0.37  &      -   &    1 \\
\hline
\end {tabular}
\begin {tabular}{|cccccc|}
\multicolumn {4}{c}{CS~22949--037}    [Fe/H]=--3.97\\
\hline
     & $log \varepsilon$& [M/H]&  [M/Fe] & $\sigma$ & N\\
C    &     5.72  &   -2.80  &    1.17  &      -   &     - \\
N    &     6.52  &   -1.40  &    2.57  &      -   &     - \\
O    &     6.78  &   -1.99  &    1.98  &      -   &     - \\
Na   &     3.80  &   -2.53  &    1.44  &      -   &     2 \\
Mg   &     5.17  &   -2.41  &    1.56  &    0.19  &     7 \\
Al   &     2.34  &   -4.13  &   -0.17  &      -   &     2 \\
Si   &     4.35  &   -3.20  &    0.77  &      -   &     1 \\
K    &     1.38  &   -3.74  &    0.23  &      -   &     2 \\
Ca   &     2.73  &   -3.63  &    0.34  &    0.17  &    10 \\
Sc   &    -0.70  &   -3.87  &    0.09  &    0.14  &     5 \\
Ti I &     1.40  &   -3.62  &    0.35  &    0.09  &     8 \\
Ti II&     1.41  &   -3.61  &    0.36  &    0.15  &    21 \\
Cr   &     1.29  &   -4.38  &   -0.42  &    0.10  &     5 \\
Mn   &     0.96  &   -4.43  &   -0.47  &     -    &     2 \\
Fe I &     3.51  &   -3.99  &   -0.03  &    0.11  &    64 \\
Fe II&     3.56  &   -3.94  &    0.02  &    0.11  &     6 \\
Co   &     1.28  &   -3.64  &    0.33  &    0.07  &     4 \\
Ni   &     2.19  &   -4.06  &   -0.10  &    0.02  &     3 \\
Zn   &     1.29  &   -3.31  &    0.66  &      -   &     1 \\
\hline
\end {tabular}
\begin {tabular}{|cccccc|}
\multicolumn {4}{c}{CS~22952--015}   [Fe/H]=--3.43\\
\hline
     & $log \varepsilon$& [M/H]&  [M/Fe] & $\sigma$ & N\\
C    &      -    &     -    &     -    &     -    &     - \\
N    &      -    &     -    &     -    &     -    &     - \\
O    &      -    &     -    &     -    &     -    &     - \\
Na   &      -    &     -    &     -    &     -    &     - \\
Mg   &     4.15  &   -3.43  &    0.00  &    0.08  &     8 \\
Al   &     2.58  &   -3.89  &   -0.46  &      -   &     2 \\
Si   &     4.45  &   -3.10  &    0.33  &      -   &     1 \\
K    &     2.18  &   -2.94  &    0.49  &      -   &     1 \\
Ca   &     3.07  &   -3.29  &    0.14  &    0.14  &    16 \\
Sc   &    -0.37  &   -3.54  &   -0.11  &    0.05  &     7 \\
Ti I &     1.69  &   -3.33  &    0.10  &    0.09  &    14 \\
Ti II&     1.59  &   -3.43  &    0.00  &    0.11  &    31 \\
Cr   &     1.78  &   -3.89  &   -0.46  &    0.15  &     7 \\
Mn   &     1.63  &   -3.76  &   -0.33  &    0.16  &     3 \\
Fe I &     4.05  &   -3.45  &   -0.02  &    0.14  &   147 \\
Fe II&     4.09  &   -3.41  &    0.02  &    0.10  &    11 \\
Co   &     1.62  &   -3.30  &    0.13  &    0.10  &     4 \\
Ni   &     2.70  &   -3.55  &   -0.12  &    0.10  &     3 \\
Zn   &     1.42  &   -3.18  &    0.25  &     -    &     1 \\
\hline
\end {tabular}
\end {center}
\end{table*}

\begin {table*}[t]
\caption {Abundances of the elements in the selected stars. Na, Al, and 
K are {\it not} corrected for non-LTE effects and O is computed with 
1-D models}
\label {tab-abund-e}
\begin {center}
\begin {tabular}{|cccccc|}
\multicolumn {4}{c}{CS~22953--003}    [Fe/H]=--2.84\\
\hline
     & $log \varepsilon$& [M/H]&  [M/Fe] & $\sigma$ & N\\
C    &     6.00  &   -2.52  &    0.32  &      -   &     - \\
N    &      -    &     -    &     -    &      -   &     - \\
O    &     6.68  &   -2.09  &    0.75  &      -   &     - \\
Na   &     3.64  &   -2.69  &    0.15  &      -   &     2 \\
Mg   &     4.87  &   -2.71  &    0.13  &    0.09  &     7 \\
Al   &     2.65  &   -3.82  &   -0.99  &      -   &     2 \\
Si   &     4.96  &   -2.59  &    0.25  &      -   &     1 \\
K    &     2.59  &   -2.63  &    0.31  &      -   &     1 \\
Ca   &     3.74  &   -2.62  &    0.22  &    0.14  &    17 \\
Sc   &     0.32  &   -2.85  &   -0.02  &    0.08  &     7 \\
Ti I &     2.33  &   -2.69  &    0.15  &    0.09  &    14 \\
Ti II&     2.29  &   -2.73  &    0.11  &    0.10  &    29 \\
Cr   &     2.43  &   -3.24  &   -0.41  &    0.10  &     7 \\
Mn   &     2.19  &   -3.20  &   -0.37  &    0.05  &     3 \\
Fe I &     4.65  &   -2.85  &   -0.01  &    0.15  &   148 \\
Fe II&     4.68  &   -2.82  &    0.01  &    0.11  &    17 \\
Co   &     2.25  &   -2.67  &    0.17  &    0.10  &     4 \\
Ni   &     3.33  &   -2.92  &   -0.09  &    0.05  &     3 \\
Zn   &     1.91  &   -2.69  &    0.15  &      -   &     1 \\
\hline
\end {tabular}
\begin {tabular}{|cccccc|}
\multicolumn {4}{c}{CS~22956--050}    [Fe/H]=--3.33\\
\hline
     & $log \varepsilon$& [M/H]&  [M/Fe] & $\sigma$ & N\\
C    &     5.46  &   -3.05  &    0.28  &     -    &     - \\
N    &      -    &     -    &     -    &     -    &     - \\
O    &      -    &     -    &     -    &     -    &     - \\
Na   &      -    &     -    &     -    &     -    &     - \\
Mg   &     4.62  &   -2.96  &    0.37  &    0.11  &     8 \\
Al   &     2.49  &   -3.98  &   -0.65  &      -   &     2 \\
Si   &     5.00  &   -2.55  &    0.78  &      -   &     1 \\
K    &     2.08  &   -3.04  &    0.29  &      -   &     1 \\
Ca   &     3.49  &   -2.87  &    0.46  &    0.13  &    17 \\
Sc   &    -0.20  &   -3.37  &   -0.04  &    0.07  &     7 \\
Ti I &     2.04  &   -2.98  &    0.35  &    0.04  &    13 \\
Ti II&     1.98  &   -3.04  &    0.29  &    0.12  &    30 \\
Cr   &     2.00  &   -3.67  &   -0.34  &    0.12  &     7 \\
Mn   &     1.55  &   -3.84  &   -0.51  &    0.05  &     3 \\
Fe I &     4.18  &   -3.32  &    0.01  &    0.13  &   145 \\
Fe II&     4.16  &   -3.34  &   -0.01  &    0.10  &    17 \\
Co   &     1.98  &   -2.94  &    0.39  &    0.10  &     4 \\
Ni   &     2.93  &   -3.32  &    0.01  &    0.03  &     3 \\
Zn   &     1.57  &   -3.03  &    0.30  &      -   &     1 \\
\hline
\end {tabular}
\begin {tabular}{|cccccc|}
\multicolumn {4}{c}{CS~22966--057}    [Fe/H]=--2.62\\
\hline
     & $log \varepsilon$& [M/H]&  [M/Fe] & $\sigma$ & N\\
C    &     5.96  &   -2.56  &    0.06  &      -   &     - \\
N    &      -    &     -    &     -    &      -   &     - \\
O    &     7.14  &   -1.63  &    0.99  &      -   &     - \\
Na   &     4.18  &   -2.15  &    0.47  &      -   &     2 \\
Mg   &     5.08  &   -2.50  &    0.12  &    0.20  &     7 \\
Al   &     2.99  &   -3.48  &   -0.86  &      -   &     2 \\
Si   &     5.50  &   -2.05  &    0.57  &      -   &     1 \\
K    &     2.91  &   -2.21  &    0.41  &      -   &     2 \\
Ca   &     4.10  &   -2.26  &    0.36  &    0.10  &    16 \\
Sc   &     0.61  &   -2.56  &    0.06  &    0.08  &     7 \\
Ti I &     2.71  &   -2.31  &    0.31  &    0.04  &    12 \\
Ti II&     2.69  &   -2.33  &    0.29  &    0.08  &    27 \\
Cr   &     2.86  &   -2.81  &   -0.20  &    0.06  &     7 \\
Mn   &     2.33  &   -3.06  &   -0.45  &    0.06  &     3 \\
Fe I &     4.89  &   -2.61  &    0.00  &    0.12  &   148 \\
Fe II&     4.88  &   -2.62  &    0.00  &    0.12  &    16 \\
Co   &     2.67  &   -2.25  &    0.37  &    0.06  &     4 \\
Ni   &     3.76  &   -2.49  &    0.13  &    0.01  &     3 \\
Zn   &     2.24  &   -2.36  &    0.26  &      -   &     1 \\
\hline
\end {tabular}
\begin {tabular}{|cccccc|}
\multicolumn {4}{c}{CS~22968--014}    [Fe/H]=--3.56\\
\hline
     & $log \varepsilon$& [M/H]&  [M/Fe] & $\sigma$ & N\\
C    &     5.26  &   -3.26  &    0.30  &      -   &     - \\
N    &       -   &     -    &      -   &      -   &     - \\
O    &     6.11  &   -2.66  &    0.90  &      -   &     - \\
Na   &     2.41  &   -3.92  &   -0.37  &      -   &     2 \\
Mg   &     4.22  &   -3.36  &    0.19  &    0.13  &     8 \\
Al   &     2.13  &   -4.34  &   -0.79  &      -   &     2 \\
Si   &     4.20  &   -3.35  &    0.21  &      -   &     1 \\
K    &     1.78  &   -3.34  &    0.22  &      -   &     2 \\
Ca   &     2.83  &   -3.53  &    0.02  &    0.17  &    17 \\
Sc   &    -0.35  &   -3.52  &    0.03  &    0.07  &     7 \\
Ti I &     1.57  &   -3.45  &    0.11  &    0.12  &    14 \\
Ti II&     1.47  &   -3.55  &    0.00  &    0.12  &    28 \\
Cr   &     1.66  &   -4.01  &   -0.46  &    0.16  &     7 \\
Mn   &     1.60  &   -3.79  &   -0.24  &    0.05  &     3 \\
Fe I &     3.93  &   -3.57  &   -0.02  &    0.15  &   151 \\
Fe II&     3.96  &   -3.54  &    0.01  &    0.10  &    10 \\
Co   &     1.92  &   -3.00  &    0.56  &    0.11  &     4 \\
Ni   &     2.91  &   -3.34  &    0.22  &    0.05  &     3 \\
Zn   &     1.46  &   -3.14  &    0.42  &      -   &     1 \\
\hline
\end {tabular}
\begin {tabular}{|cccccc|}
\multicolumn {4}{c}{CS~29491--053}    [Fe/H]=--3.04\\
\hline
     & $log \varepsilon$& [M/H]&  [M/Fe] & $\sigma$ & N\\
C    &     5.20  &   -3.32  &   -0.28  &      -   &     - \\
N    &       -   &     -    &      -   &      -   &     - \\
O    &     6.49  &   -2.28  &    0.76  &      -   &     - \\
Na   &     3.54  &   -2.79  &    0.25  &      -   &     2 \\
Mg   &     5.07  &   -2.51  &    0.53  &    0.16  &     6 \\
Al   &     2.94  &   -3.53  &   -0.50  &      -   &     2 \\
Si   &     5.05  &   -2.50  &    0.54  &      -   &     2 \\
K    &     2.66  &   -2.46  &    0.58  &      -   &     1 \\
Ca   &     3.72  &   -2.64  &    0.40  &    0.10  &    16 \\
Sc   &     0.25  &   -2.92  &    0.12  &    0.08  &     7 \\
Ti I &     2.22  &   -2.80  &    0.24  &    0.08  &    14 \\
Ti II&     2.24  &   -2.78  &    0.26  &    0.10  &    29 \\
Cr   &     2.22  &   -3.45  &   -0.42  &    0.13  &     7 \\
Mn   &     1.83  &   -3.56  &   -0.53  &    0.06  &     3 \\
Fe I &     4.47  &   -3.03  &    0.00  &    0.16  &   146 \\
Fe II&     4.46  &   -3.04  &    0.00  &    0.09  &    17 \\
Co   &     2.07  &   -2.85  &    0.19  &    0.12  &     4 \\
Ni   &     3.11  &   -3.14  &   -0.11  &    0.08  &     3 \\
Zn   &     1.83  &   -2.77  &    0.27  &      -   &     1 \\
\hline
\end {tabular}
\begin {tabular}{|cccccc|}
\multicolumn {4}{c}{CS~29495--041}    [Fe/H]=--2.82\\
\hline
     & $log \varepsilon$& [M/H]&  [M/Fe] & $\sigma$ & N\\
C    &     5.66  &   -2.86  &   -0.04  &      -   &     - \\
N    &       -   &     -    &      -   &      -   &     - \\
O    &     6.63  &   -2.14  &    0.68  &      -   &     - \\
Na   &     3.75  &   -2.58  &    0.24  &      -   &     2 \\
Mg   &     5.09  &   -2.49  &    0.33  &    0.13  &     7 \\
Al   &     2.99  &   -3.48  &   -0.66  &      -   &     2 \\
Si   &     5.30  &   -2.25  &    0.57  &      -   &     1 \\
K    &     2.84  &   -2.28  &    0.54  &      -   &     2 \\
Ca   &     3.92  &   -2.44  &    0.38  &    0.09  &    16 \\
Sc   &     0.46  &   -2.71  &    0.11  &    0.09  &     7 \\
Ti I &     2.43  &   -2.59  &    0.23  &    0.04  &    12 \\
Ti II&     2.44  &   -2.58  &    0.24  &    0.09  &    29 \\
Cr   &     2.50  &   -3.17  &   -0.35  &    0.07  &     7 \\
Mn   &     2.08  &   -3.31  &   -0.49  &    0.07  &     3 \\
Fe I &     4.69  &   -2.81  &    0.01  &    0.15  &   147 \\
Fe II&     4.67  &   -2.83  &   -0.01  &    0.10  &    17 \\
Co   &     2.32  &   -2.60  &    0.22  &    0.14  &     4 \\
Ni   &     3.38  &   -2.87  &   -0.05  &    0.02  &     3 \\
Zn   &     1.93  &   -2.67  &    0.15  &      -   &     1 \\
\hline
\end {tabular}
\end {center}
\end{table*} 

\begin {table*}[t]
\caption {Abundances of the elements in the selected stars. Na, Al, and 
K are {\it not} corrected for non-LTE effects and O is computed with 
1-D models}
\label {tab-abund-f}
\begin {center}
\begin {tabular}{|cccccc|}
\multicolumn {4}{c}{CS~29502--042}    [Fe/H]=--3.19\\
\hline
     & $log \varepsilon$& [M/H]&  [M/Fe] & $\sigma$ & N\\
C    &     5.56  &   -2.96  &    0.23  &      -   &     - \\
N    &       -   &     -    &      -   &      -   &     - \\
O    &       -   &     -    &      -   &      -   &     - \\
Na   &     2.69  &   -3.64  &   -0.45  &      -   &     2 \\
Mg   &     4.62  &   -2.96  &    0.23  &    0.10  &     8 \\
Al   &     2.47  &   -4.00  &   -0.81  &      -   &     2 \\
Si   &     4.65  &   -2.90  &    0.29  &      -   &     1 \\
K    &     2.23  &   -2.89  &    0.30  &      -   &     1 \\
Ca   &     3.39  &   -2.97  &    0.22  &    0.13  &    17 \\
Sc   &     0.17  &   -3.00  &    0.19  &    0.13  &     7 \\
Ti I &     2.09  &   -2.93  &    0.26  &    0.06  &    13 \\
Ti II&     2.07  &   -2.95  &    0.24  &    0.09  &    30 \\
Cr   &     2.11  &   -3.56  &   -0.37  &    0.09  &     7 \\
Mn   &     1.64  &   -3.75  &   -0.56  &    0.01  &     3 \\
Fe I &     4.31  &   -3.19  &    0.00  &    0.11  &   150 \\
Fe II&     4.31  &   -3.19  &    0.00  &    0.15  &    11 \\
Co   &     2.09  &   -2.83  &    0.36  &    0.07  &     4 \\
Ni   &     3.00  &   -3.25  &   -0.06  &    0.07  &     3 \\
Zn   &     1.60  &   -3.00  &    0.19  &      -   &     1 \\
\hline
\end {tabular}
\begin {tabular}{|cccccc|}
\multicolumn {4}{c}{CS~29516--024}    [Fe/H]=--3.06\\
\hline
     & $log \varepsilon$& [M/H]&  [M/Fe] & $\sigma$ & N\\
C    &     5.45  &   -3.07  &   -0.01  &      -   &     - \\
N    &       -   &     -    &      -   &      -   &     - \\
O    &     6.33  &   -2.44  &    0.62  &      -   &     - \\
Na   &     3.63  &   -2.70  &    0.36  &      -   &     2 \\
Mg   &     5.00  &   -2.58  &    0.48  &    0.11  &     7 \\
Al   &     2.82  &   -3.65  &   -0.59  &      -   &     2 \\
Si   &     5.05  &   -2.50  &    0.56  &      -   &     1 \\
K    &       -   &     -    &      -   &      -   &     0 \\
Ca   &     3.85  &   -2.51  &    0.55  &    0.10  &    16 \\
Sc   &     0.17  &   -3.00  &    0.06  &    0.10  &     7 \\
Ti I &     2.21  &   -2.81  &    0.25  &    0.03  &    12 \\
Ti II&     2.18  &   -2.84  &    0.22  &    0.09  &    29 \\
Cr   &     2.30  &   -3.37  &   -0.31  &    0.08  &     7 \\
Mn   &     1.89  &   -3.50  &   -0.44  &    0.05  &     3 \\
Fe I &     4.43  &   -3.07  &   -0.01  &    0.10  &   140 \\
Fe II&     4.45  &   -3.05  &    0.01  &    0.15  &    16 \\
Co   &     2.11  &   -2.81  &    0.25  &    0.10  &     4 \\
Ni   &     3.04  &   -3.21  &   -0.15  &    0.15  &     3 \\
Zn   &     1.74  &   -2.86  &    0.20  &      -   &     1 \\
\hline
\end {tabular}
\begin {tabular}{|cccccc|}
\multicolumn {4}{c}{CS~29518--051}    [Fe/H]=--2.78\\
\hline
     & $log \varepsilon$& [M/H]&  [M/Fe] & $\sigma$ & N\\
C    &     5.50  &   -3.02  &   -0.27  &      -   &     - \\
N    &       -   &     -    &      -   &      -   &     - \\
O    &     6.88  &   -1.89  &    0.89  &      -   &     - \\
Na   &     3.85  &   -2.48  &    0.30  &      -   &     2 \\
Mg   &     5.01  &   -2.57  &    0.21  &    0.11  &     7 \\
Al   &     2.88  &   -3.59  &   -0.82  &      -   &     2 \\
Si   &     5.28  &   -2.27  &    0.51  &      -   &     1 \\
K    &     2.76  &   -2.36  &    0.42  &      -   &     2 \\
Ca   &     3.99  &   -2.37  &    0.41  &    0.09  &    16 \\
Sc   &     0.50  &   -2.67  &    0.11  &    0.08  &     7 \\
Ti I &     2.55  &   -2.47  &    0.31  &    0.04  &    12 \\
Ti II&     2.61  &   -2.41  &    0.37  &    0.09  &    30 \\
Cr   &     2.64  &   -3.03  &   -0.25  &    0.09  &     7 \\
Mn   &     2.28  &   -3.11  &   -0.34  &    0.04  &     3 \\
Fe I &     4.75  &   -2.75  &    0.03  &    0.13  &   143 \\
Fe II&     4.70  &   -2.80  &   -0.02  &    0.13  &    10 \\
Co   &     2.44  &   -2.48  &    0.30  &    0.08  &     4 \\
Ni   &     3.53  &   -2.72  &    0.06  &    0.03  &     3 \\
Zn   &     2.10  &   -2.50  &    0.28  &      -   &     1 \\
\hline
\end {tabular}
\begin {tabular}{|cccccc|}
\multicolumn {4}{c}{CS~30325--094}    [Fe/H]=--3.30\\
\hline
     & $log \varepsilon$& [M/H]&  [M/Fe] & $\sigma$ & N\\
C    &     5.20  &   -3.32  &   -0.02  &      -   &     - \\
N    &       -   &     -    &      -   &      -   &     - \\
O    &     6.19  &   -2.58  &    0.72  &      -   &     - \\
Na   &     3.12  &   -3.21  &    0.09  &      -   &     2 \\
Mg   &     4.66  &   -2.92  &    0.38  &    0.14  &     7 \\
Al   &     2.59  &   -3.88  &   -0.58  &      -   &     2 \\
Si   &     5.06  &   -2.49  &    0.81  &      -   &     1 \\
K    &     2.55  &   -2.57  &    0.73  &      -   &     1 \\
Ca   &     3.44  &   -2.92  &    0.38  &    0.11  &    17 \\
Sc   &     0.20  &   -2.97  &    0.33  &    0.09  &     7 \\
Ti I &     1.99  &   -3.03  &    0.27  &    0.05  &    13 \\
Ti II&     2.01  &   -3.01  &    0.29  &    0.10  &    29 \\
Cr   &     1.93  &   -3.74  &   -0.44  &    0.09  &     7 \\
Mn   &     1.54  &   -3.85  &   -0.55  &    0.05  &     3 \\
Fe I &     4.20  &   -3.30  &    0.00  &    0.14  &   140 \\
Fe II&     4.20  &   -3.30  &    0.01  &    0.13  &     8 \\
Co   &     1.91  &   -3.01  &    0.29  &    0.08  &     3 \\
Ni   &     3.03  &   -3.22  &    0.08  &    0.04  &     3 \\
Zn   &     1.54  &   -3.06  &    0.24  &      -   &     1 \\
\hline
\end {tabular}
\begin {tabular}{|cccccc|}
\multicolumn {4}{c}{CS~31082--001}    [Fe/H]=--2.91\\
\hline
     & $log \varepsilon$& [M/H]&  [M/Fe] & $\sigma$ & N\\
C    &     5.82  &   -2.70  &    0.21  &      -   &     - \\
N    &     5.22  &  <-2.70  &   <0.21  &      -   &     - \\
O    &     6.46  &   -2.31  &    0.60  &      -   &     - \\
Na   &     3.70  &   -2.63  &    0.28  &      -   &     2 \\
Mg   &     5.04  &   -2.54  &    0.37  &    0.13  &     7 \\
Al   &     2.83  &   -3.64  &   -0.73  &      -   &     1 \\
Si   &     4.89  &   -2.66  &    0.25  &      -   &     1 \\
K    &     2.87  &   -2.25  &    0.66  &      -   &     2 \\
Ca   &     3.87  &   -2.49  &    0.42  &    0.11  &    15 \\
Sc   &     0.28  &   -2.89  &    0.02  &    0.07  &     7 \\
Ti I &     2.37  &   -2.65  &    0.26  &    0.09  &    14 \\
Ti II&     2.43  &   -2.59  &    0.32  &    0.14  &    28 \\
Cr   &     2.43  &   -3.24  &   -0.33  &    0.11  &     7 \\
Mn   &     1.98  &   -3.41  &   -0.50  &    0.03  &     3 \\
Fe I &     4.60  &   -2.90  &   -0.01  &    0.13  &   120 \\
Fe II&     4.58  &   -2.92  &   -0.01  &    0.11  &    18 \\
Co   &     2.28  &   -2.64  &    0.27  &    0.11  &     4 \\
Ni   &     3.37  &   -2.88  &    0.03  &    0.02  &     3 \\
Zn   &     1.88  &   -2.72  &    0.19  &      -   &     2 \\
\hline
\end {tabular}
\end {center}
\end{table*}

\end {document}